\documentclass[fleqn,usenatbib]{mnras}

\usepackage[T1]{fontenc}

\DeclareRobustCommand{\VAN}[3]{#2}
\let\VANthebibliography\thebibliography
\def\thebibliography{\DeclareRobustCommand{\VAN}[3]{##3}\VANthebibliography}

\usepackage{graphicx}
\usepackage{float}
\usepackage{amsmath}	

\usepackage{threeparttable}
\graphicspath{{img_arxiv/}}
\usepackage[normalem]{ulem}

\usepackage{xcolor}

\newcommand{\cm}{cm$^{-1}$}

\newcommand{\ket}[1]{|#1\rangle}

\newcommand{\um}{$\mu$m}

\newcommand{\exocross}{\textsc{ExoCross}}

\newcommand{\Duo}{{\sc Duo}}

\newcommand{\LEVEL}{{\sc LEVEL}}

\newcommand{\ai}{\textit{ab initio}}

\newcommand{\XS}{$X\,{}^{1}\Sigma^{+}$}
\newcommand{\AS}{$A\,{}^{1}\Pi$}
\newcommand{\oneS}{$1\,{}^{1}\Pi$}

\newcommand{\AX}{\mbox{${A}\,^{1}\Pi-{X}\,^{1}\Sigma^{+}$}}

\newcommand{\X}{\mbox{${X}\,^{1}\Sigma^{+}$}}
\newcommand{\A}{\mbox{${A}\,^{1}\Pi$}}

\newcommand{\alh}[1]{$^{#1}$AlH}
\newcommand{\ald}[1]{$^{#1}$AlD}

\newcommand{\name}{AloHa}

\title[ExoMol line lists -- {LIV}. AlH]{ExoMol line lists -- {LIV}: Empirical line lists for AlH and AlD and experimental emission spectroscopy of AlD in $A$~$^1\Pi$  ($v=0, 1, 2$)}

\author[Yurchenko et al.]{
Sergei N. Yurchenko,$^{1}$
Wojciech Szajna,$^{2}$
Rafa\l\ Hakalla,$^{2}$
Mikhail Semenov,$^{1}$
Andrei Sokolov,$^{1}$
\newauthor{
Jonathan Tennyson,$^{1}$\thanks{The corresponding author: j.tennyson@ucl.ac.uk}
Robert R. Gamache,$^{3}$
Yakiv Pavlenko,$^{4,5}$
Mirek R. Schmidt$^6$}
\vspace*{4mm}\
\\
$^1$ Department of Physics and Astronomy, University College London, Gower Street, WC1E 6BT London, UK\\
$^2$ Materials Spectroscopy Laboratory, Institute of Physics, University of Rzesz\'{o}w, Pigonia 1 Street, 35-310 Rzesz\'{o}w, Poland\\
$^3$Department of Environmental, Earth, and Atmospheric Sciences, University of Massachusetts Lowell, Lowell, MA 01854 USA\\
$^4$ Instituto de Astrof\'isica de Canarias (IAC), Calle V\'ia L\'actea s/n, E-38200 La Laguna, Tenerife, Spain \\
$^5$ Main Astronomical Observatory, Academy of Sciences of the Ukraine, 27 Zabolotnoho, Kyiv 03143, Ukraine\\
$^6$ Nicolaus Copernicus Astronomical Center, Polish Academy of Sciences, Rabianska 8, PL-87-100 Toru{\'n}, Poland}

\date{Accepted XXXX. Received XXXX; in original form XXXX}

\date{\today}

\begin{document}

\label{firstpage}

\maketitle

\pagerange{\pageref{firstpage}--\pageref{lastpage}}

\begin{abstract}

New ExoMol line lists \name\ for AlH and AlD are presented improving the previous line lists WYLLoT (Yurchenko et al., MNRAS 479, 1401 (2018)). The revision is motivated by the recent experimental measurements and astrophysical findings involving the highly excited rotational states of AlH in its \AX\ system. A new high-resolution emission spectrum of ten bands from the \AX\ system of AlD, in the region $17300 - 32000$~cm$^{-1}$ was recorded with a Fourier transform spectrometer, which probes the predissociative $A\,^1\Pi$  $v=2$ state. The AlD new line positions are combined with all available experimental data on AlH and AlD  to construct a comprehensive set of empirical rovibronic energies  of AlH and AlD covering the $X\,^1\Sigma^+$ and $A\,^1\Pi$  electronic states using the MARVEL approach. We then refine the spectroscopic model WYLLoT to our  experimentally derived energies using the nuclear-motion code \textsc{Duo} and use this fit to produce improved line lists for $^{27}$AlH,  $^{27}$AlD and $^{26}$AlH with a better coverage of the rotationally excited states of $A\,^1\Pi$ in the predissociative energy region. The lifetimes of the predissociative states are estimated and are included in the line list using the new ExoMol data structure, alongside the temperature-dependent continuum contribution to the photo-absorption spectra of AlH. The new line lists are  shown to  reproduce the experimental spectra of both AlH and AlD well, and to describe the AlH absorption in the recently reported Proxima Cen spectrum, including the strong predissociative line broadening. The line lists are included into the ExoMol database \url{www.exomol.com}.

\end{abstract}

\begin{keywords}
line: profiles - molecular data - exoplanets - stars: atmospheres - stars: low-mass
\end{keywords}



\section{Introduction}

Aluminium hydride (AlH) has been been observed in the Mira-variable $o$ Ceti \citep{16KaWoSc.AlH}, in the  photospheres of $\chi$ Cygni, a Mira-variable S-star \citep{56Herbig.AlH} as well as in the spectrum of Proxima Cen \citep{jt874}.

Accurate ExoMol line lists, called WYLLoT, for AlH and AlD  were  reported by \citet{jt732} to cover transitions within the \XS\ and \AS\ systems. These  line lists  were  included into a number of atmospheric studies of exoplanets \citep{20ChMiKa.AlO,21BrVaCh.CrH,23RaBuMe,23ZiMiBu}
and opacity compilations ÆSOPUS \citep{22MaBeGi}, ExoMolOP \citep{jt801}, ARCiS \cite{22ChMi}, EXOPLINES \citep{21GhIyLi}, HELIOS-K \citep{jt819}, Stellar studies \citep{22LyYuPa.NaH,jt874,23SiSrSh.AlH}.
AlH is yet to be observed in exoplanetary atmospheres.

The WYLLoT line lists (also known as AlHambra on the ExoMol website) were based on empirical potential energy curves (PECs), Born-Oppenheimer breakdown (BOB) curves, electronic angular momentum curves (EAMC) and \ai\ (transition) dipole moment curves that made up the WYLLoT spectroscopic model. The PECs, EAMCs and BOBs curves were obtained by fitting to experimental data on AlH and AlD collected by \citet{jt732}, who also provide a detailed  review of the literature on AlH spectroscopy  up to 2018. The AlH and AlD curves were fitted separately.

Very recently, the AlH WYLLoT line list was used to identify AlH lines in the spectra of cool star Proxima Centauri (M6 V) by \citet{jt874}. This study showed the limitations of WYLLoT for description of the high $J$ predissociative  states of AlH ($J>8$) in the  \AX\ ($v'=1$) system as well as the associated transitions in the \AS--\XS\ $(v'=0,1)$. In particular, the lines $J'>9$, $v'=1$, \AS--\XS, which appeared increasingly shifted, were also increasingly  broadened through the predissociation of \AS\ in the spectrum of Proxima Centauri thus indicating that an additional mechanism to describe the predisssociation in AlH is required in addition to the radiative, Doppler and collisional effects, included in WYLLoT spectra simulations. The limitations in the accuracy of the line positions of these lines were attributed to the limitations of the underlying experimental data used in WYLLoT, while the limitations of the  WYLLoT line shapes are due to the absence of the predissociative effects in the model. Significantly, \citet{jt874} were unable to establish the abundance of AlH in Proxima Centauri using standard bound-bound transitions as they were all saturated, and it was only by using the heavily-broadened predissociative transitions was it possible to retrieve abundances. Up until now ExoMol line lists have lacked any information on line broadening due to predissociation; this has necessitated development of a new data model \citep{jt898} allowing inclusion of  predissociation into the ExoMol data base. This paper presents our first calculations of lifetime broadening due to predissociaiton. It should be noted that state-resolved  photo-dissociation cross sections of AlH were recently computed \ai\ by \citet{21QiBaLi.AlH}
using an  \ai\ icMRCI+Q model.

Another  key, recent  study for this work is by \citet{23SzKePa.AlH}, who reported an extended  high-resolution Fourier transform-visible (FT-VIS) spectrum of  \AS\ -- \XS\ system ($v'=0, 1$) now  covering rotational excitations up to $J' = 20$ ($v'=0$) and $J'\le 9$ ($v'=1$). 

Apart from the rotational excitations in the \AX\ system of AlH, this work  also aims to improve the description of the vibrational excitations in the \AS\ state. To this end, here we present a new high-resolution emission study of ten bands of the AlD in the \AX\  system recorded with a Fourier transform spectrometer  with the (2--1), (2--2) bands reported for the first time. The previous (lower resolution) conventional studies of these bands go back to  \citet{34HoHuxx.AlH}  and  \citet{48Nilsson.AlH}, which were not included into the WYLLoT study due to their limited quality.  As a result of this exclusion, the $v'=2$ quasi-bound state of AlD was not predicted by the  WYLLoT model at all. Being quasi-bound and predissociative, the \AS\ $(v=2)$ vibronic level is especially important for modelling the \AS\ state as it samples higher energies of the  AlH potential energy curve (PEC), closer to the potential barrier.

Here we use the extended experimental data of \alh{27} and \ald{27} to improve the spectroscopic model WYLLoT for AlH and AlD and to produce new high-temperature line lists for \alh{27}, \alh{26}   and \ald{27} which accurately represent the vibrational
states of the shallow \AS\ state: $v =0,1$ for AlH and $v =0,1,2$ for AlD. Special attention is paid to the treatment of the predissociative states of AlD and AlH and the reproduction of the experimental predissociative spectra and lifetimes. We also compute a pure continuum contribution to the photo-absorption spectrum of AlH and AlD, which is included into the line list data following the recently proposed extension of the ExoMol data format \citep{jt898}.

This work illustrates the importance of experimental data for characterising complex potential energy curves, especially those with low dissociation limits or barriers, where extrapolations of the model can lead to inadequate or incorrect results.

\section{Experimental information}\label{experimental}

High-resolution emission spectra of the AlD, \AX\ system were observed in the the $17300 - 32000$~\cm\ region using a Fourier transform spectrometer (Bruker IFS-125HR) installed at the University of Rzesz\'{o}w~\citep{16NiHaTr.CO,17HaNiFi.CO} and operated in vacuum conditions $(p<0.01~\textrm{hPa})$. A water-cooled discharge lamp equipped with an aluminum hollow-cathode~\citep{23SzKePa.AlH}, filled with a mixture of Ne gas (2.5 Torr) and trace amount of ND$_3$ (0.5 Torr), was used to produce the spectrum of AlD. The lamp was operating at 1~kV and 200~mA DC. The (0-0), (0-1), (0-2), (1-0), (1-1), (1-2), (1-3), (1-4), (2-1), (2-2) bands were recorded with an instrumental resolution of 0.03 \cm\ and the best signal-to-noise ratio (SNR) ca. 4000:1 for the strongest (0-0) band. In contrast to our previous studies on AlD, \AX\ system~\citep{15SzZaHa.AlH,17SzMoLa.AlH} four new bands (0-2), (1-4), (2-1), (2-2) were recorded.

About 50 standard Ne line positions~\citep{83PaEnxx} were used for the frequency axis calibration. A linear calibration function of 0.999999697x + 0.002987348 was used. The absolute accuracy of the calibration ($\textrm{U}_{\textrm{cal.}}$) was estimated as $0.0020$~\cm, and is limited by the accuracy of the Ne line measurements~\citep{83PaEnxx}. Molecular line positions were determined by fitting Voigt profiles to each measured contour using commercial Bruker software Opus\textsuperscript{TM}~\citep{19opusbruker}.  Line position uncertainties $(\textrm{U}_{\textrm{fitt.}})$ were evaluated using an empirical relation similar to that given by~\citet{87Brault}
\begin{equation*}
\textrm{U}_{\textrm{fitt.}}=\frac{f}{\sqrt{\textrm{N}}}\cdot\frac{\textrm{FWHM}}{\textrm{SNR}},
\end{equation*}
where $f=1$ is used for the Voigt profile, FWHM is the full-width at half-maximum of the line, $N$ is the true number of statistically independent points in a line width (taking into account the zero filling factor
commonly used to interpolate FT spectra), and SNR is the signal-to-noise ratio. The profile fitting uncertainty was significantly smaller than $0.0020$~\cm\ for the lines with a typical FWHM~ca.~0.08~\cm\ and a $\textrm{SNR}\geq50$.

Almost 500 rovibronic frequencies of AlD, \AX\ system bands were measured, see Tables~\ref{t:FTSwavenumbers1} and~\ref{t:FTSwavenumbers2} in Appendix. The total uncertainty in the measured line positions $(\textrm{U})$, calculated from the $\textrm{U}=\sqrt{\textrm{U}_{\textrm{cal.}}^2+\textrm{U}_{\textrm{fitt.}}^2}$ relation, is about $0.0020$~\cm\ for most strong and isolated lines. However, accuracy is lower for the a few weakest and/or blended ones.

\subsection{Description of the spectra}

\begin{figure}
    \centering
    \includegraphics[width=1\linewidth]{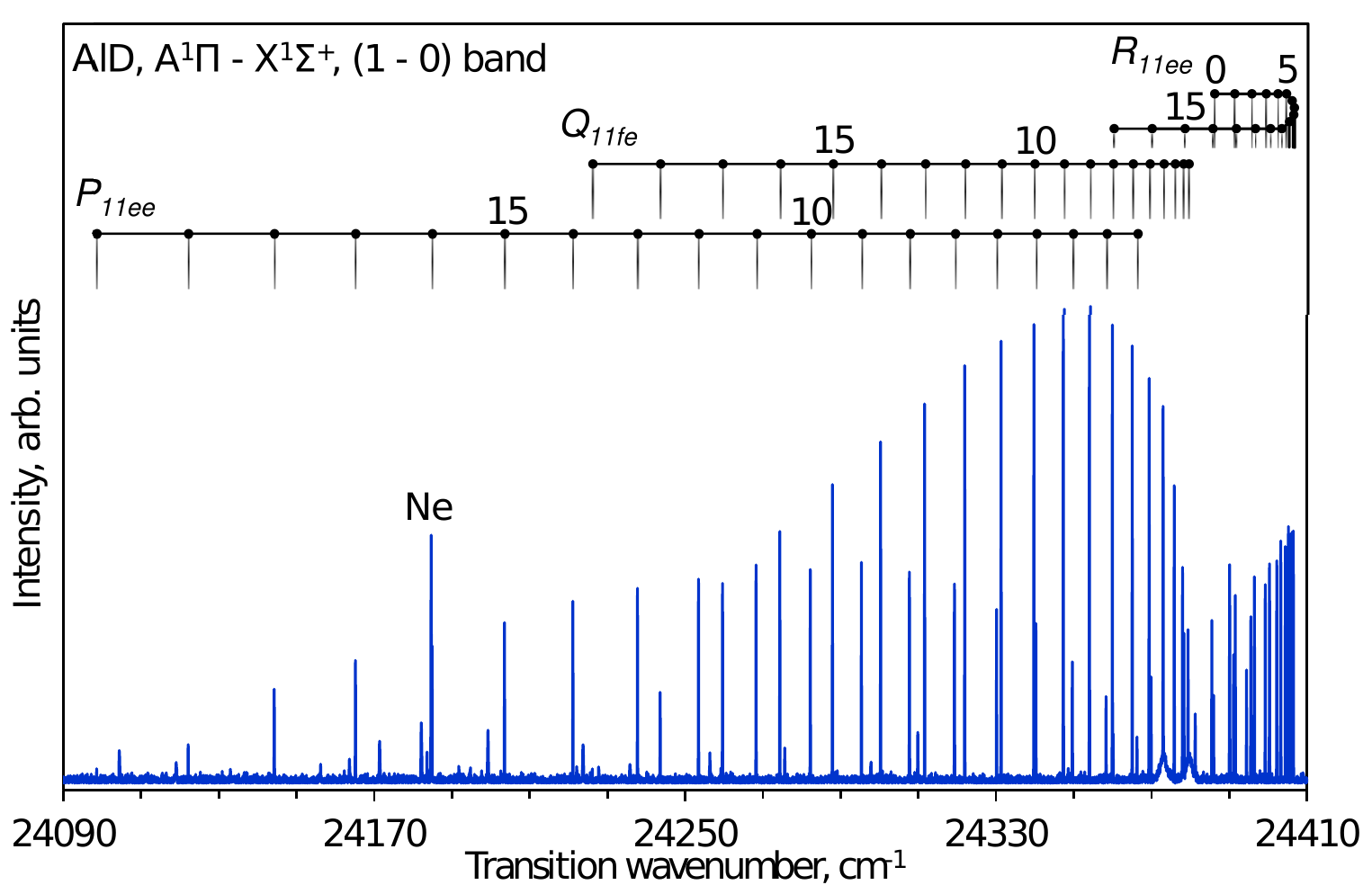}
\caption{High-resolution FT-VIS emission spectrum of the AlD, \AX\ (1-0) band recorded with SNR of about $200:1$ and FWHM of lines about~$0.08$ \cm. Bands of the $1-v''$ progression are observed up to the $J'_{\max}=19$ lines due to the predissociation in the \A~$(v=1)$ level. 
}
    \label{f:ftvis10band}
\end{figure}

The AlD,~\AX\ bands form a simple and regular structure: single $R$, single $Q$ and single $P$ branch (see Fig.~\ref{f:ftvis10band} and~\ref{f:FT:22800}). Predissociation in the \A\ state~\citep{34HoHuxx.AlH,40HeMuxx,48Nilsson.AlH} has limited the number of lines observed within current low pressure emission experiment up to $J'=29\ (v'=0)$, $J'=19\ (v'=1)$ and $J'=4\ (v'=2)$, respectively (see Table~\ref{t:FTSwavenumbers1} and~\ref{t:FTSwavenumbers2}). A part of the high quality spectrum of the AlD, \AX\ (1-0) band is shown in Fig.~\ref{f:ftvis10band}, where well resolved lines are rotationally interpreted.

\section{MARVEL procedures}

All currently available experimental transition frequencies (both extracted from literature and as part of this work) for AlH and AlD were analysed using the ``Measured Active Rotational-Vibrational Energy Levels'' (MARVEL) algorithm \citep{jt412,07CsCzFu.method,12FuCsxx.methods,jt750}. This algorithm consists of inverting a set of transition frequencies with their respective uncertainties into a consistent set of energy levels with the uncertainties propagated from all relevant transitions.

The AlH and AlD experimental data extracted is primarily concentrated around the first two singlet states \X, \A\ and their respective \XS--\XS\ and \XS--\AS\ bands. The work description of the sources is divided between AlH and AlD.

\subsection{AlH}

Many of the sources used in this work for AlH have been previously discussed by \citet{jt732,09SzZaxx.AlH} and \citet{19VoVoxx.AlH}, with the addition of a few older papers not used in our previous MARVEL study. The quantitative description of the data used in MARVEL from the sources can be seen in Table \ref{tab:alh_sources}. The qualitative descriptions of the sources for AlH denoted by their MARVEL tags are as follows:\\

\noindent \textbf{23YuSzHa\_astro} (Current work): as part of the current work 79 rovibronic transitions of AlH in the \AX\ system are reported in the current work as part of a reanalysis of the stellar (Proxima Cen) spectra from the HARPS ESO public data archive \citep{03MaPeQu}, for the (0--0), (1--0), (1--1) bands (see below). The transition data is included as part of the MARVEL input data and can be seen separately in the Table \ref{tab:Proxima_cen_AlH} of the Appendix.

\noindent \textbf{23SzKePa}: \citet{23SzKePa.AlH} reported FT-VIS emission spectra of AlH with 259 transitions in the \AS--\XS\ system for the (0--0), (0--1),(0--2), (1--0), (1--1), (1--2), (1--3), (1--4) bands.

\noindent \textbf{22PaTeYu}: \citet{jt874} reported 133 rovibronic transitions of AlH from their stellar (Proxima Cen) spectra in the \AS--\XS\ system for the (0-0), (0-1), (1-0), (1-1), (1-2) bands. The transitions were predominantly reported as $\lambda_{\rm air}$ in \AA\ and had to be converted to $\nu_{\rm  vacuum}$ in \cm\ for consistency. This was done using the method described in \citet{VALD3} and attributed to N. Piskunov.

\noindent \textbf{16HaZi}: \citet{16HaZixx.AlH} reported hyperfine rotational transitions of AlH between $J=2 \leftarrow 1$ in the \XS\ state measured using the terahertz direct absorption spectroscopy. The pure rotational frequency from this work was calculated using the common expression for the total hyperfine energy \citep{01GeWaxx.AlH,84GoCo} as a sum of the electric quadrupole and nuclear spin-rotation terms for $^{27}\text{Al}$. The weighted average values of the hyperfine corrected pure rotational frequency of the $R(1)$ line is  $755,211.403(90)$~MHz. The  methodology use to calculated pure rotational frequency is described in further detail in Appendix~\ref{line positions from hyperfine transitions}.



\noindent \textbf{14HaZi}: \citet{14HaZixx.AlH} reported hyperfine rotational transitions of AlH between $J=2 \leftarrow 1$ in the \XS\ state measured using the terahertz direct absorption spectroscopy. The pure rotational transition value from this work data is $R(1)=755,198.117(87)$~MHz (see method description in \textbf{16HaZi} and Appendix).


\noindent \textbf{11SzZaHa}: \citet{11SzZaHa.AlH}  reported  47 AlH emission transitions from  the \AS--\XS\ system and the (0-2) band.

\noindent \textbf{09SzZa}: \citet{09SzZaxx.AlH} reported emission spectra of AlH with 183 transitions in the \AS--\XS\ system for the (0-0), (0-1), (1-0), (1-1), (1-2), (1-3) bands. A local minor perturbation in the \AS\ $v'=1$, $J'=5$ was reported and attributed to the a~$^{3}\Pi$ state.

\noindent \textbf{04HaZi}: \citet{04HaZixx.AlH} reported hyperfine rotational transitions in AlH between $J'=1$ and $J''=0$ in the \XS\ state measured using submillimeter direct absorption spectroscopy.  The pure rotational transition value from this work data is $R(0)=377,737.022(90)$~MHz (see method description in \textbf{16HaZi}).


\noindent \textbf{96RaBe}: \citet{96RaBexx.AlH} reported emission spectra of AlH with 66 rovibronic transitions in the \AS--\XS\ system for the (0-0), (1-1) bands.

\noindent \textbf{95GoSa}: \citet{95GoSaxx.AlH} reported hyperfine rotational transitions in AlH between $J'=1$ and $J''=0$ in the \XS\ state measured using the submillimeter-wave spectrometer.  The pure rotational transition value from this work data is $R(0)=377,738.10(72)$~MHz (see method description in \textbf{16HaZi}).

\noindent \textbf{94ItNaTa}: \citep{94ItNaTa.AlH} reported Fourier Transform Infrared (FTIR) absorption spectra of AlH with 87 rovibrational transitions in the \XS--\XS\ system for the (1-0), (2-1), (3-2), (4-3) bands.

\noindent \textbf{93WhDuBe}: \citet{93WhDuBe.AlH} reported FTIR emission spectra of AlH with 260 rovibrational transitions observed in the \XS--\XS\ system for the (1-0), (2-1), (3-2), (4-3), (5-4) bands.

\noindent \textbf{92YaHi}: \citet{92YaHixx.AlH} reported infrared diode laser absorption spectra with 22 rovibrational transitions in the \XS--\XS\ system for the (1-0), (2-1), (3-2), (4-3) bands.

\noindent \textbf{87DeNeRa}: \citet{87DeNeRa.AlH} reported emission spectra of AlH with 333 rovibrational transitions in the \XS--\XS\ system with the $\Delta v=2$ sequence for the  (2-0), (3-1), (4-2), (5-3), (6-4), (7-5), (8-6) bands.

\noindent \textbf{54ZeRi}: \citet{54ZeRixx.AlH} reported both emission and absorption bands of AlH with 162 rovibronic transitions observed in the  \AS--\XS\ system for the (0-2), (0-3), (1-3), (1-4) bands.

\noindent \textbf{34Holst}: \citep{34Holst.AlH} reported an absorption band with 25 rovibronic transitions for the AlH as a  $^{1}\Sigma^{***}$--$^{1}\Pi$ system. Through matching with the newer experiments this band was identified as (0-2) of the \AS--\XS\ system.

\noindent \textbf{30BeRy}: \citet{30BeRyxx.AlH} reported two emission bands with 76 rovibronic transitions for the AlH as a $^{1}\Pi $ $\rightarrow$ ${}^{1}\Sigma$ system. Through matching with the newer experiments these bands were identified as (0-1) and (1-2) of the \AS--\XS\ system.

\begin{figure}
    \centering
    \includegraphics[width=0.44\textwidth]{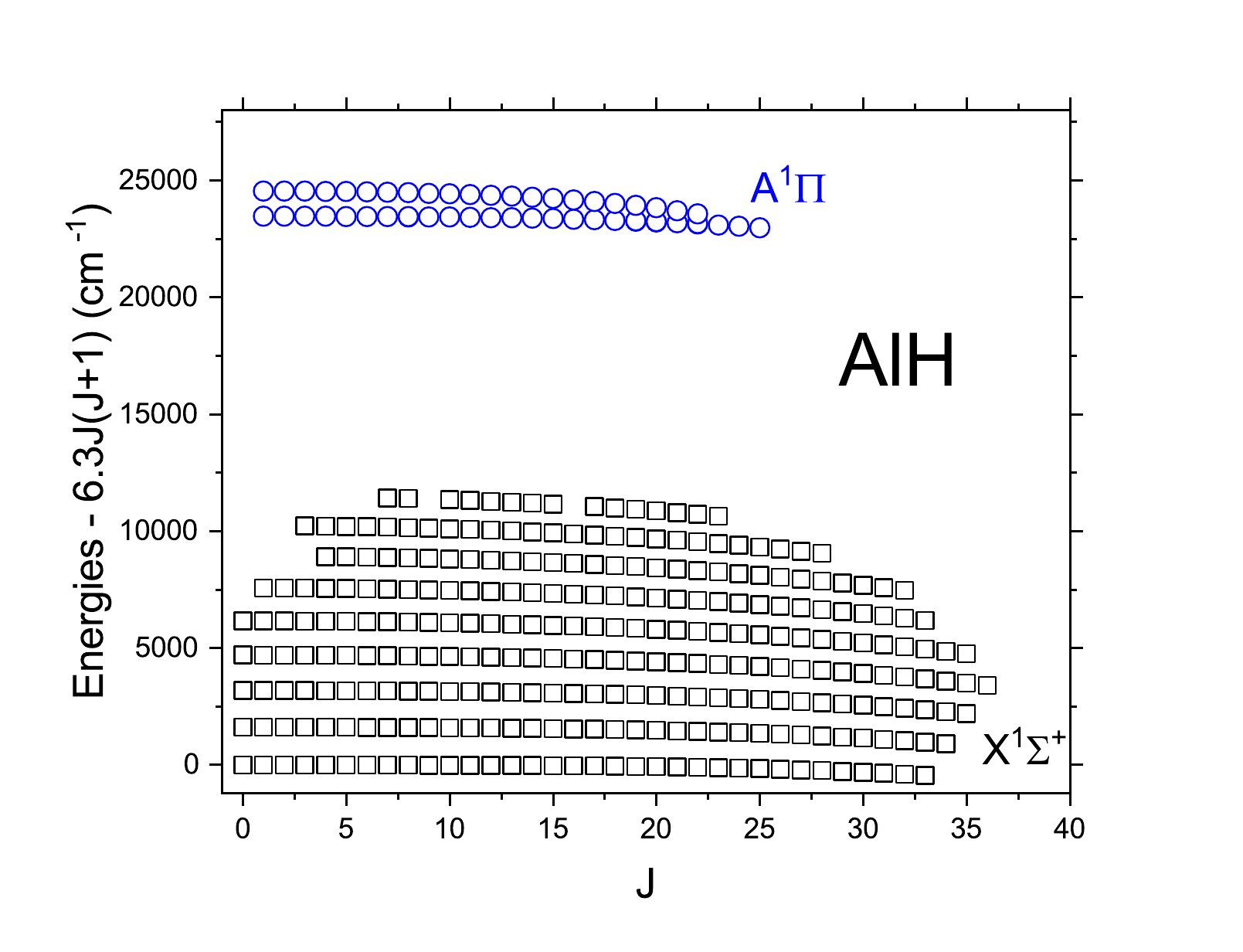}
    \caption{Experimentally derived reduced energy term values for the AlH in the rotational decomposition. }
    \label{f:AlH_en_levels}
\end{figure}

A complete set of experimentally derived energy term values for the AlH represented in the rotational decomposition can be seen in Figure \ref{f:AlH_en_levels}. The quantitative description of MARVEL-derived AlH term values can be found in Table \ref{t:alh_energy_levels}.

\begin{table*}
\centering
\caption{List of transition lines used in MARVEL procedure for AlH grouped by source.} \label{tab:alh_sources}
\begin{tabular}{llccccc}\hline \hline
Segment tag & Source & Range \cm & A/V & MSU \cm  & LSU \cm  & ASU \cm \\ \hline
\textbf{23YuSzHa\_astro} & current work & 22618.80 - 24583.50 & 79/79 & 1.600e-2 & 1.142e+0 & 1.165e-1  \\
\textbf{23SzKePa} & \cite{23SzKePa.AlH}  & 18275.458 - 24585.496 & 259/259 & 2.000e-3 & 4.360e-2 & 4.209e-3  \\
\textbf{22PaTeYu} & \citep{jt874} & 22742.0 - 24556.1 & 24/24 & 1.000e-2 & 1.038e+0 & 1.443e-1  \\
\textbf{16HaZi} & \cite{16HaZixx.AlH} & 25.1911 - 25.1911 & 1/1 & 2.681e-4 & 2.681e-4 & 2.681e-4  \\
\textbf{14HaZi} & \cite{14HaZixx.AlH} &  25.1907 - 25.1907 & 1/1 & 3.729e-4 & 3.729e-4 & 3.729e-4  \\
\textbf{11SzZaHa} & \cite{11SzZaHa.AlH} & 20067.602 - 20468.352 & 47/47 & 3.000e-3 & 1.958e-2 & 4.714e-3  \\
\textbf{09SzZa} & \cite{09SzZaxx.AlH}  & 19724.41 - 24585.49 & 183/183 & 3.000e-2 & 3.000e-2 & 3.000e-2  \\
\textbf{04HaZi} & \cite{04HaZixx.AlH} & 12.59995 - 12.59995 & 1/1 & 2.643e-5 & 2.643e-5 & 2.643e-5  \\
\textbf{96RaBe} & \cite{96RaBexx.AlH} & 22782.590 - 23572.451 & 66/66 & 3.000e-3 & 2.377e-2 & 4.619e-3  \\
\textbf{95GoSa} & \cite{95GoSaxx.AlH} & 12.59999 - 12.59999 & 1/1 & 1.722e-5 & 1.722e-5 & 1.722e-5  \\
\textbf{93WhDuBe} & \cite{93WhDuBe.AlH} & 1225.5735 - 1802.7058 & 260/260 & 2.000e-4 & 5.953e-3 & 3.204e-4  \\
\textbf{94ItNaTa} & \cite{94ItNaTa.AlH} & 1400.4802 - 1793.1351 & 87/87 & 1.000e-4 & 2.300e-3 & 6.969e-4  \\
\textbf{92YaHi} & \cite{92YaHixx.AlH} & 1432.09 - 1781.54 & 22/22 & 3.000e-3 & 1.000e-1 & 1.129e-2  \\
\textbf{87DeNeRa} & \cite{87DeNeRa.AlH} &  2405.968 - 3292.072 & 333/329 & 3.000e-3 & 2.490e-1 & 7.084e-3  \\
\textbf{54ZeRi} & \cite{54ZeRixx.AlH} & 18195.18 - 20477.47 & 162/148 & 1.000e-2 & 1.810e-1 & 2.852e-2  \\
\textbf{34Holst} & \cite{34Holst.AlH} &  20034.93 - 20281.98 & 25/14 & 5.000e-2 & 1.855e-1 & 8.748e-2  \\
\textbf{30BeRy} & \cite{30BeRyxx.AlH} & 21226.3 - 21988.9 & 76/71 & 5.000e-1 & 4.348e+1 & 1.149e+0  \\
\hline \hline
\end{tabular}\\ \vspace{2mm}
\rm
\noindent
A/V -  Available lines vs Verified \\
MSU - Minimal uncertainty \\
LSU - Largest uncertainty \\
ASU - Average uncertainty \\
\end{table*}

\begin{table*}
    \centering
    \caption{Description of AlH energy levels derived from MARVEL. }
    \label{t:alh_energy_levels}
    \begin{tabular}{cccccc} \hline \hline
    State & $v$  & $J$ Range &Unc. Range \cm & Avg. of Unc. \cm &  Range of energy levels \cm  \\ \hline
\AS  & 0 & 1 - 24 & 0.0040 - 1.9836 & 0.1066&23482.94 -- 26810.43  \\
&1 & 0 - 13 & 0.0040 - 1.7813 & 0.1662&24553.99 -- 25468.85  \\
\XS & 0 & 0 - 33 & 0.0000 - 0.0663 & 0.0136&0.0000 -- 6629.52  \\
&1 &0 - 34 & 0.0004 - 0.0659 & 0.0136&1625.07 -- 8419.93  \\
&2 & 0 - 35 & 0.0008 - 0.0663 & 0.0136&3194.21 -- 10143.16  \\
&3 & 0 - 36 & 0.0012 - 0.0667 & 0.0138&4708.82 -- 11800.00  \\
&4 & 0 - 35 & 0.0016 - 0.0654 & 0.0128&6170.19 -- 12697.60  \\
&5 & 1 - 33 & 0.0020 - 0.0495 & 0.0123&7590.40 -- 13256.44  \\
&6 & 4 - 32 & 0.0101 - 0.0486 & 0.0202&9042.94 -- 14131.90  \\
&7 & 3 - 28 & 0.0080 - 0.0555 & 0.0174&10307.65 -- 14170.79  \\
&8 & 7 - 23 & 0.0161 - 0.0546 & 0.0277&11781.03 -- 14129.84  \\


    \hline \hline
    \end{tabular}
\end{table*}

\subsection{AlD}

As for AlH, there have been previous discussions of the sources used in \citet{jt732,09SzZaxx.AlH} and \citet{19VoVoxx.AlH}. The quantitative description of the data used in MARVEL from the sources can be seen in Table \ref{tab:ald_sources}. The qualitative descriptions of the sources for AlD denoted by their MARVEL tags are as follows: \\

\noindent \textbf{23YuSzHa} (Current work): 491 rovibronic transitions of AlD in the \AX\ system are reported using the FT-VIS spectroscopy, as described in Section 2, for the (0-0), (0-1), (0-2), (1-0), (1-1), (1-2), (1-3), (1-4), (2-1), (2-2) bands. The full list of transitions can be found in Appendix (see Table~\ref{t:FTSwavenumbers1} and~\ref{t:FTSwavenumbers2}).

\noindent \textbf{22ShKaRa}: \citet{22ShKaRa.AlH} reported 76 rovibronic  spectra transitions of AlD in the \AS--\XS\ for the (0-0), (1-2), (1-3) bands observerd in sunspot umbra. The MARVEL uncertainty was set to 0.1~\cm, the same as the tolerance of wavenumber reported for line identification in the sunspot spectra.

\noindent \textbf{17SzMoLa}: \citet{17SzMoLa.AlH} reported 379 rovibronic transitions of AlD in the \AS--\XS\ for the (0-0), (0-1), (1-0), (1-1), (1-2), (1-3) bands using the FT-VIS spectroscopy. The accuracy of the lines is assigned based on whether the line was distorted or not. Distorted lines had an accuracy of 0.01 \cm and clear lines were reported with uncertainty of  0.002~\cm.

\noindent \textbf{17SzMoLa\_hyperfine}: \citet{17SzMoLa.AlH} also reported the pure rotational $R(1)=393661.660(36)$~MHz, $R(2)=590314.932(66)$~MHz and $R(3)=786755.135(122)$~MHz frequencies of the \XS\ $v=0$ provided by Halfen via private communication as calculated from their experimental data \citep{10HaZixx.AlH,14HaZixx.AlH}.


\noindent \textbf{15SzZaHa}: \citet{15SzZaHa.AlH}  reported an emission spectrum of AlD with 133 rovibronic transitions in the \AS--\XS\ for the (0-0), (1-1) bands. Like \textbf{17SzMoLa}, the accuracy reported depended on whether the lines were blended or not. Blended lines were reported with an accuracy of 0.005~\cm\ and not blended with 0.003~\cm.

\noindent \textbf{14HaZi}: \citet{14HaZixx.AlH} reported hyperfine rotational transitions of AlD between $J=4 \leftarrow 3$ in the \XS\, measured using the terahertz direct absorption spectroscopy. The pure rotational frequency from this work was calculated using the common expression for the total hyperfine energy \citep{01GeWaxx.AlH,84GoCo} as a sum of the electric quadrupole and nuclear spin-rotation terms for $^{27}\text{Al}$. The weighted average value of the hyperfine corrected pure rotational frequency of the $R(3)$ line is  $786,755.195(77)$~MHz. The pure rotational value calculation is described in further detail in Appendix~\ref{line positions  from hyperfine transitions}.

\noindent \textbf{10HaZi}: \citet{10HaZixx.AlH} reported hyperfine rotational transitions in AlD between $J=2\leftarrow1$ and $J=3\leftarrow2$ using submillimeter direct absorption spectroscopy. Additionally, predictions for the $J=1\leftarrow0$ and $J=4\leftarrow3$ transitions of AlD have been made. The pure rotational transition values from this  data are $R(2)=590,314.920(43)$~MHz and $R(1)=393,661.697(52)$~MHz (see method description in \textbf{14HaZi}). Contaminated lines were excluded from the weighted average calculation.

\noindent \textbf{04HaZi}: \citet{04HaZixx.AlH} reported hyperfine rotational transitions in AlD between $J=2\leftarrow1$ using submillimeter direct absorption spectroscopy. The pure rotational transition value from this data is $R(1)=393,661.772(55)$~MHz  (see method description in \textbf{14HaZi}).


\noindent \textbf{93WhDuBe}: \citet{93WhDuBe.AlH} reported FTIR emission spectra of AlD with 465 rovibrational transitions observed in the \XS--\XS\ system for the (1-0), (2-1), (3-2), (4-3), (5-4), (6-5), (7-6) bands.

\noindent \textbf{92UrJo}: \citet{92UrJoxx.AlH} reported infrared spectra of AlD with 114 rovibrational transitions observed in the \XS--\XS\ system for the (1-0), (2-1), (3-2), (4-3), (5-4), (6-5), (7-6) bands.

\noindent \textbf{48Nilsson}: \citet{48Nilsson.AlH} reported emission spectra of AlD with 240 rovibronic transitions observed in the \AS--\XS\ system for the (0-0), (0-1), (1-0), (1-1), (1-2), (1-3), (2-1), (2-2) bands. The uncertainties were not reported as part of the work; based on how close the values were to the more recent experiments the original minimum uncertainty was set at 0.3~\cm. Additionally, the R branch transition for $J=0$ in the (1-0) band is off by 20~\cm\ and is excluded from the analysis.

\begin{figure}
    \centering
    \includegraphics[width=0.44\textwidth]{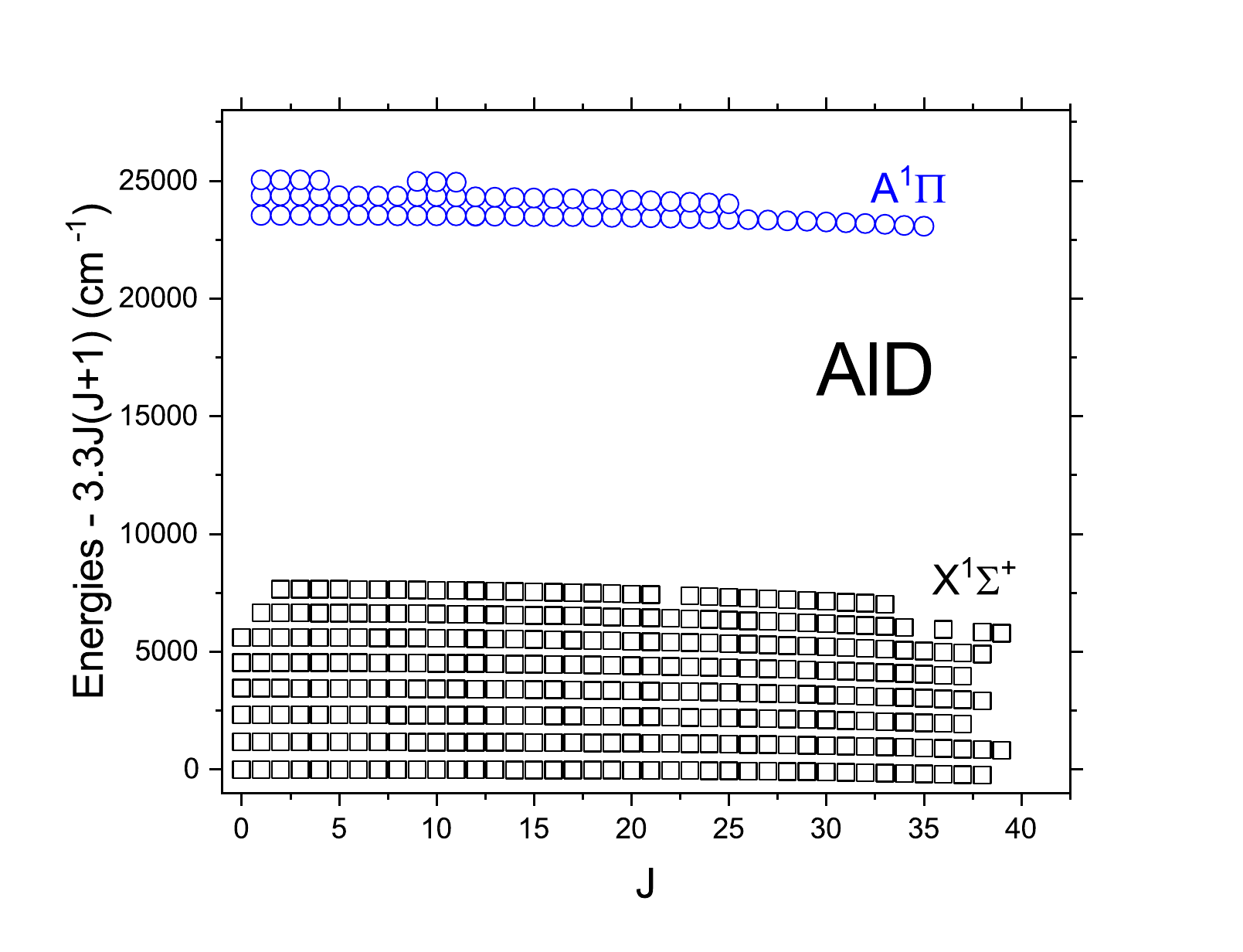}
    \caption{ Experimentally derived reduced energy term values for the AlD in the rotational decomposition. }
    \label{f:AlD_en_levels}
\end{figure}

The complete set of experimentally derived energy term values for the AlH represented in the rotational decomposition, can be seen in Figure \ref{f:AlD_en_levels}. The quantitative description of MARVEL-derived AlD term values can be found in Table \ref{t:ald_energy_levels}.

The MARVEL input, transtion files, and output, energy files, for both AlH and AlD are given in the supporting material.

\begin{table*}
\centering
\caption{List of transition lines used in MARVEL procedure for AlD grouped by source.} \label{tab:ald_sources}
\begin{tabular}{llccccc}\hline \hline
Segment tag & Source&  Range \cm & A/V & MSU \cm  & LSU \cm  & ASU \cm \\ \hline
\textbf{23YuSzHa} &  current work &  19749.782 - 24406.512 & 491/491 & 2.000e-3 & 3.890e-2 & 5.589e-3  \\
\textbf{22ShKaRa} & \cite{22ShKaRa.AlH} & 20805.3 - 23588.2 & 76/74 & 1.000e-1 & 1.000e-1 & 1.000e-1  \\
\textbf{17SzMoLa} & \cite{17SzMoLa.AlH} & 20755.112 - 24406.511 & 379/379 & 2.000e-3 & 2.133e-2 & 4.031e-3  \\
\textbf{17SzMoLa\_hyperfine} b& \cite{17SzMoLa.AlH} & 13.131140 - 26.243326 & 3/3 & 1.201e-6 & 4.069e-6 & 2.491e-6  \\
\textbf{15SzZaHa} & \cite{15SzZaHa.AlH} &  22945.366 - 23604.065 & 133/133 & 3.000e-3 & 2.307e-2 & 4.229e-3  \\
\textbf{14HaZi} & \cite{14HaZixx.AlH} & 26.243328 - 26.243328 & 1/1 & 4.253e-6 & 4.253e-6 & 4.253e-6  \\
\textbf{10HaZi} & \cite{10HaZixx.AlH}  & 13.131141 - 19.690786 & 2/2 & 1.745e-6 & 1.945e-6 & 1.845e-6  \\
\textbf{04HaZi} & \cite{04HaZixx.AlH} & 13.131143 - 13.131143 & 1/1 & 2.398e-6 & 2.398e-6 & 2.398e-6  \\
\textbf{93WhDuBe} & \cite{93WhDuBe.AlH} & 850.8496 - 1311.3449 & 465/465 & 2.000e-4 & 1.037e-2 & 3.540e-4  \\
\textbf{92UrJo} & \cite{92UrJoxx.AlH} & 945.746 - 1193.703 & 114/114 & 5.000e-3 & 5.000e-3 & 5.000e-3  \\
\textbf{48Nilsson} & \cite{48Nilsson.AlH} & 20708.8 - 24406.4 & 240/215 & 2.000e-1 & 2.000e-1 & 2.000e-1  \\
\hline \hline
\end{tabular}\\ \vspace{2mm}
\rm
\noindent
A/V -  Available lines vs Verified \\
MSU - Minimal uncertainty \\
LSU - Largest uncertainty \\
ASU - Average uncertainty \\
\end{table*}

\begin{table*}
    \centering
    \caption{Description of AlD energy levels derived from MARVEL.}
    \label{t:ald_energy_levels}
    \begin{tabular}{cccccc} \hline \hline
    State & $v$  & $J$ Range &Unc. Range \cm & Avg. of Unc. \cm &  Range of energy levels \cm  \\ \hline
\AS & 0 & 1 - 35 & 0.0044 - 0.4236 & 0.0783&23543.17 - 27229.71  \\
&1 & 1 - 25 & 0.0040 - 0.4124 & 0.0963&24385.98 - 26175.82  \\
&2 & 1 - 11 & 0.0064 - 0.4132 & 0.1473&25055.97 - 25381.24  \\
\XS & 0 & 0 - 38 & 0.0000 - 0.0244 & 0.0134&0.00 - 4655.89  \\
&1 & 0 - 39 & 0.0004 - 0.0248 & 0.0131&1181.94 - 5967.09  \\
&2 & 0 - 37 & 0.0008 - 0.0236 & 0.0125&2334.58 - 6574.10  \\
&3 & 0 - 38 & 0.0012 - 0.0240 & 0.0132&3458.43 - 7820.30  \\
&4 & 0 - 37 & 0.0024 - 0.0236 & 0.0135&4554.01 - 8611.26  \\
&5 & 0 - 38 & 0.0028 - 0.0240 & 0.0147&5621.82 - 9794.48  \\
&6 & 1 - 39 & 0.0032 - 0.0276 & 0.0171&6668.10 - 10946.56  \\
&7 & 2 - 33 & 0.0036 - 0.0349 & 0.0197&7692.91 - 10730.06  \\


    \hline \hline
    \end{tabular}

\end{table*}

\section{Spectroscopic model and refinement}

We use the variational diatomic nuclear-motion code \Duo\ \citep{Duo} to solve the coupled system of  Schr\"{o}dinger equations for a set of curves defining the spectroscopic model of the \XS\ and \AS\ system of AlH and AlD. We used the Sinc DVR method for the vibrational degree of freedom on a grid of 1601 points ranging from 0.5 to 13.5~\AA.

\begin{figure*}
    \centering
    \includegraphics[width=0.44\textwidth]{ 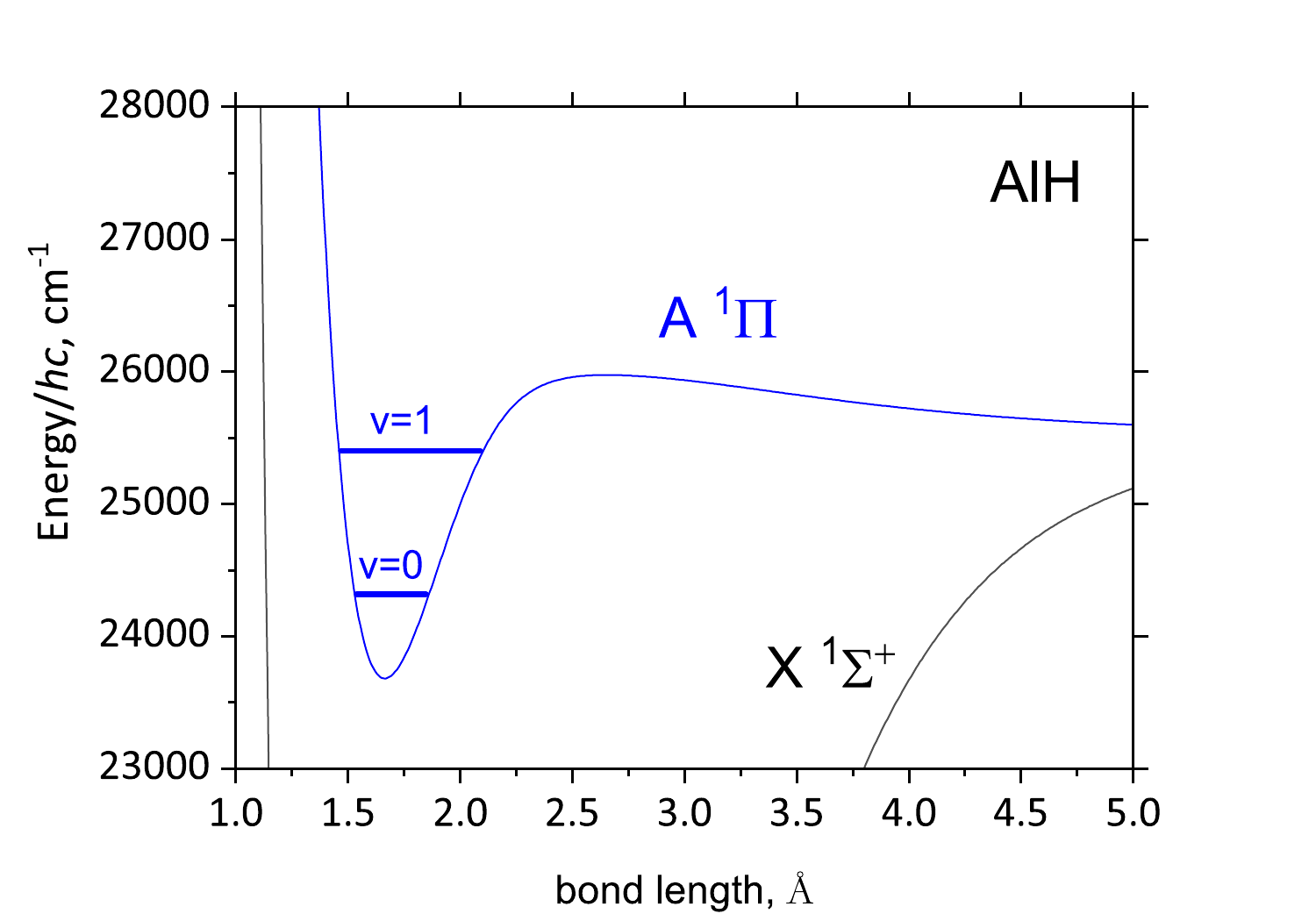}
    \includegraphics[width=0.44\textwidth]{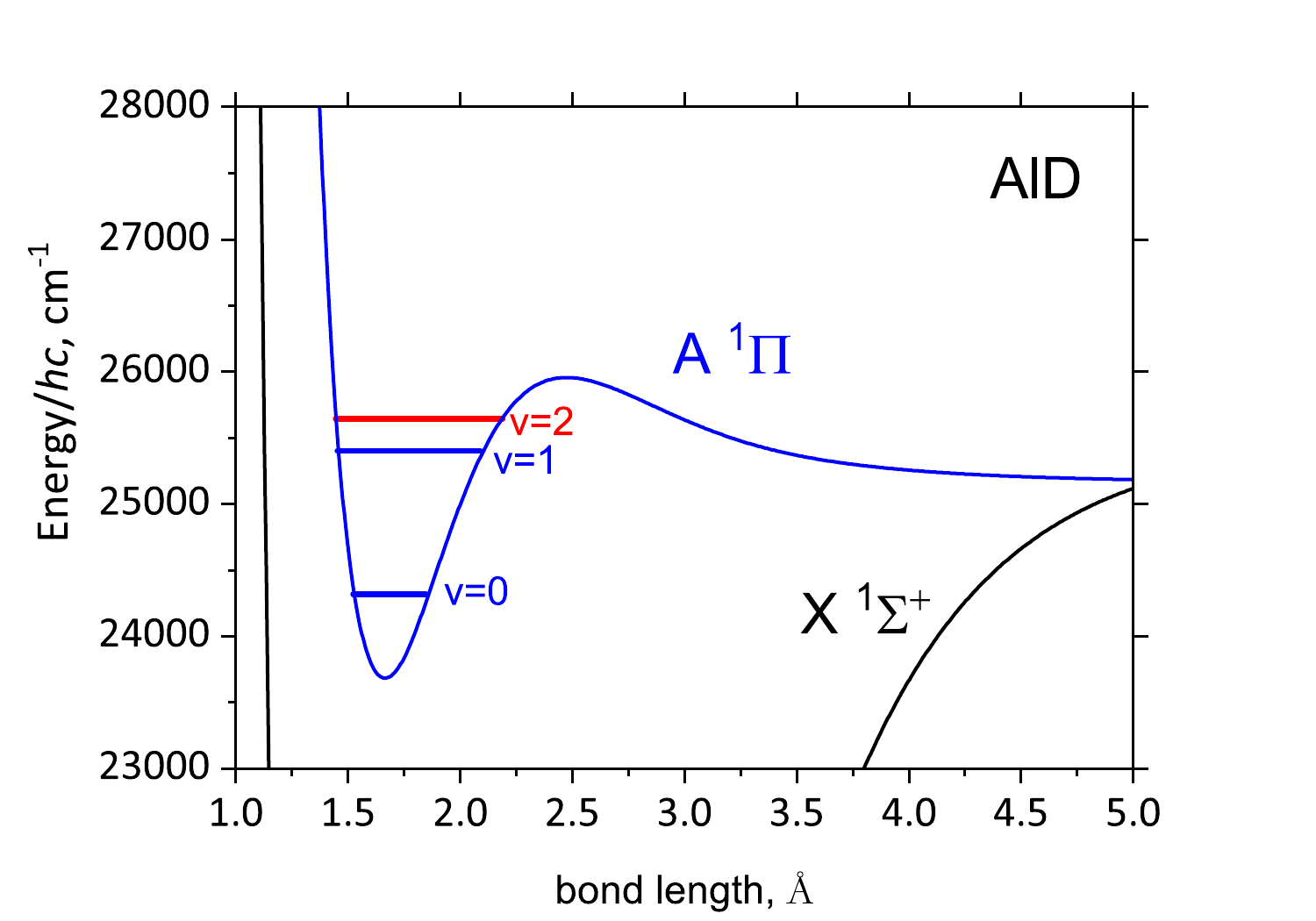}
\caption{Refined  potential energy curves of  \XS\ and \AS\ of AlH and AlD and the corresponding (quasi-)bound vibrational energy term values.}
    \label{f:PECs}
\end{figure*}

The AlH/AlD PEC in its the \AS\ state has a shallow minimum with a small barrier to the dissociation,  which can hold only two bound vibrational states in AlH ($v=0,1$) and three ($v=0,1,2$) in AlD as illustrated in Fig.~\ref{f:PECs}. Furthermore, the  highest vibrational states  ($v=1$ and $v=2$, respectively)  exhibit strong predissociative characters.

\begin{figure}
    \centering
    \includegraphics[width=0.44\textwidth]{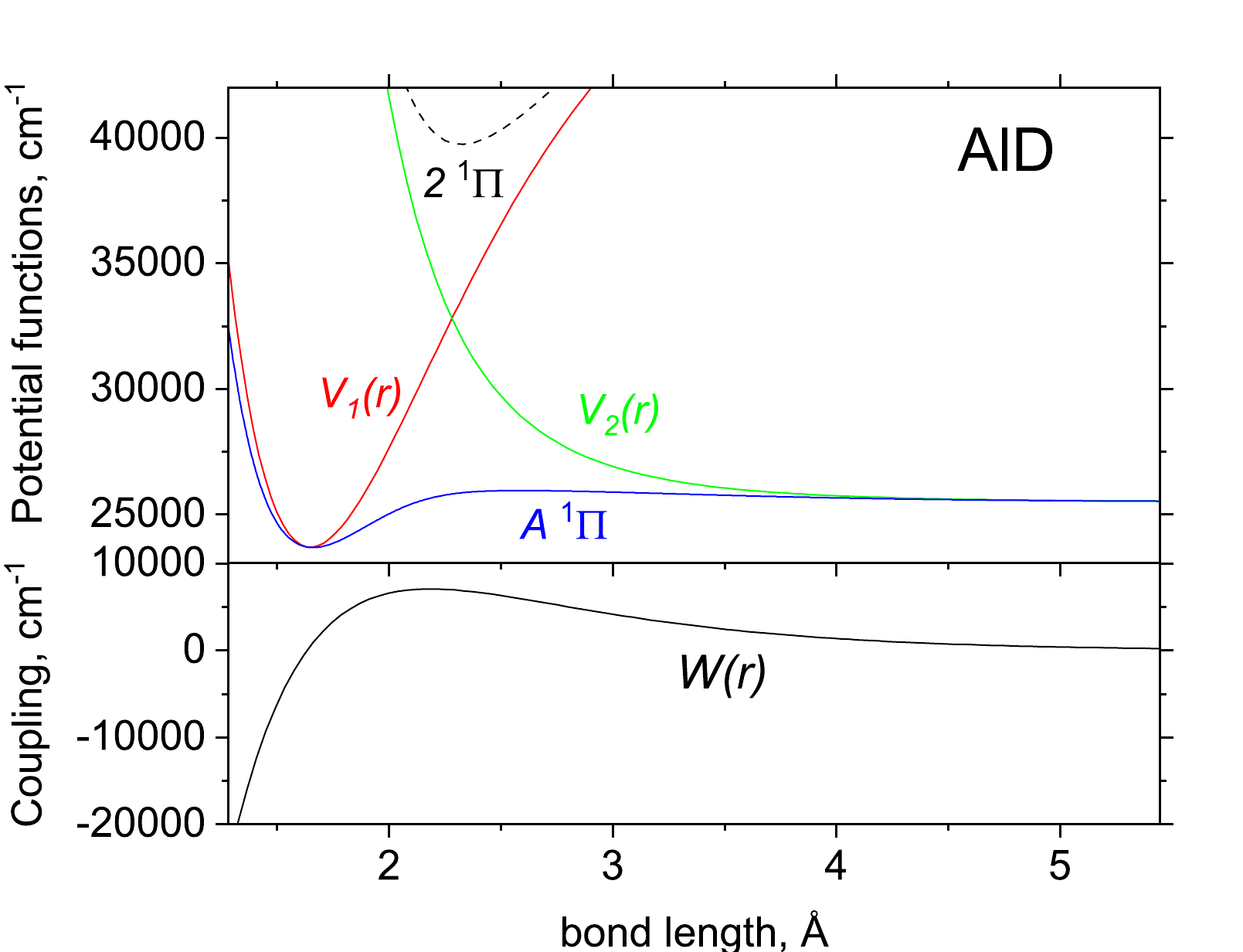}
\caption{Adiabatic PECs of \AS\  (blue dashed curve) and $2~^1\Pi$ (black dash) of AlD and the corresponding curves in the diabatic representation, PECs $V_1(r)$ (red) and $V_2(r)$ (green) and the diabatic coupling curve $W(r)$ (lower display). }
    \label{f:diabatic}
\end{figure}

Following \citet{jt732},   we use a diabatic representation to model the shallow \AS\ PECs of AlH and AlD, with  two diabatic PECs $V_1(r)$  and $V_2(r)$ coupled with a  term $W(r)$  via a $2\times 2$ diabatic matrix
\begin{equation}\label{e:V1W/WV2}
\bf{A} = \left(
\begin{array}{cc}
  V_1(r) & W(r) \\
  W(r) & V_2(r)
\end{array}
\right).
\end{equation}
The functions $V_1(r)$, $V_2(r)$ and $W(r)$ are illustrated in Fig.~\ref{f:diabatic} in the case of AlD. The diabatic PEC $V_1(r)$ is modelled with an EMO (Extended Morse Oscillator) function \citep{06LeHuJa.fu} as given by
\begin{equation}\label{e:EMO}
V(r)=V_{\rm e}\;\;+\;\;(A_{\rm e} - V_{\rm
e})\left[1\;\;-\;\;\exp\left(-\sum_{k=0}^{N} B_{k}\xi_p^{k}(r-r_{\rm e})
\right)\right]^2,
\end{equation}
where $A_{\rm e} $ is a dissociation asymptote,  $A_{\rm e} - V_{\rm e}$ is the dissociation energy, $r_{\rm e}$ is an equilibrium distance of the diabatic PEC, and $\xi_p$ is the \v{S}urkus variable given by:
\begin{equation}
\label{e:surkus:2}
\xi_p= \frac{r^{p}-r^{p}_{\rm e}}{r^{p}+r^{p}_{\rm e }}.
\end{equation}
$V_2(r)$  in Eq.~\eqref{e:V1W/WV2} is modelled by a repulsive curve playing  a role of a dummy state (called here \oneS) and represented by:
\begin{equation}\label{e:V(r)}
V_2(r)=A_{\rm e}^{A} + \frac{a_6}{r^6},
\end{equation}
with the asymptote $A_{\rm e}^{A}$ fixed   to  the dissociation asymptote  of  the \AS\ state, $A_{\rm e}$ = 25500~\cm.

For the coupling function $W(r)$, an inverted EMO PEC with an asymptote of $W(r) \to 0$ at $r\to \infty$  was used
\begin{equation}\label{e:diab}
W(r)=W_{0}-W_{0}\left[1\;\;-\;\;\exp\left(- w_0 (r-r_{0})
\right)\right]^2,
\end{equation}
where $W_{0}$ is the height of the coupling at $r_0$, see Fig.~\ref{f:diabatic}.

The  adiabatic PEC of \AS\ is then given by the lower eigenvalue  of the diabatic matrix $\bf{A}$ in Eq.~\eqref{e:V1W/WV2} as
\begin{equation}
  V_{A\, ^1\Pi}(r) = \frac{V_1(r)+V_2(r)}{2}-\frac{\sqrt{[V_1(r)-V_2(r)]^2+4 \, W^2(r)}}{2}.
\end{equation}
The upper diabatic component is disregarded in the rest of the calculations.

The expansion parameters defining the diabatic curves were obtained in the fit to the MARVEL energies of AlH/AlD and are given in the supplementary material (see also below).

We used the EMO function to represent the PEC of the \XS\ state with the corresponding expansion parameters taken from and constrained to the values of \citet{jt732}. A  Born-Oppenheimer Breakdown (BOB) correction curve was added and modelled using the following function:
\begin{equation}
\label{e:bob}
F(r)=\sum^{N}_{k=0}B_{k}\, z^{k} (1-\xi_p) + \xi_p\, B_{\infty},
\end{equation}
where $z$ is taken as a damped-coordinate given by
\begin{equation}\label{e:damp}
z = (r-r_{\rm ref})\, e^{-\beta_2 (r-r_{\rm ref})^2-\beta_4 (r - r_{\rm ref})^4},
\end{equation}
see also \citet{jt703} and \citet{jt711}.  Here $r_{\rm ref}$ is a
reference position equal to $r_{\rm e}$ by default and $\beta_2$ and $\beta_4$ are damping factors.  In order to model the deviation of PEC of AlD from the AlH PEC, a diabatic correction term $\Delta V(r)$ was added to the adiabatic PEC $V_{A\,^1\Pi}(r)$, which was modelled with the same form as in Eq.~\eqref{e:bob}.

A $\Lambda$-doubling empirical curve  $q(r)$ was also included in the fit modelled using Eq.~\eqref{e:bob} with a single expansion term
$$
q(r)=q_{0}  (1-\xi_p) + \xi_p\, q_{\infty},
$$
where $\xi_{p}$ as in Eq.~\eqref{e:surkus:2}.

The AlD PEC has an extra vibrational state, $v=2$  (see Fig.~\ref{f:PECs}), which samples a larger range of the PEC than that of AlH. We, therefore, decided to process the AlD curves first by fitting to the experimentally derived (MARVEL) energies, and then refine the AlD spectroscopic model for AlH  by fitting to the corresponding MARVEL energies (see above).

Because of the limited amount of experimental data and high complexity of the diabatic model, the fit is highly degenerate. As a work around we applied a rather subjective criterion of physically sensible shapes of the diabatic curves. During this user-guided fit, attention was paid to the predissociative lifetimes of the \AS\ $v=2$ states of AlD and $v=1$ states of AlH, which had to be consistent with the experimental data:  predissociative line shapes as in \citet{jt874} as well as lifetimes (see discussion below).

\begin{figure}
    \centering
    \includegraphics[width=0.97\columnwidth]{ 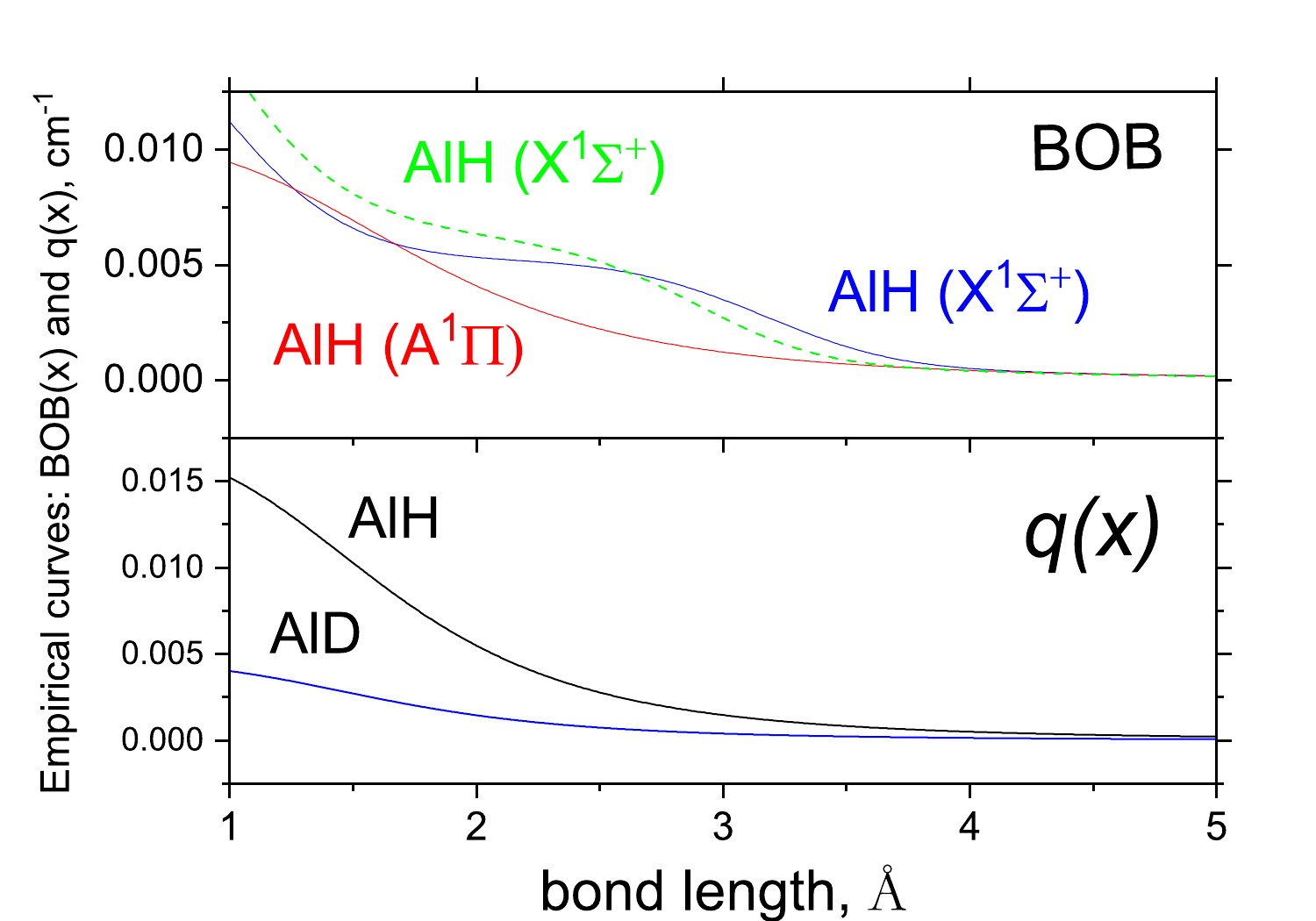}    
\caption{Curves for AlH and AlD: Upper panel, Born-Oppenheimer breakdown (BOB); lower panel, $\Lambda$ doubling, $q(r)$ .}
    \label{f:DMCs:BOB:q}
\end{figure}

\begin{figure*}
    \centering
\includegraphics[width=0.44\textwidth]{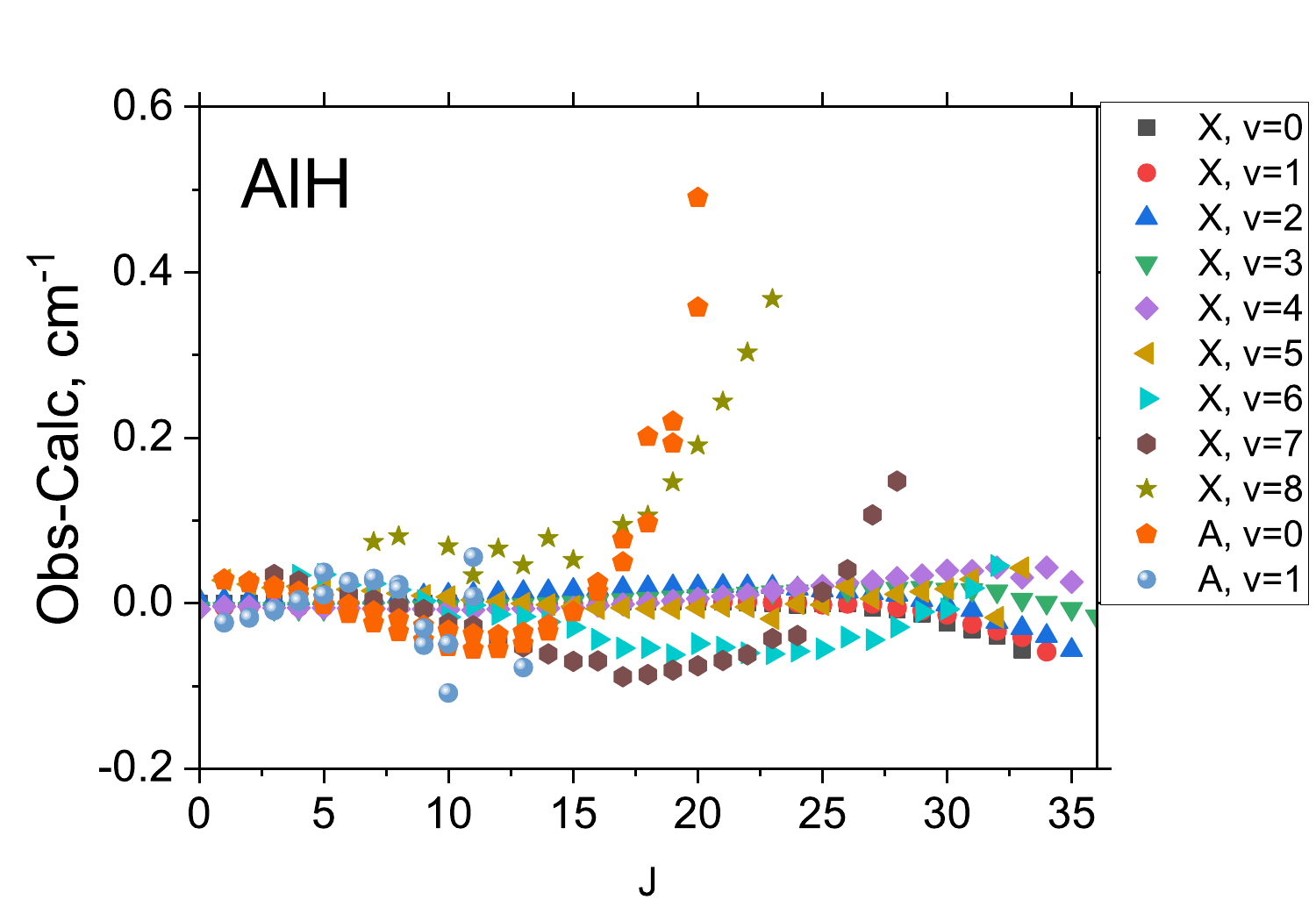}
\includegraphics[width=0.44\textwidth]{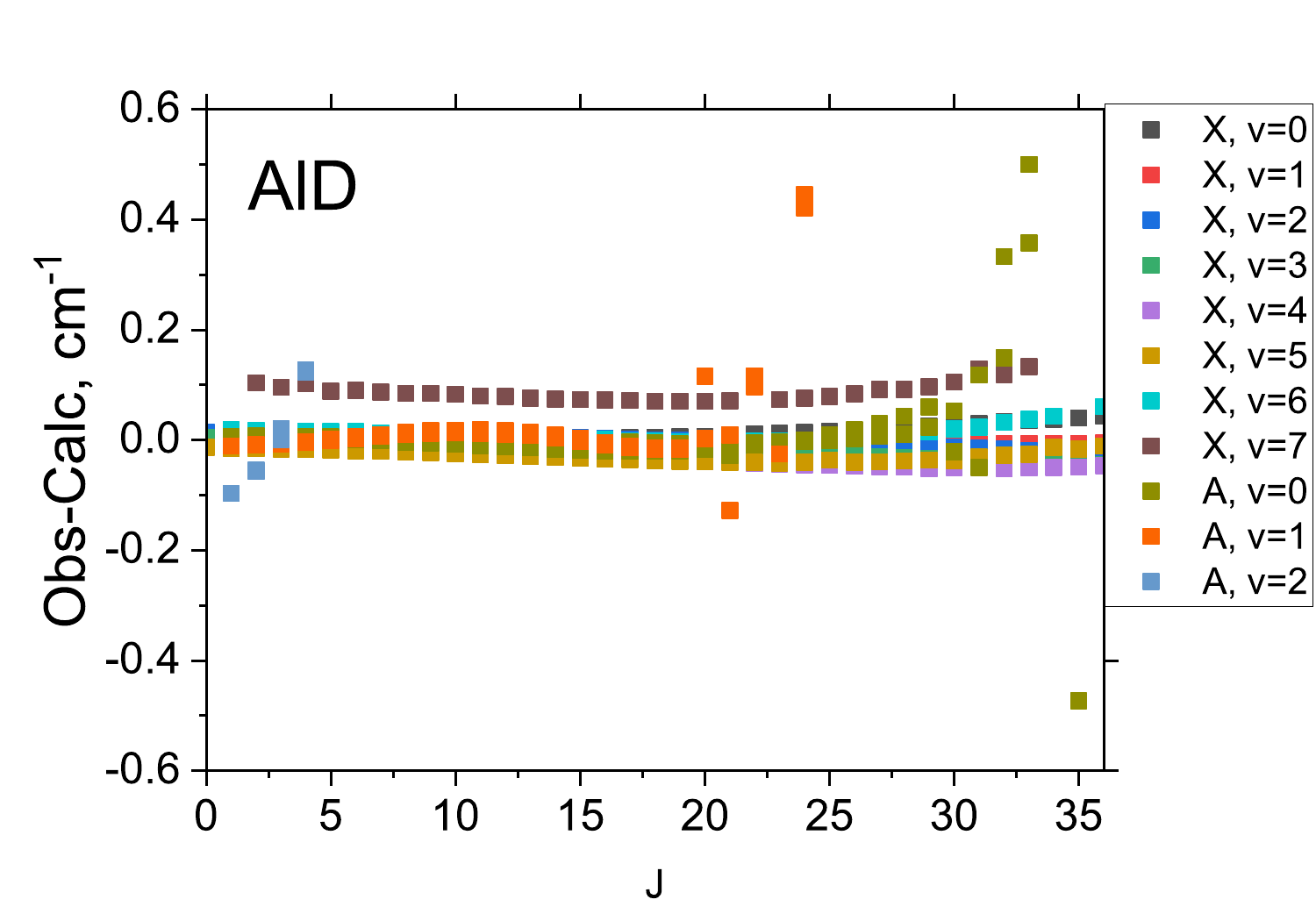}
\caption{Obs.-calc. residuals between the MARVEL and calculated energies of AlH  and AlD using the corresponding refined models.}
    \label{f:obs-calc}
\end{figure*}

The final spectroscopic model of AlD consists of 14 varying parameters reproducing the AlD 423 MARVEL energies with the root-mean-square (rms) error of 0.06~\cm. The corresponding curves are illustrated in Figs.~\ref{f:PECs} and \ref{f:DMCs:BOB:q}.  The residuals are shown in Fig.~\ref{f:obs-calc}.

In the AlH fit, the \AS\ PEC was constrained to that of AlD. In order to allow for variation in the shapes of the corresponding curves, an `adiabatic' potential correction term was added to the model using   Eq.~\eqref{e:bob}. We also introduced a BOB term for \AS\ of AlH and varied the parameter $q_0$ of the $\Lambda$-doubling  curve $q(x)$. The \XS\ PEC parameters were still constrained to the values from  \citet{jt732}, but we refitted the \XS\ BOB term to improve the quality of the model. The AlH spectroscopic model consists of 8 parameters reproducing 346 MARVEL energies (see above) with an rms error of 0.08~\cm.

The dipole moment and transition dipole moment curves were taken from \citet{jt732}.

All curves or parameters defining the AlH and AlD spectroscopic models are given as part of the supplementary material to the paper in the form of \Duo\ input files.

\subsection{Lifetimes and predissociation line broadening}
\label{s:lifetime}

As part of the ExoMol States files,  the lifetimes of species are usually included (see Table~\ref{t:states}).  In  most cases of negligible predissociation effects, the radiative lifetime (of a state $i$) is computed via
\begin{equation}
  \tau_i^{\rm rad} = \frac{1}{\sum_{j<i} A_{ij}},
\end{equation}
where $ A_{ij}$ are the Einstein $A$ coefficients for all states $j$ lower than $i$. According to the recent changes to the ExoMol format~\citep{jt898},  predissociative lifetimes $\tau_{\rm prediss}$ are to be included into the line list with the radiative lifetimes, if  non-negligible, which is the case for many \AS\ rovibronic states of AlH and AlD. Here we used the \LEVEL\ program \citep{LEVEL} to estimate lifetimes for the predissociative \AS\ states of AlH ($v=1$) and AlD ($v=2$) with the our new PECs. \LEVEL\ uses the uniform semiclassical procedure of \citet{81CoSmxx} to compute the  widths $\gamma$ (\cm) of the predissociative states, which we converted to lifetimes via
\begin{equation}
\label{e:tau:prediss}
\tau_{\rm prediss} = \frac{1}{2\pi c \gamma_{\rm prediss}},
\end{equation}
where $c$ is the speed of light in cm s$^{-1}$.  These are shown in Table~\ref{t:lifetime}; our lifetimes show reasonable
agreement with the laboratory values obtained by \citet{79BaNexx.AlH} using in a hollow cathode discharge by dye laser excitation as well as the astrophysical estimates of \citep{jt874} from analysis of Proxima Cen predissociative spectrum of  AlH. The \LEVEL\ predissociative  values $\tau_{\rm prediss}$  are then added to the radiative lifetime to give total lifetime in the States file:
\begin{equation}
\frac{1}{\tau_{\rm total}}= \frac{1}{\tau_{\rm rad}}+\frac{1}{\tau_{\rm prediss}}.
\end{equation}
The  lifetimes can be then used to evaluate the line broadening of the the predissociated lines by inverting Eq.~\eqref{e:tau:prediss} for the HWHM $\gamma_{\rm prediss}$ and apply alone side the collisional value of $\gamma_{\rm col}$:
\begin{equation}
   \gamma_{\rm total} \approx \gamma_{\rm col} + \gamma_{\rm prediss}.
\end{equation}
This feature is now implemented in the spectrum simulator \exocross\ \citep{ExoCross,jt914}.

\begin{table*}
\centering
\caption{Lifetimes (pico second) of AlH and AlD in their \AS\ state: $\tau_{\rm rad}$, $\tau_{\rm prediss}$, by 79BaNe \citep{79BaNexx.AlH} and 22PaTeYu \citep{jt874}.}
\label{t:lifetime}
\centering
\begin{tabular}{rrrrrrcrrrr}
\hline\hline
$v$& $J$ & $\tau_{\rm rad}$ & $\tau_{\rm prediss}$ & 79BaNe & 22PaTeYu &&  $v$& $J$ & $\tau_{\rm rad}$ & $\tau_{\rm prediss}$ \\
\hline
\multicolumn{6}{c}{AlH} && \multicolumn{4}{c}{AlD} \\
\hline
  0 &  17 &    88770 &  339418.03 &       &                                &&    2 &   1 &  130610 &  5044.98 \\
  0 &  18 &    91120 &   11817.11 &       &                                &&    2 &   2 &  131170 &  3649.69 \\
  0 &  19 &    94209 &     786.81 &       &                                &&    2 &   3 &  131870 &  2295.71 \\
  0 &  20 &    97872 &      85.74 &       &                                &&    2 &   4 &  132920 &  1284.75 \\
  0 &  21 &   130660 &      14.05 &   9.9 &    10($\frac{+10}{-5}$)        &&    2 &   5 &  134310 &   656.50 \\
  0 &  22 &   347810 &       3.24 &       &                                &&    2 &   6 &  136450 &   314.52 \\
  0 &  23 &   166570 &       0.99 &  0.92 &  0.63($\frac{+0.26}{-0.18}$)   &&    2 &   7 &  138430 &   144.85 \\
  0 &  24 &   268810 &       0.39 &  0.45 &  0.32                   $\pm 0.07$  &&    2 &   8 &  160270 &    65.56 \\
  0 &  25 &          &       0.18 &       &                                &&    2 &   9 &  168650 &    29.73 \\
  0 &  26 &          &       0.10 &       &                                &&    2 &  10 &  167640 &    13.72 \\
  1 &   6 &   113580 &  375687.32 &       &                                &&    2 &  11 &  168570 &     6.53 \\
  1 &   7 &   115640 &   26541.53 &       &                                &&    2 &  12 &  217830 &     3.25 \\
  1 &   8 &   118230 &    2651.90 &       &                                &&    2 &  13 &  303440 &     1.74 \\
  1 &   9 &   121560 &     342.35 &       &                                &&    2 &  14 &  362560 &     0.76 \\
  1 &  10 &   126400 &      55.23 &       &                                &&    2 &  15 &  581010 &     0.46 \\
  1 &  11 &   148800 &      11.09 &       &                                &&    2 &  16 &         &     0.25 \\
  1 &  12 &   159980 &       2.80 &       &                                &&    2 &  17 &         &     0.15 \\
  1 &  13 &   228510 &       0.89 &       &                                &&                                 \\
  1 &  14 &   508120 &       0.35 &       &                                &&                                 \\
  1 &  15 &          &       0.16 &       &                                &&                                 \\
  1 &  16 &          &       0.09 &       &                                &&                                 \\
    \hline\hline
\end{tabular}
\end{table*}

\section{Line Lists}
\label{s: Line List}

Using the new empirical spectroscopic models of AlH and AlD,  line lists \name\ for the \XS, \AS\ system were computed with \Duo. In intensity calculations, we distinguish bound-to-bound and bound-to-free transitions and compute two line lists, bound-bound and continuum (bound-free). The transitions to the quasi-bound states, especially important in the \AS-\XS\ band, are included in the bound-bound line list.
In order to improve the resolution of the continuum spectrum, we use a significantly larger calculation box, with the bond length ranging from 0.5 to 60~\AA.  Since \Duo\ is a pure bound state variational method, it produces both (quasi-)bound and continuum eigenfunctions  $\psi_{\lambda}(r)$ as part of the same variational calculations. All eigenfunctions are ortho-normal, including the continuum ones, and all satisfy the boundary conditions that they vanish exactly, together with their first derivatives, at both edges of the box.

\begin{figure}
    \centering
    \includegraphics[width=0.97\columnwidth]{ 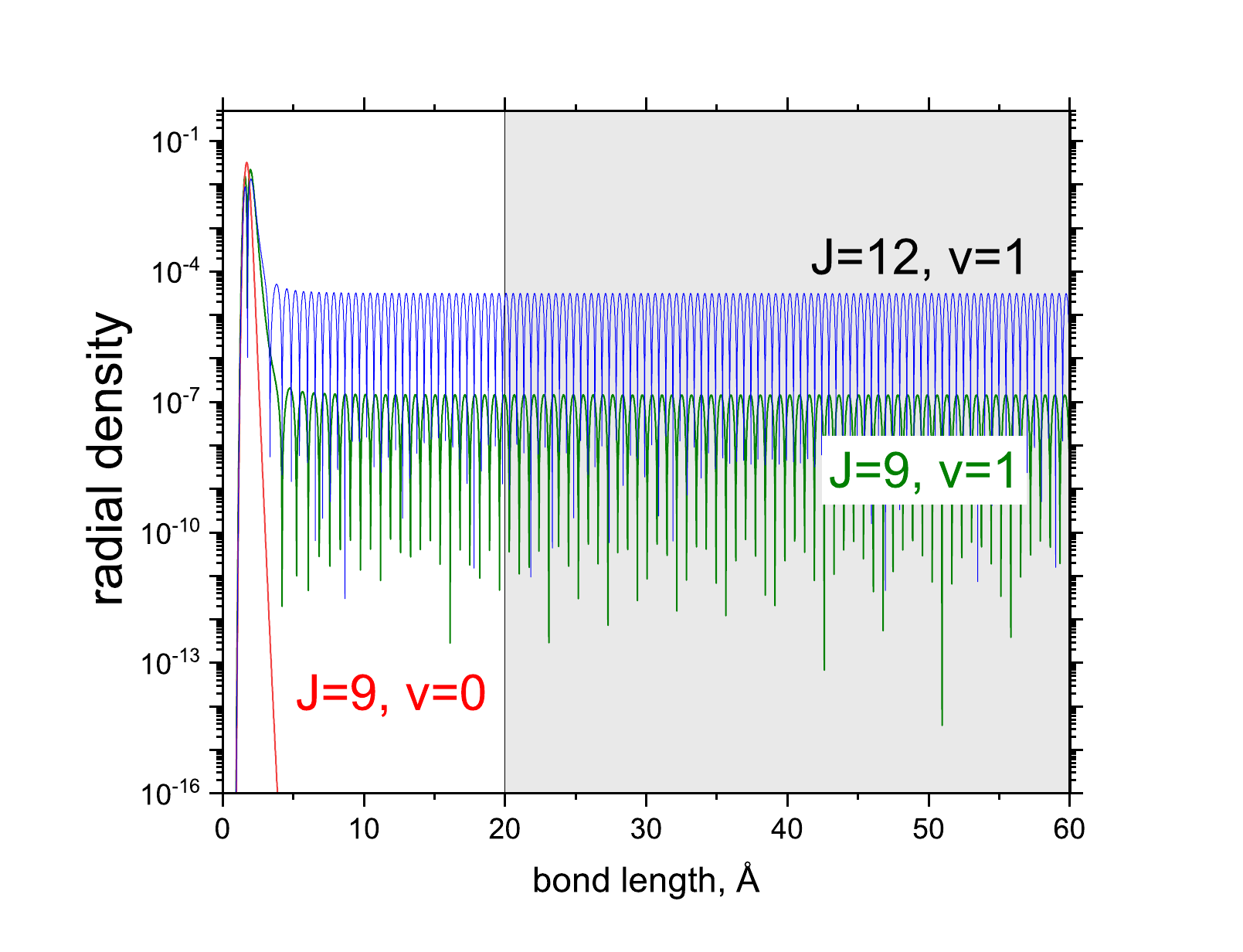}
    \caption{Reduced densities of AlH $\rho(r)$ for $J=9$ ($v=0$), $J=9$  ($v=1$) and $J=12$ ($v=1$) together with an integration box used to disentangle (quasi-)bound and continuum states.}
    \label{f:dens}
\end{figure}

In order to identify continuum states and then separate them from the (quasi-)bound states,  we check if they have non-zero density across a region $\Delta r$ adjacent  to the outer border $r_{\rm max}$ against some threshold value $\epsilon_{\rm max}$ as given by (see  \citet{jt887}):
\begin{equation}
\epsilon = \int_{r_{\rm max - \Delta r}}^{r_{\rm max}} |\psi_{\lambda}(r)|^2 dr > \epsilon_{\rm max} ,
\end{equation}
where the value of $\epsilon_{\rm max}$ must be tuned to the specific case.
For the box size $L$ of 59.5~\AA, the integration region was chosen as 40~\AA. Figure~\ref{f:dens} shows examples of reduced radial densities for  the bound state \AS, $v=0$, $J=9$ and two quasi-bound states \AS, $v=1$, $J=9$ and $J=12$ together with an integration box used. The corresponding values of integrated densities $\epsilon$ are 0, $2.4\times 10^{-4}$ and $4.8\times 10^{-2}$, respectively and of the average densities $\epsilon/L$ of  0, $3.7\times 10^{-6}$\AA$^{-1}$ and $7.8\times 10^{-4}$/\AA$^{-1}$. Here we adopted the threshold value of $\epsilon=0.46$, which was tuned to allow  the $J=25$, $v=1$, \AS\ state,  the highest $J$ observed for \AS\ $v=1$ \citep{30BeRyxx.AlH}, to be included in the  AlH line list.

According to the new ExoMol data structure \citep{jt898},  bound and quasi-bound states and the corresponding Einstein $A$ coefficients ($X$--$X$, $X$--$A$) are stored in the bound ExoMol line lists, while continuum \AS\ `states' and the corresponding bound-free transitions to/from the bound \XS\ states form temperature-dependent photo-absorption cross sections, see also \citet{jt865}.

\subsection{(Quasi-)bound line lists of AlH/AlD}

A bound ExoMol line list consists of a States file, Transition file  and  Partition function file computed using bound and quasi-bound wavefunctions. The AlH/AlD line lists cover the wavenumber range up to 30~000~\cm\ ($<0.3333$~\um), $J = 0\ldots 60$ of the \XS\ state, $J_{\rm max} =  25$ (\AS) of AlH  \AS\  and $J_{\rm max} =  36$ (\AS) of AlD. The vibrational excitations of the \XS\ are limited to $v=22$ for both AlH and AlD, which is just below the AlH dissociation limit, while for the \AS\ state these are $v_{\rm max} =1$ (AlH) and $v_{\rm max} = 2$ (AlD).

A States file (see an extract in Table~\ref{t:states}) consists of state IDs, energy term values (\cm), total degeneracies, quantum numbers, energy uncertainties (\cm) and lifetimes (s$^{-1}$). The calculated energies are replaced with the MARVEL values where available. The uncertainties are taken as the MARVEL uncertainties for the substituted values. Otherwise, we use the following empirical and rather conservative expression as an estimate for uncertainties of the calculated energies:
$$
{\rm unc.} = \left\{
\begin{array}{cc}
    0.01\, v + 0.008\, v^2 + 0.0002\, J (J+1), &  X\\
    0.05\, v + 0.008\, v^2 + 0.002\, J (J+1), &  A
\end{array}
\right.
$$

The Transition files (see an extract in Table~\ref{t:trans}) consists of the  IDs of the upper $f$ and lower $i$ states and  Einstein $A_{fi}$ coefficients. The latter are the calculated values, i.e. not modified using the MARVEL energies and are given as reference only. We always recommend using energies from the State file for any practical purposes.

The partition function of AlH has been recomputed with the new line list but is very close to the one computed using the WYLLoT line list. This is unsurprising as the main contribution to the partition function is from the ground electronic state, and we, therefore, do not expect any significant changes from the current model of AlH.
As before the partition function agrees well with the ones derived by  \citet{84SaTaxx.partfunc} and \citet{16BaCoxx.partfunc}.

As a part of the \name\ data set, a set of bound-free temperature-dependent photo-absorption cross sections of AlH and AlD are provided. The cross sections are generated on a wavenumber grid of 0.01~\cm\ ranging from 0 to 30~000~\cm\ for a set of 50 temperatures,  100~K, 200~K, \ldots, 5000~K. The AlH cross sections should be considered as add-ons for the spectra produced using the bound-bound line lists of AlH, see \citet{jt898}.

A line list and photo-absorption data for the minor isotopologue \alh{26} was also produced using the \alh{27} spectroscopic model and the same calculation parameters but for a different mass of Al. $^{26}$Al is radioactive with a half-life
of about 710~000 years and has been clearly detected in the Milky Way \citep{05DiHaKr.Al}. We are not aware of any experimental data on the spectroscopy of\alh{26}.

\begin{table*}
\centering
\caption{ Extract from the states file of the line list for  AlH .}
\label{t:states}
{\tt  \begin{tabular}{rrrrrcclrrrrrrr} \hline \hline
$i$ & $\tilde{E}$ (\cm) & $g_i$ & $J$ & unc. (\cm) & $\tau$ (s$^{-1}$) & \multicolumn{2}{c}{Parity} 	& State	& $v$	&${\Lambda}$ &	${\Sigma}$ & $\Omega$ & Ma/Ca & $\tilde{E}$ (\cm) \\
\hline
     292&   24124.935533&      252 &       10 &     0.009064 &   7.9391E-08 &  -   &  f   & A1Pi     &    0 &   -1 &    0 &   -1 & Ma   &  24124.987854   \\
     293&   25119.784093&      252 &       10 &     0.318000 &   2.0419E-11 &  -   &  f   & A1Pi     &    1 &   -1 &    0 &   -1 & Ca   &  25119.784093   \\
     294&   24253.846482&      276 &       11 &     0.014672 &   8.0291E-08 &  +   &  f   & A1Pi     &    0 &    1 &    0 &    1 & Ma   &  24253.897241   \\
     295&   25228.674665&      276 &       11 &     0.025744 &   5.9852E-12 &  +   &  f   & A1Pi     &    1 &    1 &    0 &    1 & Ma   &  25228.646969   \\
     296&     825.362379&      276 &       11 &     0.010672 &   2.6026E+01 &  -   &  e   & X1Sigma+ &    0 &    0 &    0 &    0 & Ma   &    825.362457   \\
     297&    2426.330942&      276 &       11 &     0.005744 &   4.9291E-03 &  -   &  e   & X1Sigma+ &    1 &    0 &    0 &    0 & Ma   &   2426.334029   \\
     298&    3971.846039&      276 &       11 &     0.010172 &   2.6279E-03 &  -   &  e   & X1Sigma+ &    2 &    0 &    0 &    0 & Ma   &   3971.835720   \\
     299&    5463.280105&      276 &       11 &     0.005910 &   1.8702E-03 &  -   &  e   & X1Sigma+ &    3 &    0 &    0 &    0 & Ma   &   5463.284608   \\
     300&    6901.924290&      276 &       11 &     0.010172 &   1.4991E-03 &  -   &  e   & X1Sigma+ &    4 &    0 &    0 &    0 & Ma   &   6901.933377   \\
     301&    8288.978591&      276 &       11 &     0.005916 &   1.2831E-03 &  -   &  e   & X1Sigma+ &    5 &    0 &    0 &    0 & Ma   &   8288.975579   \\
\hline
\hline
\end{tabular}}
\mbox{}\\
{\flushleft
$i$:   State counting number.     \\
$\tilde{E}$: State energy term values in \cm, MARVEL or Calculated (\textsc{Duo}). \\
$g_i$:  Total statistical weight, equal to ${g_{\rm ns}(2J + 1)}$.     \\
$J$: Total angular momentum.\\
unc: Uncertainty, \cm.\\
$\tau$: Lifetime (s$^{-1}$).\\
$+/-$:   Total parity; e/f: rotationless parity. \\
State: Electronic state.\\
$v$:   State vibrational quantum number. \\
$\Lambda$:  Projection of the electronic angular momentum. \\
$\Sigma$:   Projection of the electronic spin. \\
$\Omega$:   Projection of the total angular momentum, $\Omega=\Lambda+\Sigma$. \\
Label: `Ma' is for MARVEL and `Ca' is for Calculated. \\
$\tilde{E}$: State energy term values in \cm, Calculated (\textsc{Duo}). \\
}
\end{table*}

\begin{table}
\centering
\caption{Extract from the transitions file of the line list for  AlH.}
\tt
\label{t:trans}
\centering
\begin{tabular}{rrrr} \hline\hline
\multicolumn{1}{c}{$f$}	&	\multicolumn{1}{c}{$i$}	& \multicolumn{1}{c}{$A_{fi}$ (s$^{-1}$)}	\\ \hline
         416    &      424 & 2.9549E+04\\
         391    &      395 & 2.8823E+04\\
         362    &      370 & 2.8164E+04\\
         883    &      861 & 1.8092E+05\\
         998    &      953 & 1.9332E-11\\
         337    &      341 & 2.7560E+04\\
         308    &      316 & 2.7001E+04\\
         835    &      838 & 1.6168E+05\\
         282    &      287 & 2.6475E+04\\
    \hline\hline
\end{tabular} \\ \vspace{2mm}
\rm
\noindent
$f$: Upper  state counting number;\\
$i$:  Lower  state counting number; \\
$A_{fi}$:  Einstein-$A$ coefficient in s$^{-1}$. \\
\end{table}

\subsection{Temperature-dependent photo-absorption cross sections of AlH/AlD}

Using energies and Einstein coefficients from the bound (\XS) and continuum (\AS) solutions, a set of temperature-dependent cross sections of AlH and AlD are computed, using a wavenumber grid of 0.01~\cm\ and a temperature grid ranging from 100~K to 5000~K in steps of 100~K. Here we use the procedure established in \citet{jt840}, where all discrete transition intensities to the continuum states are re-distributed in their vicinity to form continuum photo-absorption cross sections using a Gaussian line profile
$$
f(\tilde\nu ) = \sqrt{\frac{\ln2}{\pi}} \frac{1}{\alpha_{\rm G}} \exp\left(-\frac{(\tilde\nu-\tilde\nu_{fi})^2\ln2}{\alpha_{\rm G}^2}\right),
$$
where $\alpha_{\rm G}$ is the Gaussian half-width-at-half-maximum (HWHM). For the size box of $\sim$ 60~\AA, the distance between the `continuum' lines does not exceed  26~\cm, which adopt as the values of $\alpha_{\rm G}$.

\begin{figure}
    \centering
    \includegraphics[width=0.97\columnwidth]{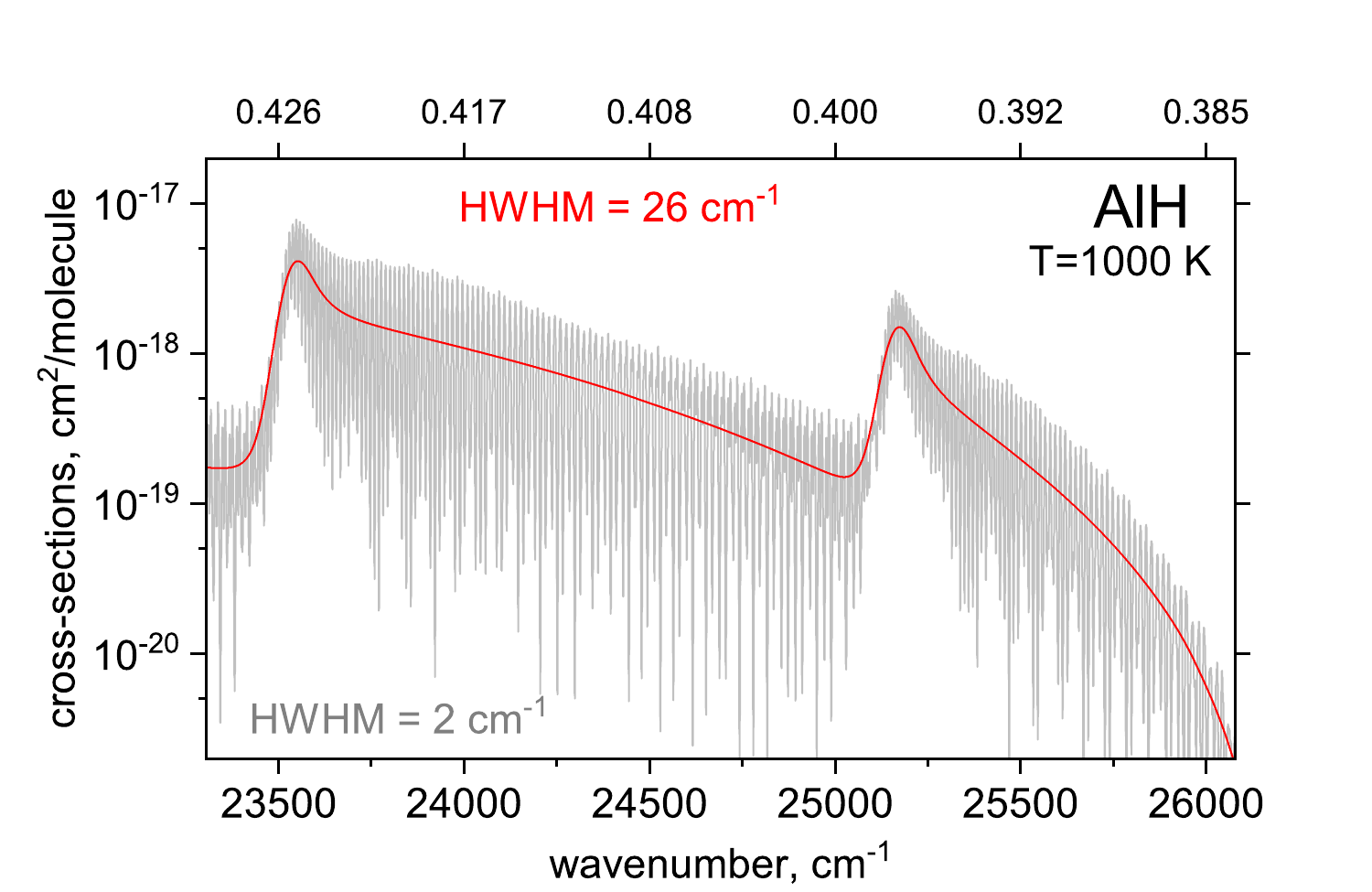}
    \caption{Example of the photo-absorption continuum cross sections of AlH at $T=1000$~K generated using a Gaussian line profile of HMWM = 26~\cm\ (red line), overlaid with the same spectrum but using HMWM = 2~\cm\ (grey line). }
    \label{f:smoothing}
\end{figure}

Figure~\ref{f:smoothing} shows the continuum (bound-unbound) spectrum of AlH at $T=1000$~K generated using the Gaussian profile smoothing with HWHM of 26~\cm. As an illustration, the original separation between the `unbound' discrete absorption lines before the smoothing applied can be seen in the same spectrum generated using \mbox{HWHM = 2~\cm}.

When computing the total cross sections of a molecule using the extended ExoMol format \citep{jt898}, we first compute cross sections for a given temperature and pressure using the \mbox{(quasi-)bound} line list and then add them to the photo-absorption cross section for the temperature in question. The pressure dependence of the continuum transitions is ignored.

\begin{figure*}
    \centering
\includegraphics[width=0.44\textwidth]{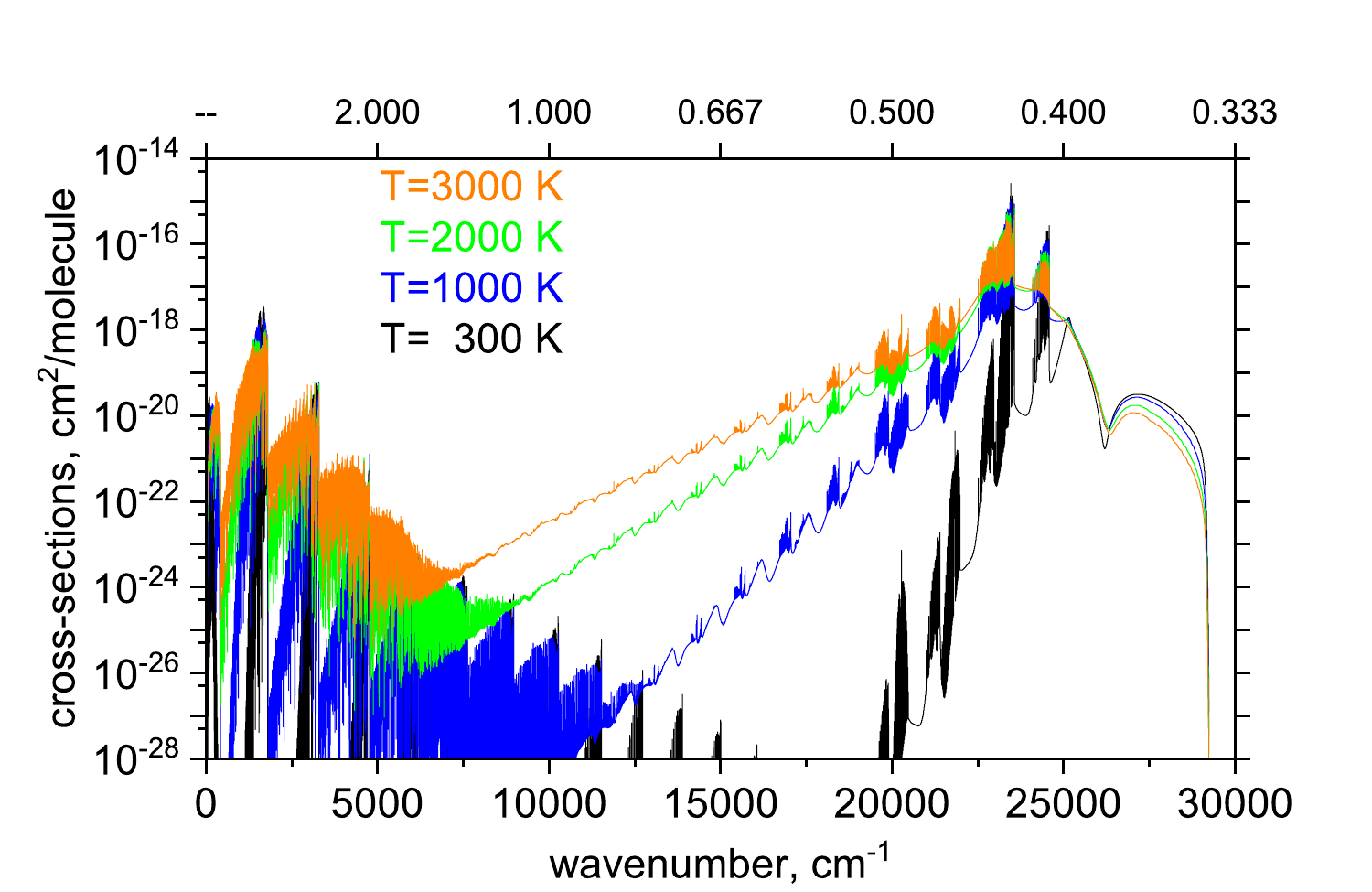}
\includegraphics[width=0.44\textwidth]{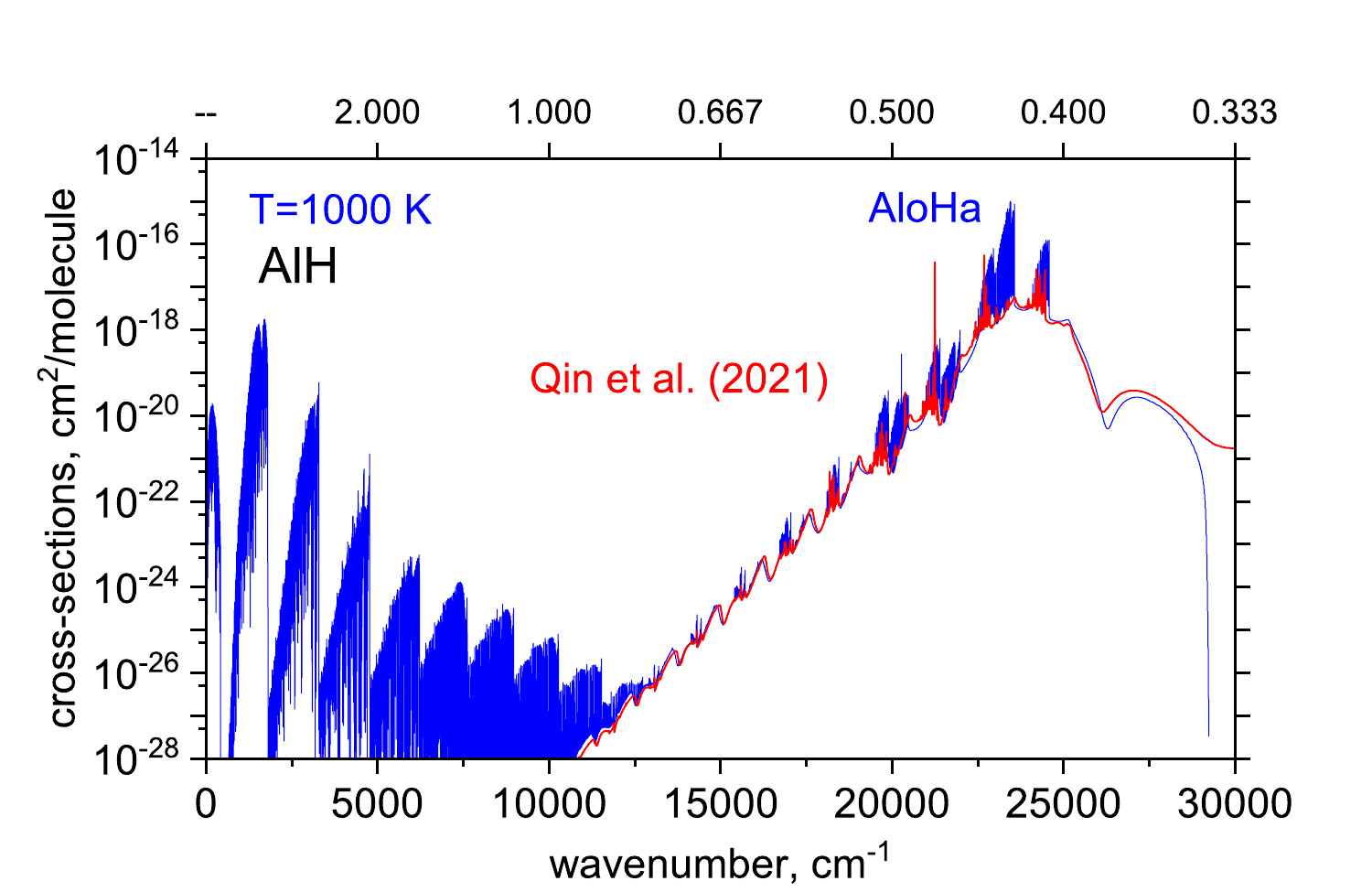}
\caption{Temperature dependence of the AlH absorption spectrum using the Gaussian profile with HWHM = 1 \cm\ (left) and a comparison with the $T=1000$~K cross-sections of AlH by \citet{21QiBaLi.AlH} }
    \label{f:Temp:AlH}
\end{figure*}

Figure~\ref{f:Temp:AlH} (left) shows total (bound+continuum) cross sections of AlH for four temperatures and zero pressure computed using the procedure described above. In the same figure, where the 1000~K spectrum is also compared to the \ai\ cross sections by \citet{21QiBaLi.AlH}. Despitte a generally good agreement between the continuum contributions,  our semi-empirical model provides more accurate data for high-resolution applications.

\begin{figure*}
    \centering
\includegraphics[width=0.44\textwidth]{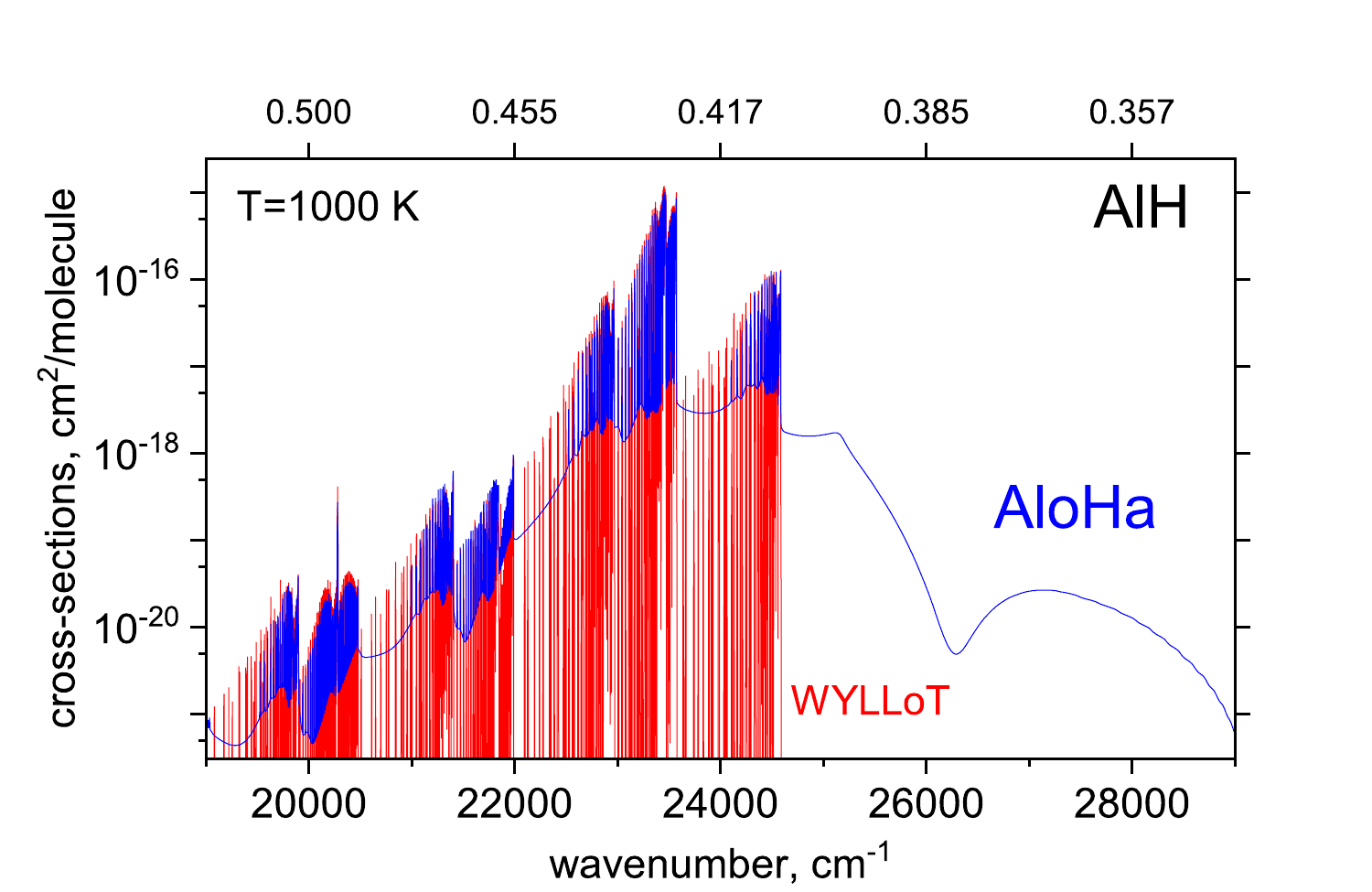}
\includegraphics[width=0.44\textwidth]{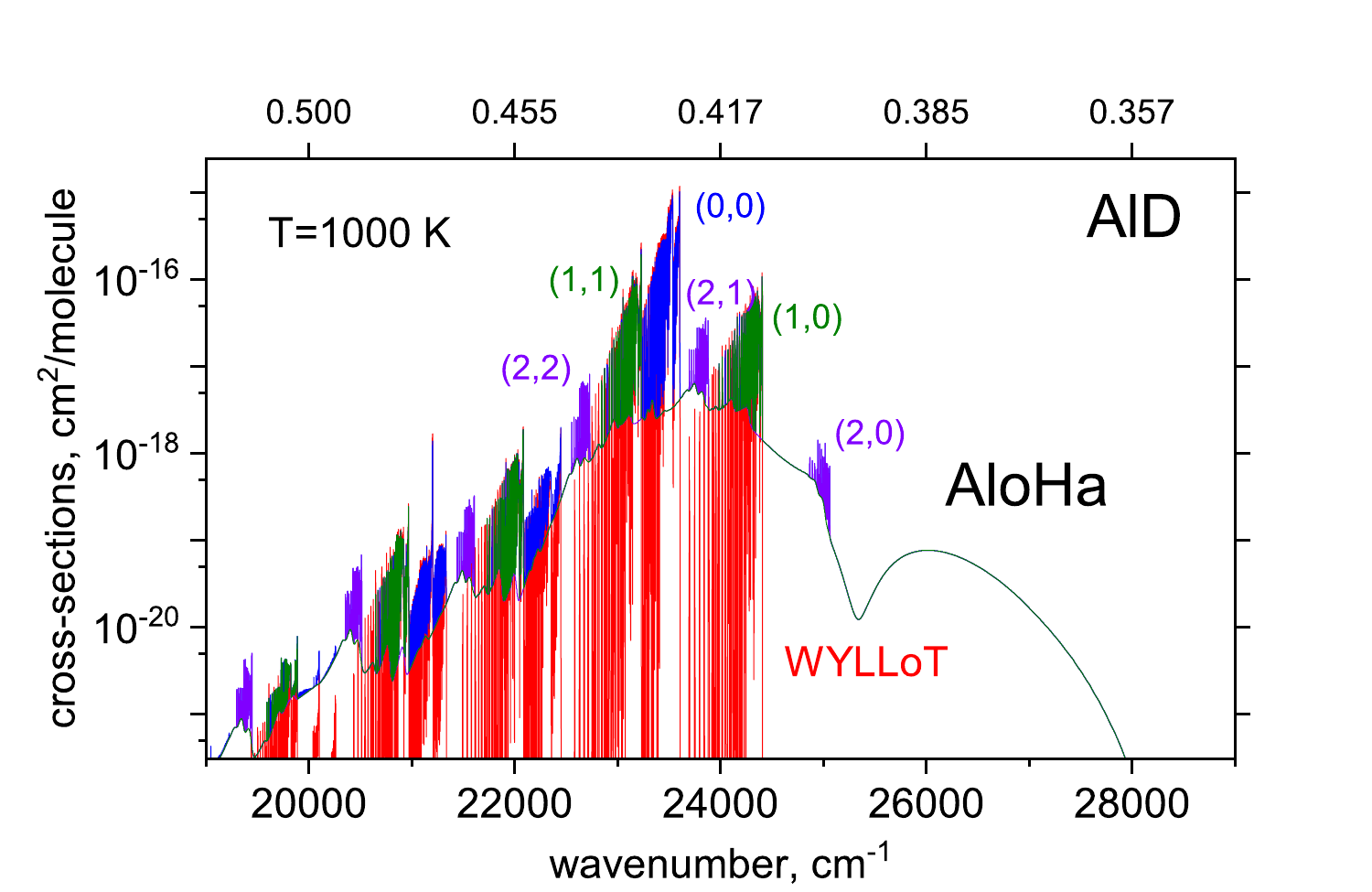}
\caption{Comparison of the WYLLoT and \name\ simulations of $T= $ 1000~K absorption spectra of AlH (Left display) and AlD (Right display) using the Gaussian line profile with  HWHM of 1~\cm. }
    \label{f:old:new:AlH}
\end{figure*}

In Figure~\ref{f:old:new:AlH}, we compare absorption spectra of AlH and AlD simulated using the WYLLoT and \name\ line lists at $T=1000$~K. The main differences are (i) the continuum contributions in the spectra and (ii) the $v'=2$ bands in the spectrum of AlD, missing in the WYLLoT simulations.

\subsection{Simulations of spectra of AlH and AlD}

\begin{figure}
    \centering
\includegraphics[width=0.97\columnwidth]{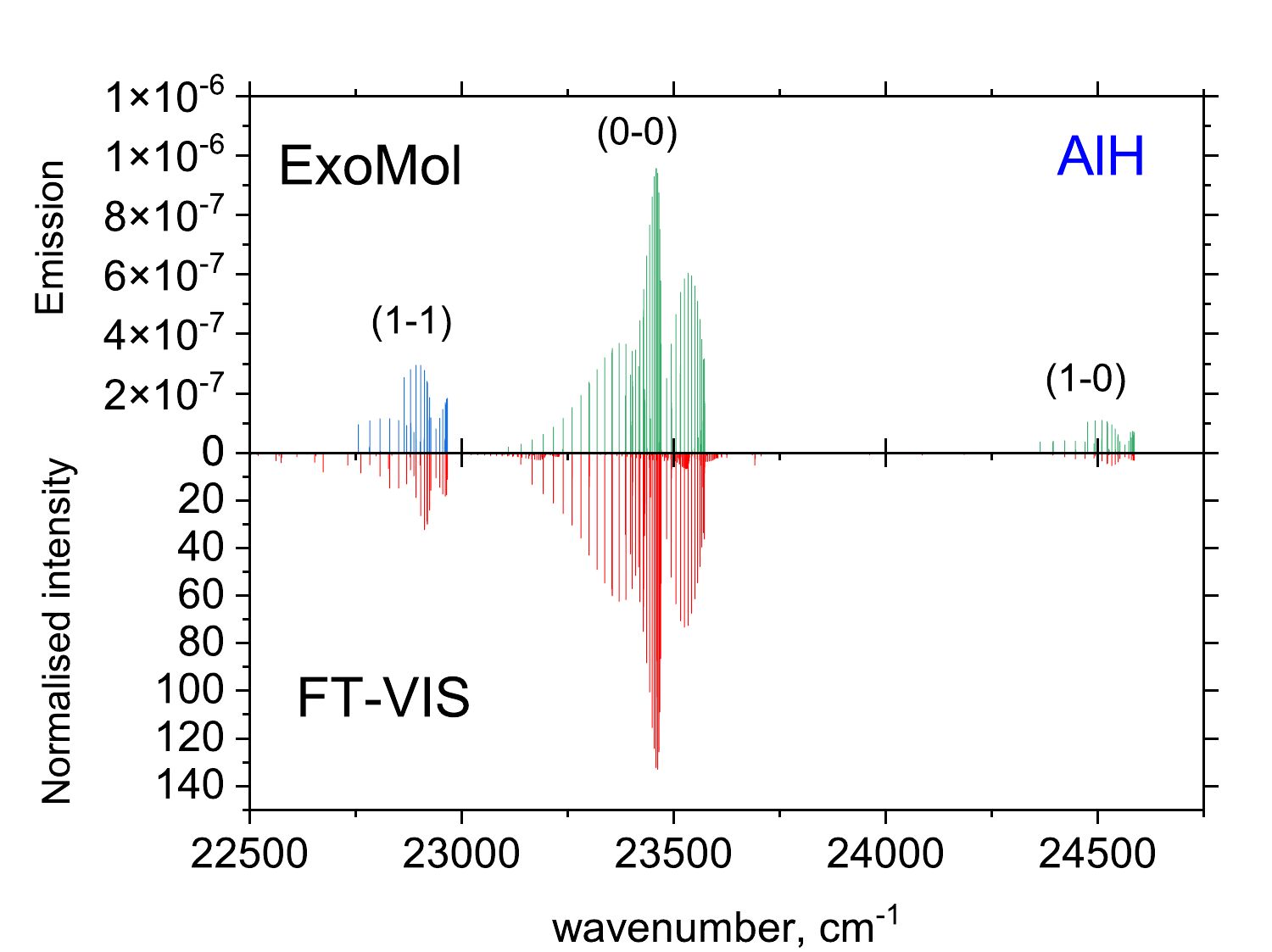}
\caption{Comparison of the experimental FT-VIS by \citet{23SzKePa.AlH} and simulated \name\  \AX\ spectrum of AlH in the region of the (1-1), (0-0) and (1-0) bands. For the theoretical spectrum, a rotational temperature of $T=750$~K and vibrational temperature $T = 4500$~K were used; and a Gaussian line profile with FWHM of 0.08~\cm\ was assumed}
    \label{f:FT:AlH}
\end{figure}

As illustrations, here we simulate emission spectra of AlH and AlD to compare to the experimental spectra from \citet{23SzKePa.AlH} and the current work, respectively. All spectra were generated using our open-access Fortran code \exocross\ \citep{ExoCross}\footnote{\exocross\ can be obtained at \url{github.org/exomol}}. Figure \ref{f:FT:AlH} shows a general overview of the AlH emission spectra generated using a Gaussian line profile with the HWHM of 0.08~\cm\ and the rotational temperature of 750~K, where the vibrational temperature of $T_{\rm vib}=$ 4500~K was  assumed as in \citet{jt732}.

\begin{figure*}
    \centering
\includegraphics[width=0.44\textwidth]{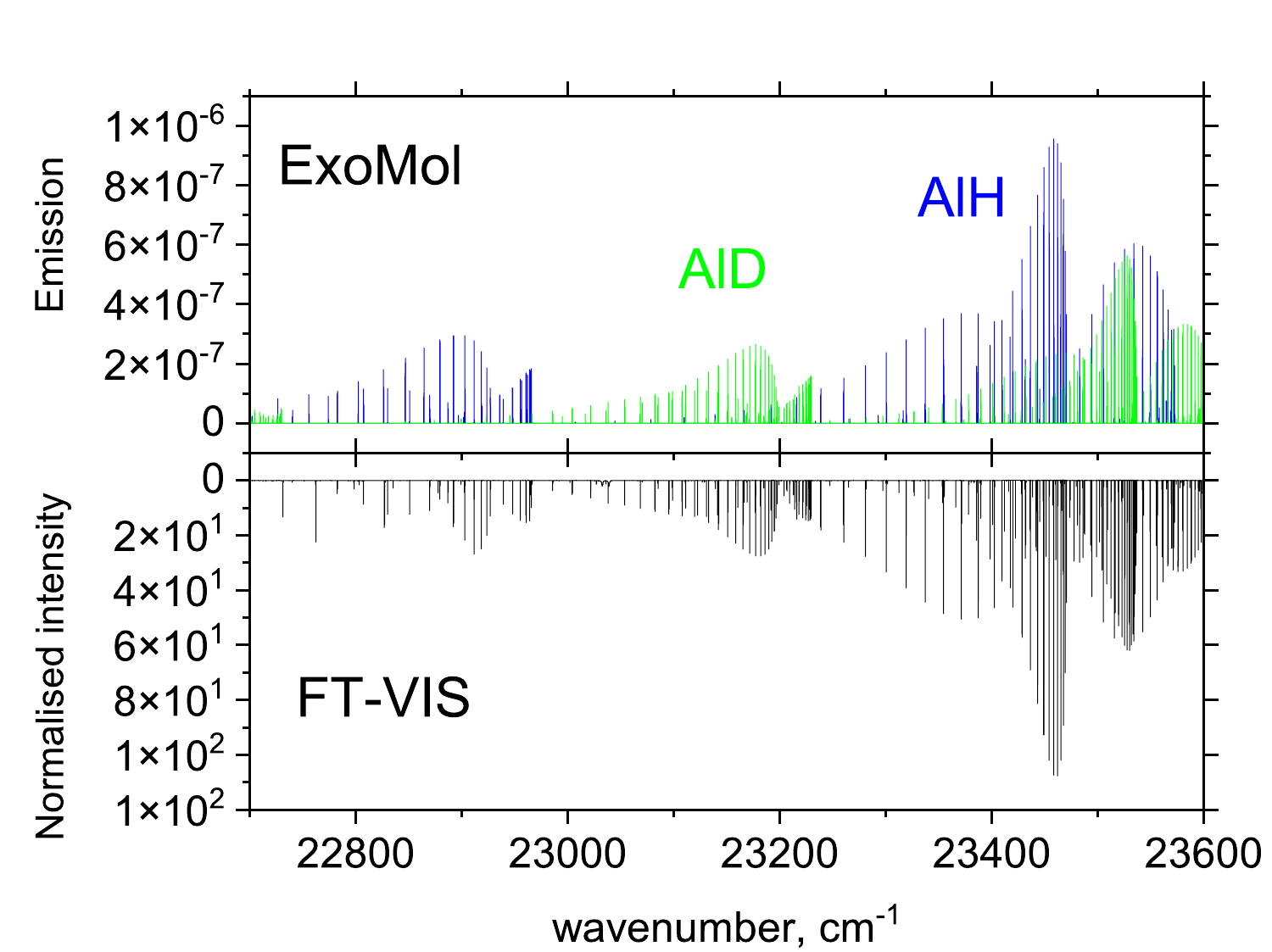}
\includegraphics[width=0.44\textwidth]{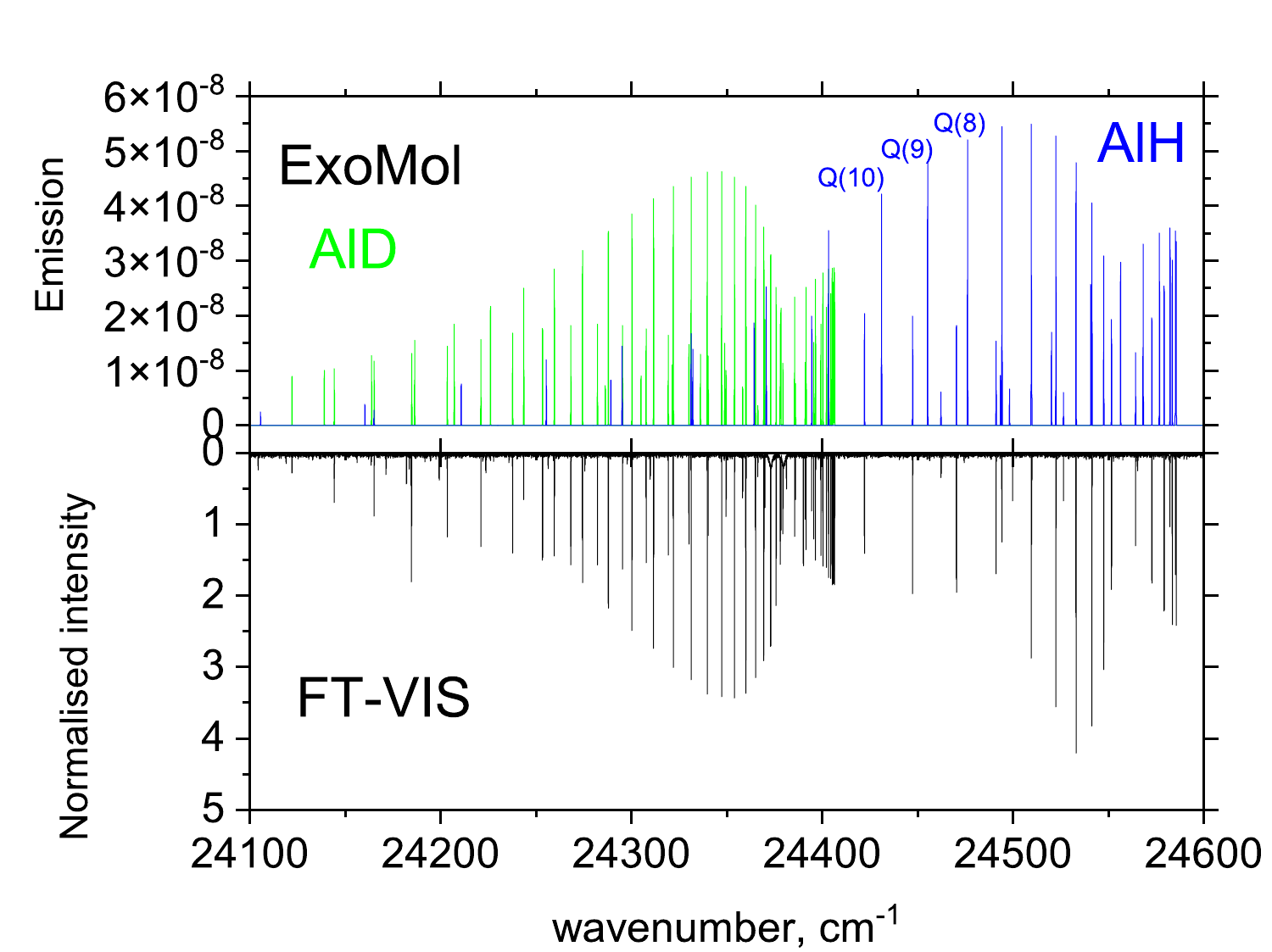}
\caption{Comparison of the current experimental FT-VIS and simulated ExoMol  \AX\ spectrum of AlD  with AlH present:
Left display, (1-1) and (0-0) bands, assuming AlH:AlD is 1:1. Right display, (1-0) band, AlH:AlD is 1:0.5. For the theoretical spectrum a rotational temperature of $T=750$~K and a Gaussian line profile with FWHM of 0.08~\cm\ were assumed.}
    \label{f:FT:22800}
\end{figure*}

Figure \ref{f:FT:22800} provide a similar illustration for AlD, where the simulations of the regions containing the (1-1), (0-0) and (1-0) bands are shown. The appearance of an extra line in the right display is due to the predissociative effects and discussed below.

For the sake of completeness, we reproduce a comparison of the IR of AlH (\XS--\XS) with the emission measurements by \citet{93WhDuBe.AlH}, see \citet{jt732}. The current line list preserves the high quality of the original ExoMol line list WYLLoT.

\begin{figure}
    \centering
\includegraphics[width=0.97\columnwidth]{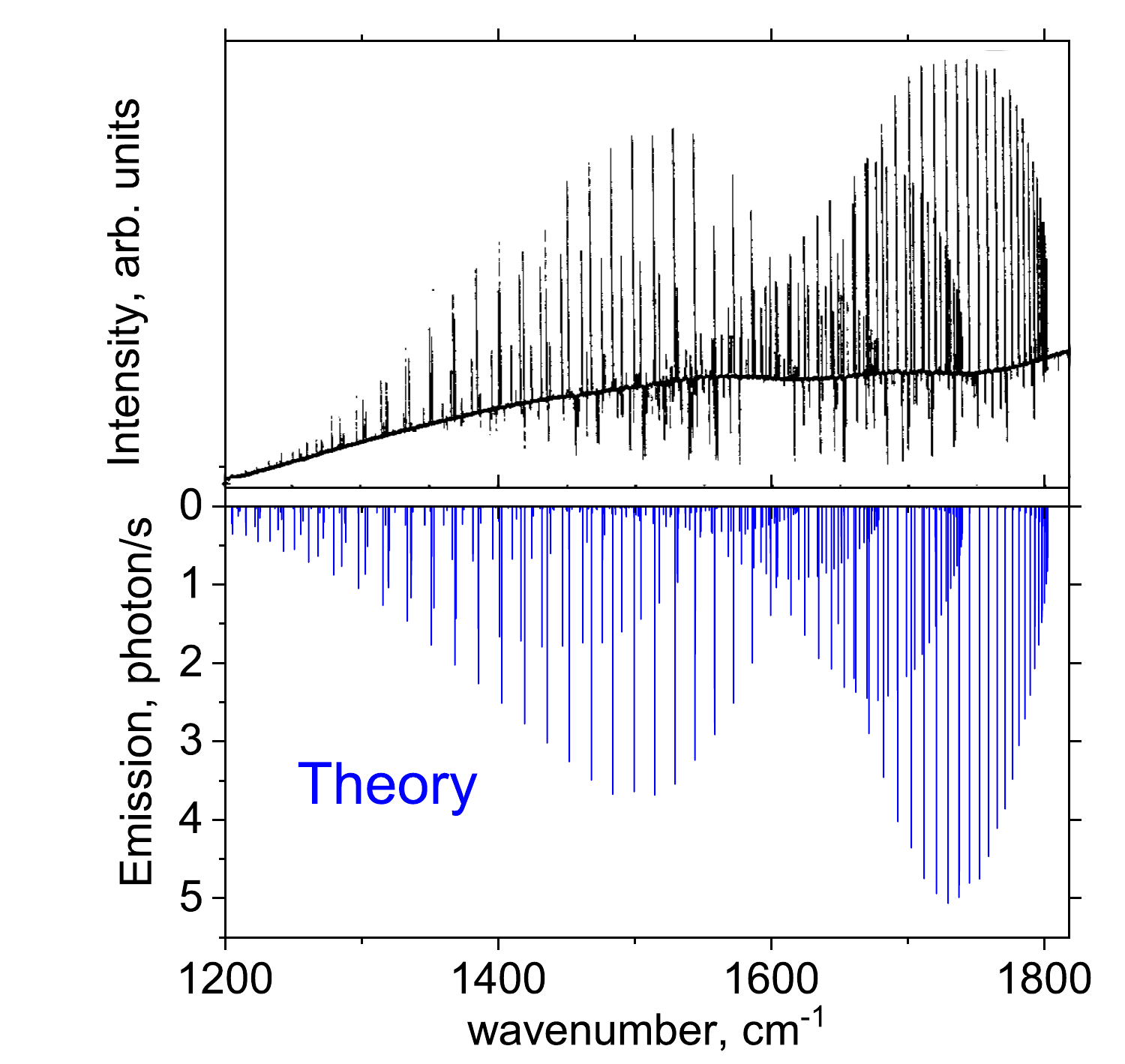}
\caption{Infrared spectrum of AlH by \citet{93WhDuBe.AlH} (upper) compared to the emission spectrum computed using our line list assuming a temperature of 1700 K and a Gaussian profile with the FWHM of 0.01 \cm. }
    \label{f:Bernath:AlH}
\end{figure}

\begin{figure*}
    \centering
\includegraphics[width=\textwidth]{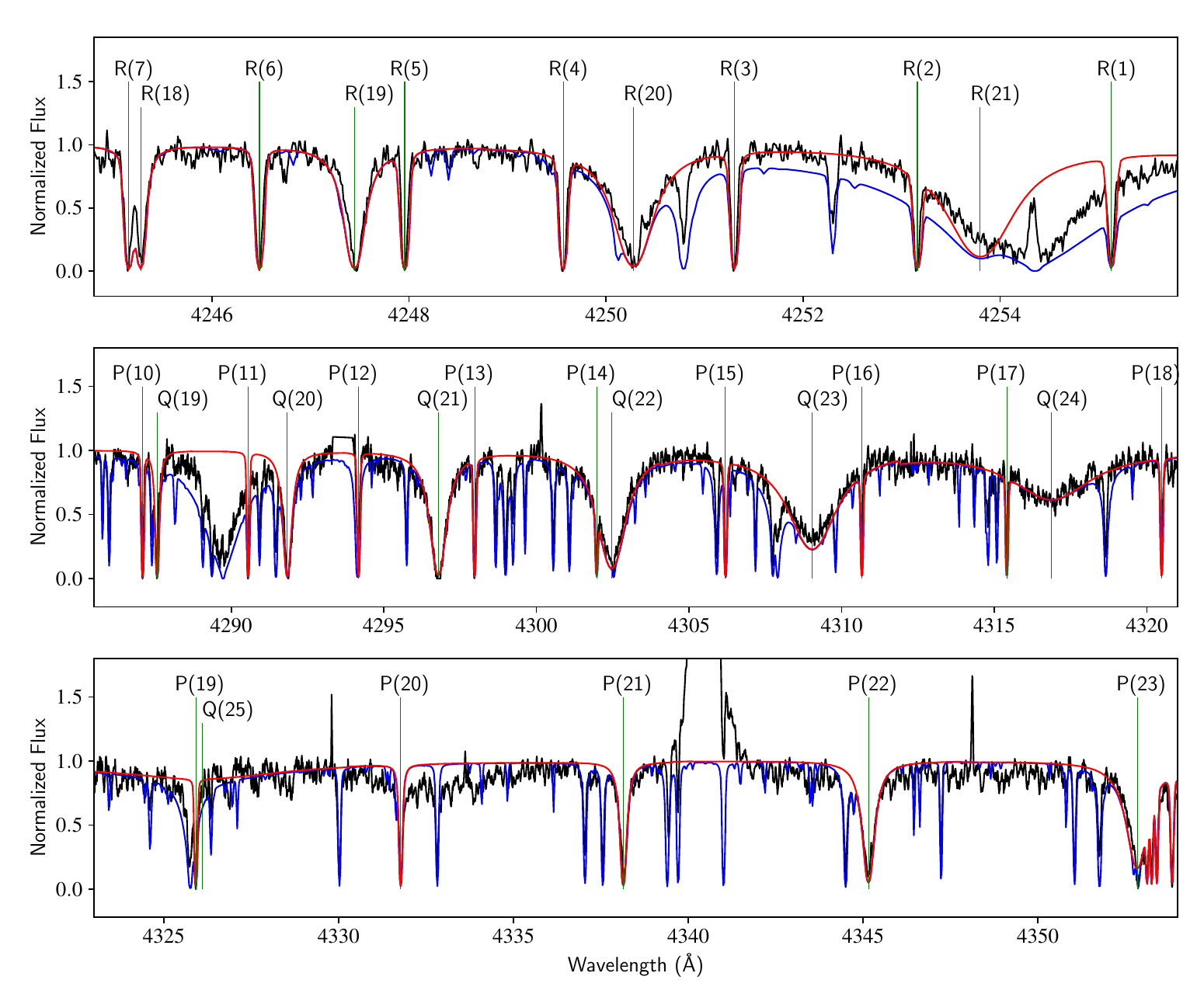}
\caption{Comparison with the observed spectrum of AlH of Proxima Cen analysed by \citet{jt874} shown by black line for the (0-0) band spectral range.
Here and onward, we have adhered to the procedure of computation and identification of spectral features outlined by \citet{jt874}. The blue line marks the synthetic spectrum including atomic and molecular species, the red line spectrum is calculated including AlH lines only. A version of this figure, with the atomic lines also indicated, is provided in the Appendix, see \ref{f:prox:3}.}
    \label{f:prox:1}
\end{figure*}

\begin{figure*}
    \centering
\includegraphics[width=\textwidth]{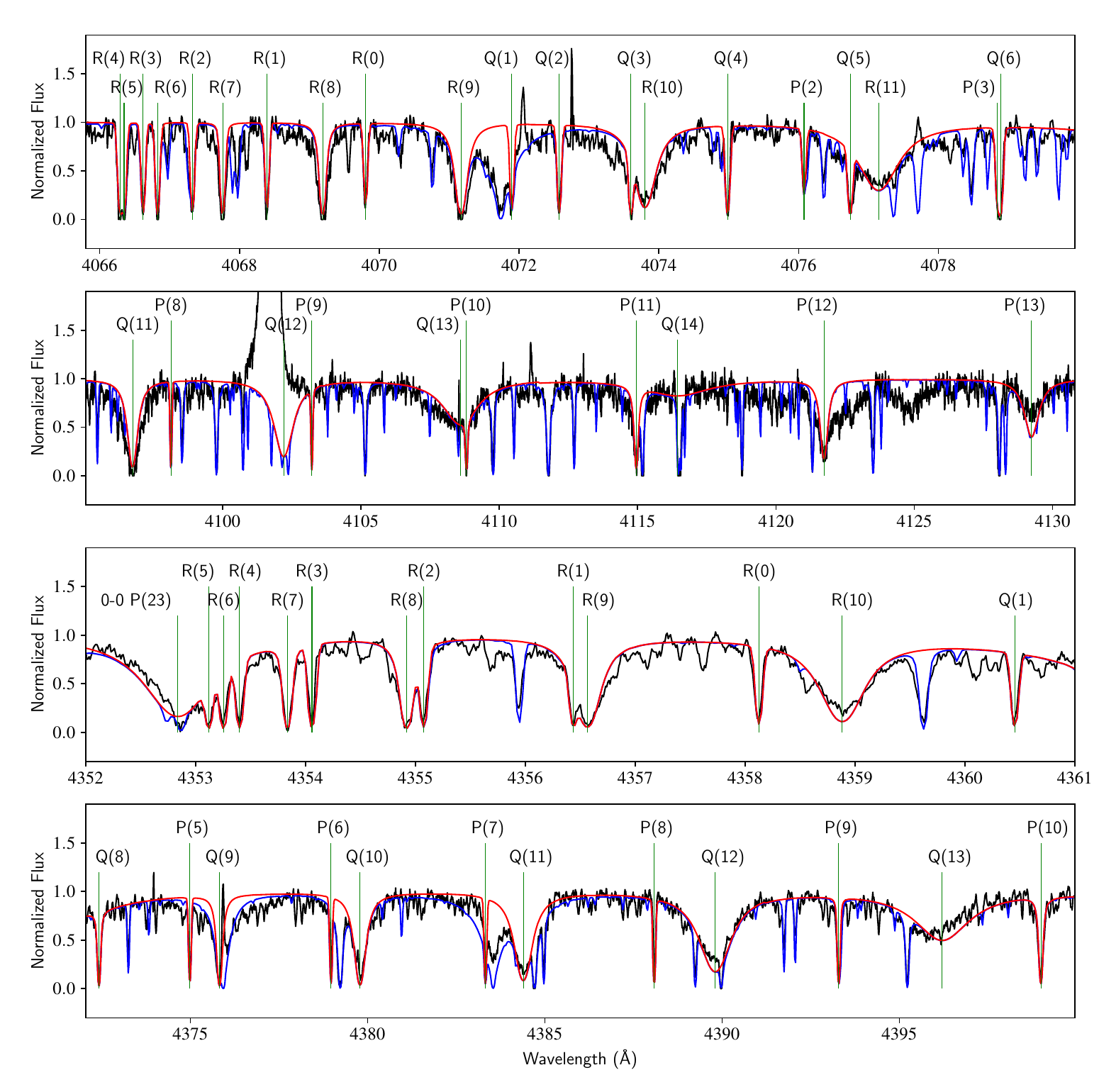}
\caption{Comparison of the spectrum of AlH of Proxima Cen by \citet{jt874} with the synthetic spectrum. Two upper panels show the spectral range of the 1-0 band and the two lower of the 1-1 band. The observed spectrum is shown by black line. The blue line marks the synthetic spectrum including atomic and molecular species, the red line spectrum is calculated including AlH lines only.
}
    \label{f:prox:2}
\end{figure*}

\citet{jt874} recently studied the absorption of AlH in the spectrum of  Proxima Cen  from the HARPS ESO public data archive \citep{03MaPeQu}, recorded over the spectral range from 3780 to 6810 \AA\ with a resolving power $R \sim 115 000$. This study has demonstrated the importance of the accurate description of line lists of AlH. In particular, the previous AlH line list WYLLoT was shown to deteriorate the higher $J$ spectral lines of AlH in the $v=0$ and $v=1$. It also could not describe the  predissociative broadening effects in this band for $J>19$ in $v'=0$ and $J>8$ in $v'=1$.
In Figs.~\ref{f:prox:1} and \ref{f:prox:2} we simulate the high resolution AlH spectrum  in model of stellar atmosphere appropriate for Proxima Cen in the spectral region covering (1-0), (0-0) and (1-1) bands of A$^1\Pi$ -- X$^1\Sigma^{+}$ system. For details of the calculations please consult \citet{jt874}. To underline the prominent presence of AlH molecular lines in the spectrum one of the two synthetic spectra (the red one) includes only molecular lines.
Fig.~\ref{f:prox:1} shows (0-0) band and Fig.~\ref{f:prox:1} presents bands which upper level is $v'=1$ - the first two  panels show the (1-0) band and the next two panels show (1-1) band.
The actual list of lines of AlH shows very good consistency with the observed spectrum both in line position and in profiles of diffusive lines. Some differences in line shapes and in depths of broad atomic lines and of diffusive molecular lines may be ascribed to the uncertainty in the continuum tracing of the observed spectrum before its normalization.
Simulations show a much better description of the AlH spectrum in  Proxima Cen, including the predissociative broadening effects. Even the heavily predissociated lines $Q(13)$, $Q(14)$, $Q(15)$ of $v'=1$ and $Q(23)$ and $Q(24)$ of $v'=0$ can be clearly recognised. The presence of $Q(25)$ (See the bottom panel of Fig.~\ref{f:prox:1}) is less evident in the observed spectrum. The approach used to model the predissociation line broadening is described in details in Section~\ref{s:lifetime}.


\subsection{Breaking-off of predissociation lines of AlH and AlD}

\begin{figure}
    \centering
    \includegraphics[width=0.97\columnwidth]{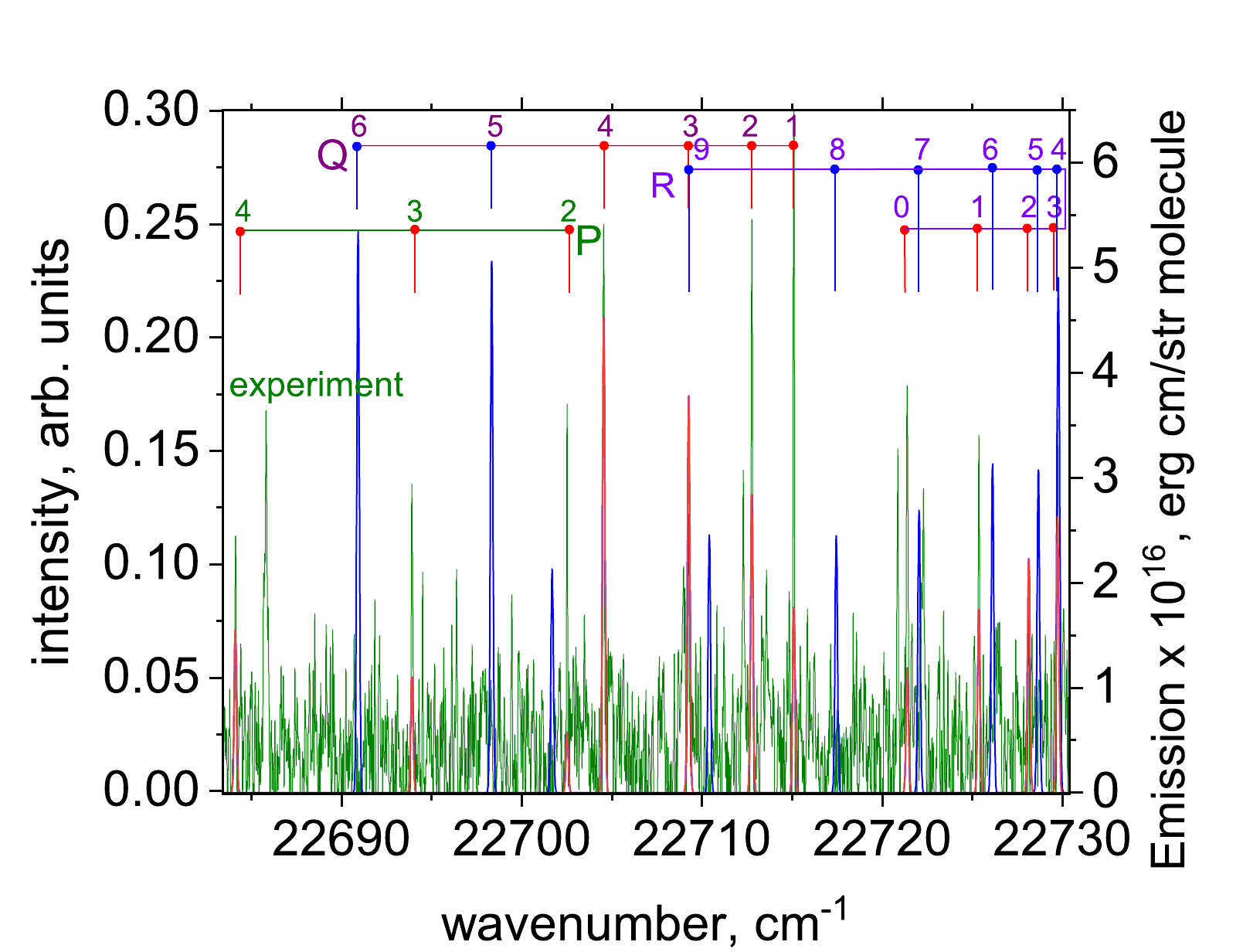}
\caption{Comparison of the experimental FT-VIS and simulated ExoMol \AS-\XS\ (2-2) emission band of AlD. For the theoretical spectrum a rotational temperature of $T=750$~K and a Gaussian line profile with HWHM of 0.08~\cm\ were assumed.}
    \label{f:FT:v=2}
\end{figure}

Figure \ref{f:FT:v=2} shows the experimental spectrum of the (2,2) band of AlD from this work and our attempt to model it using the new \name\ line list. Only the $J'\le 4$ lines appear in the experiment while the theory predicts lines with higher $J$. In fact, higher $J$ ($J\le 11$) predissociative lines were observed experimentally by \citet{48Nilsson.AlH}. The effect of ``breaking-off'' of the predissociative lines in different experimental setups was studied by \citet{30BeRyxx.AlH} and  discussed by
\citet{39Herzberg.book} and was attributed to the non-local thermal equilibrium (non-LTE) effects present in some low pressure conditions. In LTE, the number of predissociating molecules is compensated by new molecules formed by inverse predissociation.

\begin{figure}
    \centering
    \includegraphics[width=0.97\columnwidth]{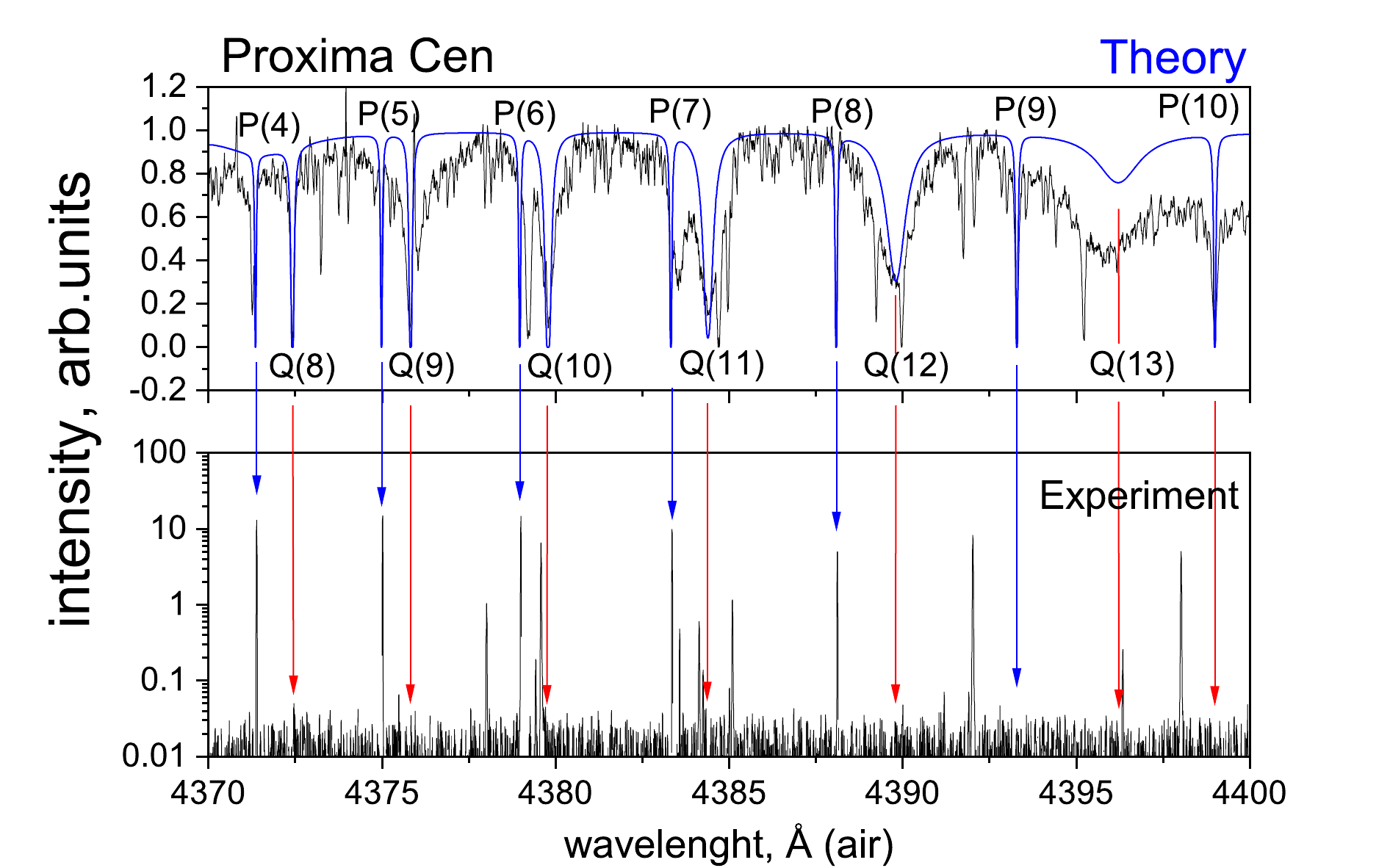}
\caption{Comparison of the experimental FT-VIS \citep{23SzKePa.AlH} of AlH  (bottom) and the Proxima Cen spectrum (top), observation and  simulation  ($T=2900$~K) in the region of \AS-\XS\ (1-1) band using the new ExoMol line list as in the bottom display of Fig.~\ref{f:prox:1}. The blue/red arrows indicate the AlH lines that are present/disappear in the experiment. }
    \label{f:exp:Proxima}
\end{figure}

This effect can be nicely demonstrated in the comparison of the experimental FT spectrum of AlH from \citep{23SzKePa.AlH} with our that of Proxima Cen as shown in Fig.~\ref{f:exp:Proxima}. This figure reproduces our simulation of the Proxima Cen from the bottom display of Fig.~\ref{f:prox:1} and the experimental spectrum is converted to air for a better comparison. It is evident how the emission lines from the FT spectrum break off for  $J'>8$  in comparison to the spectrum of Proxima Cen. It should be noted that this is not due to the lower temperature conditions of the FT spectrum. Indeed, if we assumed the LTE,  the population of the corresponding states with $J \ge 8$ is comparable to those visible in the spectrum at $T=750$~K, indicating that the breaking-off of $J \ge 8$ in the experiment is due to non-LTE effects.

The effect can be also seen in right display of Fig.~\ref{f:FT:22800}, where extra $Q(J)$ lines ($J>7$) of AlH appear compared to the experimental spectrum.

\subsection{Collisional line-broadening parameters}

Collisional line-broadening parameters of AlH for the \XS\ state with different partners (H$_2$, He, N$_2$, and AlH) have been computed using the MCRB approach \citep{06AnGaSz}.
This is a semi-classical approach where internal degrees of freedom of the radiator and the perturber are treated quantum-mechanically and their relative translational motion is described classically. Line broadening can be said to appear as a consequence of monochromatic wave-train interruption when the radiating molecule is interacting with a perturber during a collision. The magnitude of this effect for completed collisions is described with a scattering matrix, which in this approach is expanded up to the second order in perturbation theory \citep{08HaBoRo}.
The model interaction potential between the radiator (AlH) and a perturber is constructed from short-ranged and long-ranged parts. The former, repulsive part is obtained from atom-atom Lennard-Jones contributions \citep{62Svehla}, while the latter is composed from electrostatic interactions and uses molecular multipole moments from NIST \citep{22Johnson}. Trajectories are computed within the rigid rotor approximation using equilibrium geometries  of the \XS\ state and a different isotropic potential to drive them \citep{21LoShxxa}.

\begin{figure}
    \centering
    \includegraphics[width=0.97\columnwidth]{ 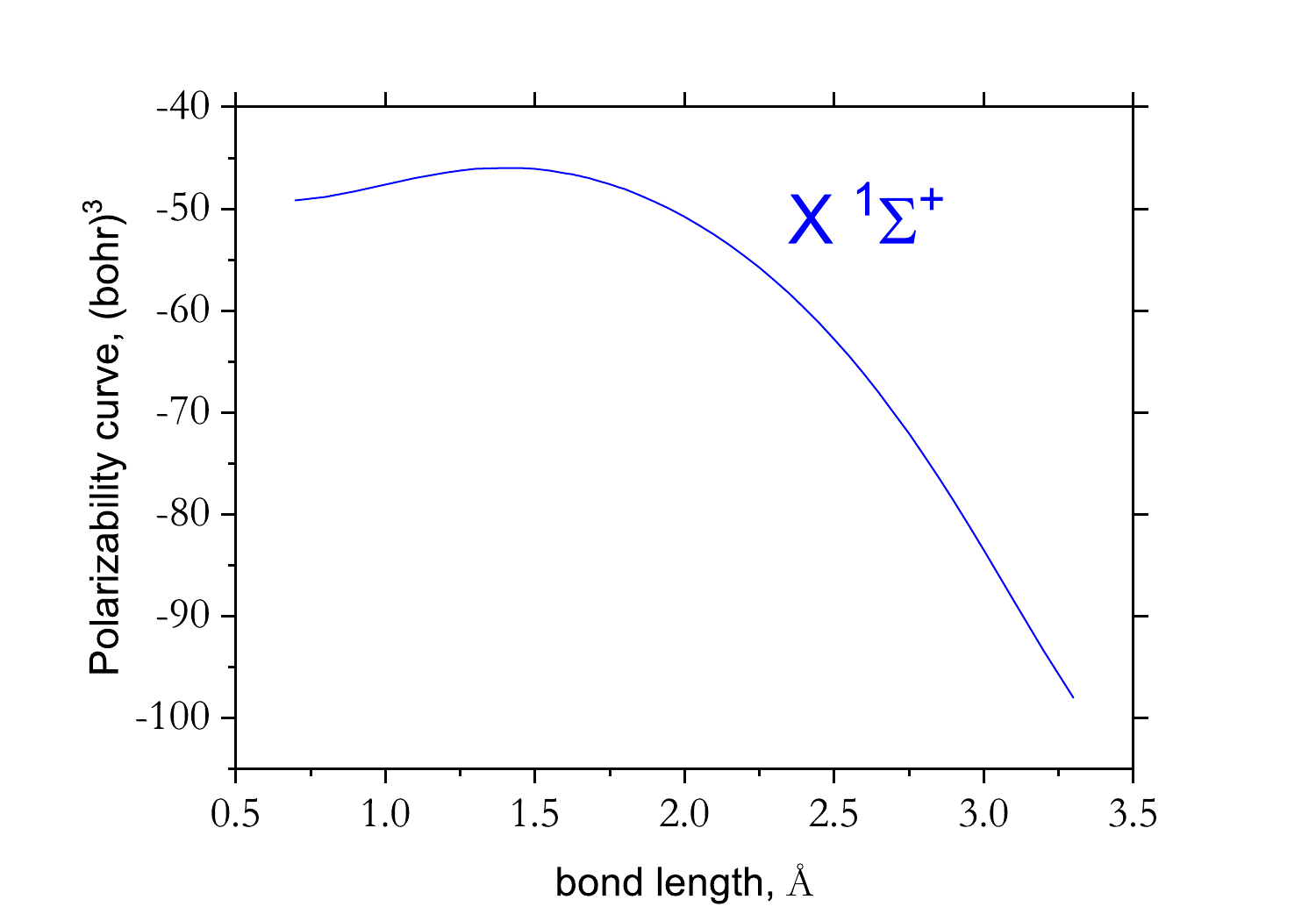}
\caption{\textit{Ab initio} isotropic electric polarizability curve of AlH.}
    \label{f:IPC}
\end{figure}

Vibrational dependence of broadening parameters  have also been modelled, assuming that only the changes in long-ranged van-der-Waals interactions with vibrational state significantly change scattering cross-sections. Diagonal rovibrational matrix elements of the \XS\ electric dipole moment $\langle vJ \!\mid\!\! \mu^2(r) \!\! \mid \! vJ \rangle$ and  isotropic polarizability  $\langle vJ \mid\! \alpha_{\text{iso}}(r) \!\mid vJ \rangle$ curves  required for this part were computed using \Duo's ro-vibrational  wavefunctions $\ket{vJ}$. The polarizability curve $\alpha_{\text{iso}}(r)$ was computed \ai\ with MOLPRO \citep{MOLPRO2020} using the CCSD(T)/aug-cc-pVQZ level of theory as  second order derivatives of the \XS\ energy with respect to the electric field. It is shown in Fig.~\ref{f:IPC}. The dipole moment curve of \XS\ was taken from \citet{jt732}.

It is worth mentioning that this semi-classical approach works best when the interaction potential is fitted to improve agreement with experimental broadening coefficients. Without this adjustment, theoretical values usually overestimate experimental ones \citep{13MaBoTi}. However, to the best of our knowledge, no experimental measurements of AlH broadening by any molecules are available, so our broadening coefficients are presented without any adjustments.

The new broadening parameters of AlH are included into the ExoMol database using the ExoMol diet format \citep{jt684}, which is based on the representation of the temperature- and pressure-dependence of the half-width-at-half-maximum $\gamma$ {(\cm/atm)} by a  single-power law:
\begin{equation}
\gamma(T,P) = \gamma_{0} \left(\frac{T_{\rm ref}}{T}\right)^n \frac{P}{P_{\rm ref}},
\end{equation}
where $T_{\rm ref}$ is the HITRAN reference temperature of 296~K and $P_{\rm ref}$ is the reference pressure of 1~atm.
The $J$-dependence is best parameterised  by the standard HITRAN $\mid\!\! m\!\!\mid$ dependence , where $m=J_\text{lower} + 1$ for the R-branch and $m=-J_\text{lower}$ for the P-branch. We have therefore introduced a new ExoMol diet type $m0$. An example of the the diet file for AlH broadened by H$_2$ is given in Table~\ref{t:diet}. Our MCRB calculations predict a mildly sloping dependence on $m$ (and therefore $J$). The \verb!m0! type is implemented and now available in \exocross.

The methodology described is currently only applicable to the ground \XS\ electronic state rovibrational transitions. The production of line shape parameters for rovibronic transitions is more complicated, see \citet{jt907}, and will be considered separately.

\begin{table}
\centering
\caption{ExoMole diet line broadening file for AlH with H$_2$ as perturber.}
\tt
\label{t:diet}
\centering
\begin{tabular}{rrrr} \hline\hline
Type & $\gamma_{0}$ & $n$ & $m$   \\
    \hline
m0         &    0.1482&   0.6179   &       1   \\
m0         &    0.1462&   0.6107   &       2   \\
m0         &    0.1443&   0.6040   &       3   \\
m0         &    0.1420&   0.5962   &       4   \\
m0         &    0.1401&   0.5903   &       5   \\
m0         &    0.1386&   0.5862   &       6   \\
m0         &    0.1375&   0.5837   &       7   \\
m0         &    0.1367&   0.5825   &       8   \\
m0         &    0.1362&   0.5824   &       9   \\
m0         &    0.1360&   0.5833   &      10   \\
m0         &    0.1360&   0.5851   &      11   \\
m0         &    0.1362&   0.5876   &      12   \\
m0         &    0.1365&   0.5904   &      13   \\
    \hline\hline
\end{tabular}
\end{table}

\section{Conclusions}

Improved line lists for AlH and AlD (\XS, \AS) are presented. They now provide a better description of the high $J$ predissociation effects in the \AS\ state and a proper description of predissociative line broadening via the inclusion of the predissociative lifetimes into the ExoMol States file. The AlD line list now contains the $v'=2$ \AS\ predissociative band, which was not present in WYLLoT. As part of the \name\ line list, we also provide temperature-dependent photo-absorption cross sections of AlH/AlD. These data are complimentary and should be added to the temperature- and pressure-dependent cross sections produced from the bound-bound line lists.
The AlH \name\ line lists are  freely available at \url{www.exomol.com}.

The new AlH/AlD line lists can be used for some modelling and analysis of non-LTE spectral effects, at least as far as the  radiative rates are concerned. As it is typical for diatomics, the hot vibronic  bands of AlH are well separated (see Fig.~\ref{f:old:new:AlH}), which helps estimate the vibrational temperatures (populations) of the corresponding (lower) states and thus to assess the presence and magnitude of non-LTE effects, see, e.g. \citet{22WrWaYu,jt903}. However, a full non-LTE treatment would also require the other contributions to the statistical population balance, including collisional rates and reaction rates  \citep{07vaBlSc}, which will need further work.

It should be noted, that there are experimental data on the higher excited singlet states \cite{17SzMoLa.AlH, 10SzZaxx.AlH, 58Khan.AlH, 62Khan.AlH, 92ZhShGr.AlH, 28Bengtsson.AlH, 39GrHuxx.AlH, 34Holst.AlH, 39GrHuxx.AlH, 70LaLuNe.AlH} and triplet states of AlH \cite{17SzHaKo.AlH, 03TaTaDa.AlH, 37ChAlxx.AlH, 33Holst.AlH, 92ZhShGr.AlH, 53Kleman.AlH}, which can be used to extend the current spectroscopic model and the line list as part of the  future work.

In AlH the lifetime broadening is due to tunneling in the \AS\ state. More commonly predissociation is caused by tunneling. Work reporting extension of \textsc{Duo} to allow for predissociation due to curve crossing will be reported elsewhere and line lists for molecules such as OH, for which this mechanism is important, will presented in this journal in due course.

\section*{Acknowledgements}

This work was supported by the European Research Council (ERC) under the European Union’s Horizon 2020 research and innovation programme through Advance Grant number 883830 and the STFC Projects No. ST/M001334/1 and ST/R000476/1. The authors acknowledge the use of the Cambridge Service for Data Driven Discovery (CSD3) as part of the STFC DiRAC HPC Facility (www.dirac.ac.uk),  funded by BEIS capital funding via STFC capital grants ST/P002307/1 and ST/R002452/1 and STFC operations grant ST/R00689X/1. WSz and RH thank European Regional Development Fund and the Polish state budget within the framework of the Carpathian Regional Operational
Programme (RPPK.01.03.00-18-001/10) through the funding of the Center for Innovation and Transfer of Natural Sciences and Engineering Knowledge of the University of Rzesz\'{o}w.
YP's work has been carried out in the framework of the MSCA4Ukraine program, Project Number: 1.4-UKR–1233448-MSCA4Ukraine.
\section*{Data Availability}

The states, transition, photo-absorption and partition function files for AlH/AlD \name\ line lists can be downloaded from \href{https://exomol.com}{www.exomol.com}.  The open access programs \Duo, \textsc{ExoCross} and \textsc{pyExoCross} are available from \href{https://github.com/exomol}{github.com/exomol}.

\section*{Supporting Information}

Supplementary data are available at MNRAS online. This includes  (i) the spectroscopic model in the form of the \Duo\ input file, containing all the curves, and parameters; (ii) the MARVEL input and output files.







\appendix

\section{Experimental measurements of AlD in the current work}

All line positions of the \AX\ system of AlD measured and analysed in this work are listed in Tables~\ref{t:FTSwavenumbers1} and \ref{t:FTSwavenumbers2}.

\begin{table*}
\centering
\caption{Measured wavenumbers (in \cm) of the FT-VIS \AX\ emission bands of AlD.} \label{t:FTSwavenumbers1}
\setlength\tabcolsep{4pt}
\begin{threeparttable}
\begin{tabular}{llrlrlrp{0.3cm}lrlrlr}\hline \hline
&\multicolumn{6}{c}{$0-0$ band}&&\multicolumn{6}{c}{$0-1$ band}\\\cline{2-7}\cline{9-14}
$J$&$R_{11ee}$&U\textsuperscript{a}&$Q_{11fe}$&U&$P_{11ee}$&U&&$R_{11ee}$&U&$Q_{11fe}$&U&$P_{11ee}$&U\\\hline
0	&	23543.1725	&	0.0020	&		&		&		&		&	&	22361.2411	&	0.0180	&		&		&		&		\\
1	&	23549.2561	&	0.0020	&	23536.6012	&	0.0020	&		&		&	&	22367.4555	&	0.0157	&	22354.8106	&	0.0121	&		&		\\
2	&	23555.0911	&	0.0020	&	23536.1114	&	0.0020	&	23523.4745	&	0.0020	&	&	22373.5647	&	0.0039	&	22354.5859	&	0.0079	&		&		\\
3	&	23560.6704	&	0.0020	&	23535.3732	&	0.0020	&	23516.4348	&	0.0020	&	&	22379.5504	&	0.0068	&	22354.2639	&	0.0053	&	22335.3213	&	0.0160	\\
4	&	23565.9859	&	0.0020	&	23534.3818	&	0.0020	&	23509.1571	&	0.0020	&	&	22385.4252	&	0.0060	&	22353.8176	&	0.0031	&	22328.6007	&	0.0106	\\
5	&	23571.0270	&	0.0020	&	23533.1314	&	0.0020	&	23501.6407	&	0.0020	&	&	22391.1454	&	0.0063	&	22353.2547	&	0.0026	&	22321.7663	&	0.0102	\\
6	&	23575.7824	&	0.0020	&	23531.6144	&	0.0020	&	23493.8816	&	0.0020	&	&	22396.7321	&	0.0046	&	22352.5649	&	0.0026	&	22314.8351	&	0.0101	\\
7	&	23580.2386	&	0.0020	&	23529.8216	&	0.0020	&	23485.8744	&	0.0020	&	&	22402.1481	&	0.0033	&	22351.7356	&	0.0024	&	22307.7851	&	0.0044	\\
8	&	23584.3810	&	0.0020	&	23527.7420	&	0.0020	&	23477.6122	&	0.0020	&	&	22407.3923	&	0.0032	&	22350.7526	&	0.0025	&	22300.6232	&	0.0037	\\
9	&	23588.1926	&	0.0020	&	23525.3631	&	0.0020	&	23469.0856	&	0.0020	&	&	22412.4384	&	0.0031	&	22349.6106	&	0.0024	&	22293.3336	&	0.0037	\\
10	&	23591.6546	&	0.0020	&	23522.6705	&	0.0020	&	23460.2861	&	0.0020	&	&	22417.2771	&	0.0029	&	22348.2915	&	0.0024	&	22285.9117	&	0.0034	\\
11	&	23594.7464	&	0.0020	&	23519.6477	&	0.0020	&	23451.2001	&	0.0020	&	&	22421.8739	&	0.0031	&	22346.7796	&	0.0025	&	22278.3248	&	0.0046	\\
12	&	23597.4446	&	0.0020	&	23516.2764	&	0.0020	&	23441.8137	&	0.0020	&	&	22426.2203	&	0.0031	&	22345.0507	&	0.0025	&	22270.5947	&	0.0038	\\
13	&	23599.7244	&	0.0020	&	23512.5361	&	0.0020	&	23432.1105	&	0.0020	&	&	22430.2777	&	0.0066	&	22343.0943	&	0.0024	&	22262.6706	&	0.0040	\\
14	&	23601.5580	&	0.0020	&	23508.4040	&	0.0020	&	23422.0726	&	0.0020	&	&	22434.0227	&	0.0073	&	22340.8751	&	0.0026	&	22254.5458	&	0.0034	\\
15	&	23602.9138	&	0.0020	&	23503.8547	&	0.0020	&	23411.6786	&	0.0020	&	&	22437.4423	&	0.0059	&	22338.3770	&	0.0026	&	22246.2046	&	0.0039	\\
16	&	23603.7675	&	0.0020	&	23498.8601	&	0.0020	&	23400.9055	&	0.0020	&	&	22440.4689	&	0.0031	&	22335.5645	&	0.0027	&	22237.6112	&	0.0034	\\
17	&	23604.0619	&	0.0020	&	23493.3893	&	0.0020	&	23389.7276	&	0.0020	&	&	22443.0861	&	0.0040	&	22332.4169	&	0.0026	&	22228.7559	&	0.0094	\\
18	&	23603.7675	&	0.0020	&	23487.4084	&	0.0020	&	23378.1159	&	0.0020	&	&	22445.2537	&	0.0039	&	22328.8839	&	0.0029	&	22219.5972	&	0.0085	\\
19	&	23602.8566	&	0.0020	&	23480.8792	&	0.0020	&	23366.0373	&	0.0020	&	&	22446.9232	&	0.0073	&	22324.9370	&	0.0030	&	22210.0996	&	0.0085	\\
20	&	23601.2620	&	0.0020	&	23473.7603	&	0.0020	&	23353.4571	&	0.0020	&	&	22448.0409	&	0.0096	&	22320.5313	&	0.0030	&	22200.2260	&	0.0145	\\
21	&	23598.9328	&	0.0020	&	23466.0048	&	0.0020	&	23340.3349	&	0.0020	&	&	22448.5557	&	0.0055	&	22315.6277	&	0.0032	&	22189.9568	&	0.0132	\\
22	&	23595.8103	&	0.0020	&	23457.5623	&	0.0020	&	23326.6256	&	0.0020	&	&	22448.4164	&	0.0053	&	22310.1640	&	0.0033	&	22179.2345	&	0.0089	\\
23	&	23591.8306	&	0.0021	&	23448.3754	&	0.0020	&	23312.2818	&	0.0020	&	&	22447.5398	&	0.0091	&	22304.0972	&	0.0063	&	22168.0097	&	0.0125	\\
24	&	23586.9164	&	0.0020	&	23438.3795	&	0.0020	&	23297.2458	&	0.0020	&	&	22445.8777	&	0.0184	&	22297.3242	&	0.0071	&	22156.1913	&	0.0051	\\
25	&	23580.9870	&	0.0020	&	23427.5037	&	0.0020	&	23281.4570	&	0.0020	&	&	22443.3030	&	0.0072	&	22289.8293	&	0.0057	&	22143.7752	&	0.0089	\\
26	&	23573.9448	&	0.0021	&	23415.6658	&	0.0020	&	23264.8441	&	0.0020	&	&	22439.7519	&	0.0122	&	22281.4787	&	0.0074	&	22130.6553	&	0.0186	\\
27	&	23565.6808	&	0.0032	&	23402.7726	&	0.0020	&	23247.3262	&	0.0020	&	&		&		&	22272.2070	&	0.0087	&	22116.7568	&	0.0126	\\
28	&		&		&	23388.7169	&	0.0022	&	23228.8127	&	0.0024	&	&		&		&	22261.9195	&	0.0100	&		&		\\
29	&		&		&	23373.3745	&	0.0040	&	23209.1996	&	0.0034	&	&		&		&		&		&		&		\\
30	&		&		&		&		&	23188.3652	&	0.0124	&	&		&		&		&		&		&		\\
\hline
&\multicolumn{6}{c}{$0-2$ band}&&\multicolumn{6}{c}{$1-0$ band}\\\cline{2-7}\cline{9-14}
$J$&$R_{11ee}$&U&$Q_{11fe}$&U&$P_{11ee}$&U&&$R_{11ee}$&U&$Q_{11fe}$&U&$P_{11ee}$&U\\\hline
0	&	21208.5898	&	0.0199	&		&		&		&		&	&	24385.9843	&	0.0023	&		&		&		&		\\
1	&	21214.9609	&	0.0120	&	21202.2910	&	0.0112	&		&		&	&	24391.1367	&	0.0022	&	24379.4111	&	0.0025	&		&		\\
2	&	21221.3308	&	0.0196	&	21202.3500	&	0.0076	&		&		&	&	24395.5730	&	0.0021	&	24377.9932	&	0.0021	&	24366.2799	&	0.0055	\\
3	&	21227.7368	&	0.0064	&	21202.4477	&	0.0099	&	21183.5021	&	0.0155	&	&	24399.2832	&	0.0021	&	24375.8584	&	0.0020	&	24358.3138	&	0.0023	\\
4	&	21234.1510	&	0.0061	&	21202.5464	&	0.0071	&	21177.3190	&	0.0202	&	&	24402.2491	&	0.0021	&	24372.9981	&	0.0021	&	24349.6382	&	0.0022	\\
5	&	21240.5634	&	0.0061	&	21202.6542	&	0.0056	&	21171.1729	&	0.0099	&	&	24404.4563	&	0.0020	&	24369.4008	&	0.0020	&	24340.2537	&	0.0021	\\
6	&	21246.9563	&	0.0051	&	21202.7825	&	0.0045	&	21165.0598	&	0.0098	&	&	24405.8864	&	0.0020	&	24365.0538	&	0.0020	&	24330.1441	&	0.0021	\\
7	&	21253.3244	&	0.0076	&	21202.9054	&	0.0046	&	21158.9568	&	0.0081	&	&	24406.5115	&	0.0022	&	24359.9378	&	0.0020	&	24319.3039	&	0.0021	\\
8	&	21259.6560	&	0.0079	&	21203.0005	&	0.0041	&	21152.8712	&	0.0091	&	&	24406.3075	&	0.0022	&	24354.0327	&	0.0020	&	24307.7165	&	0.0021	\\
9	&	21265.9056	&	0.0084	&	21203.0863	&	0.0042	&	21146.8047	&	0.0075	&	&	24405.2421	&	0.0020	&	24347.3117	&	0.0020	&	24295.3607	&	0.0021	\\
10	&	21272.0912	&	0.0114	&	21203.1094	&	0.0041	&	21140.7412	&	0.0081	&	&	24403.2777	&	0.0020	&	24339.7470	&	0.0020	&	24282.2114	&	0.0021	\\
11	&	21278.1915	&	0.0089	&	21203.0863	&	0.0041	&	21134.6487	&	0.0080	&	&	24400.3761	&	0.0021	&	24331.3041	&	0.0020	&	24268.2494	&	0.0021	\\
12	&	21284.1445	&	0.0080	&	21203.0005	&	0.0039	&	21128.5228	&	0.0092	&	&	24396.4878	&	0.0021	&	24321.9445	&	0.0020	&	24253.4368	&	0.0021	\\
13	&	21289.9670	&	0.0087	&	21202.7825	&	0.0040	&	21122.3481	&	0.0113	&	&	24391.5630	&	0.0021	&	24311.6260	&	0.0020	&	24237.7395	&	0.0021	\\
14	&	21295.6024	&	0.0090	&	21202.4477	&	0.0057	&	21116.1238	&	0.0125	&	&	24385.5386	&	0.0021	&	24300.2968	&	0.0020	&	24221.1151	&	0.0021	\\
15	&	21301.0351	&	0.0097	&	21201.9697	&	0.0040	&	21109.7989	&	0.0088	&	&	24378.3479	&	0.0025	&	24287.8988	&	0.0020	&	24203.5132	&	0.0021	\\
16	&	21306.2260	&	0.0121	&		&		&	21103.3633	&	0.0190	&	&	24369.9110	&	0.0023	&	24274.3660	&	0.0020	&	24184.8866	&	0.0025	\\
17	&	21311.1170	&	0.0166	&	21200.4557	&	0.0073	&	21096.8044	&	0.0146	&	&	24360.1346	&	0.0043	&	24259.6214	&	0.0021	&	24165.1587	&	0.0022	\\
18	&	21315.7031	&	0.0267	&	21199.3460	&	0.0088	&	21090.0587	&	0.0179	&	&		&		&	24243.5783	&	0.0031	&	24144.2642	&	0.0023	\\
19	&		&		&	21197.9452	&	0.0096	&		&		&	&		&		&	24226.1269	&	0.0085	&	24122.1082	&	0.0036	\\
20	&		&		&	21196.2229	&	0.0098	&	21075.9152	&	0.0108	&	&		&		&		&		&	24098.5986	&	0.0222	\\
21	&		&		&	21194.1315	&	0.0184	&	21068.4506	&	0.0389	&	&		&		&		&		&		&		\\
22	&		&		&	21191.6049	&	0.0095	&	21060.6570	&	0.0376	&	&		&		&		&		&		&		\\
\hline \hline
\end{tabular}
\begin{tablenotes}
\item[a] The total uncertainty of the measured spectral line position represents $1\sigma$ standard deviation being combinations of calibration ($\textrm{U}_{\textrm{cal.}}$) and fitting ($\textrm{U}_{\textrm{fitt.}}$) uncertainty (see Section~\ref{experimental}).
\end{tablenotes}
\end{threeparttable}
\end{table*}

\begin{table*}
\centering
\caption{Measured wavenumbers (in \cm) of the FT-VIS \AX\ system emission bands of AlD.} \label{t:FTSwavenumbers2}
\setlength\tabcolsep{4pt}
\begin{threeparttable}
\begin{tabular}{llrlrlrp{0.3cm}lrlrlr}\hline \hline
&\multicolumn{6}{c}{$1-1$ band}&&\multicolumn{6}{c}{$1-2$ band}\\\cline{2-7}\cline{9-14}
$J$&$R_{11ee}$&U\textsuperscript{a}&$Q_{11fe}$&U&$P_{11ee}$&U&&$R_{11ee}$&U&$Q_{11fe}$&U&$P_{11ee}$&U\\\hline

0	&	23204.0406	&	0.0020	&		&		&		&		&	&	22051.4089	&	0.0027	&		&		&		&		\\
1	&	23209.3323	&	0.0020	&	23197.6078	&	0.0020	&		&		&	&	22056.8388	&	0.0024	&	22045.1111	&	0.0024	&		&		\\
2	&	23214.0449	&	0.0020	&	23196.4648	&	0.0020	&	23184.7558	&	0.0020	&	&	22061.8207	&	0.0023	&	22044.2372	&	0.0022	&	22032.5355	&	0.0084	\\
3	&	23218.1669	&	0.0020	&	23194.7429	&	0.0020	&	23177.1996	&	0.0020	&	&	22066.3481	&	0.0022	&	22042.9250	&	0.0024	&	22025.3793	&	0.0029	\\
4	&	23221.6844	&	0.0020	&	23192.4338	&	0.0020	&	23169.0750	&	0.0020	&	&	22070.4091	&	0.0022	&	22041.1598	&	0.0021	&	22017.8040	&	0.0026	\\
5	&	23224.5810	&	0.0020	&	23189.5258	&	0.0020	&	23160.3763	&	0.0020	&	&	22073.9861	&	0.0021	&	22038.9311	&	0.0021	&	22009.7850	&	0.0023	\\
6	&	23226.8356	&	0.0020	&	23186.0044	&	0.0020	&	23151.0945	&	0.0020	&	&	22077.0550	&	0.0021	&	22036.2232	&	0.0020	&	22001.3135	&	0.0022	\\
7	&	23228.4244	&	0.0020	&	23181.8512	&	0.0020	&	23141.2173	&	0.0020	&	&	22079.5926	&	0.0022	&	22033.0196	&	0.0020	&	21992.3866	&	0.0022	\\
8	&	23229.3201	&	0.0020	&	23177.0452	&	0.0020	&	23130.7283	&	0.0020	&	&	22081.5729	&	0.0024	&	22029.2974	&	0.0020	&	21982.9819	&	0.0022	\\
9	&	23229.4909	&	0.0020	&	23171.5608	&	0.0020	&	23119.6088	&	0.0020	&	&	22082.9635	&	0.0024	&	22025.0330	&	0.0020	&	21973.0815	&	0.0022	\\
10	&	23228.8999	&	0.0020	&	23165.3687	&	0.0020	&	23107.8351	&	0.0020	&	&	22083.7253	&	0.0024	&	22020.1941	&	0.0020	&	21962.6603	&	0.0022	\\
11	&	23227.5069	&	0.0020	&	23158.4351	&	0.0020	&	23095.3810	&	0.0020	&	&	22083.8211	&	0.0025	&	22014.7501	&	0.0020	&	21951.6930	&	0.0022	\\
12	&	23225.2645	&	0.0020	&	23150.7212	&	0.0020	&	23082.2131	&	0.0020	&	&	22083.2013	&	0.0025	&	22008.6575	&	0.0020	&	21940.1498	&	0.0022	\\
13	&	23222.1195	&	0.0020	&	23142.1829	&	0.0020	&	23068.2971	&	0.0020	&	&	22081.8091	&	0.0025	&	22001.8739	&	0.0020	&	21927.9885	&	0.0022	\\
14	&	23218.0113	&	0.0020	&	23132.7691	&	0.0020	&	23053.5877	&	0.0020	&	&	22079.5926	&	0.0022	&	21994.3491	&	0.0021	&	21915.1677	&	0.0022	\\
15	&	23212.8714	&	0.0020	&	23122.4219	&	0.0020	&	23038.0387	&	0.0020	&	&	22076.4704	&	0.0031	&	21986.0223	&	0.0021	&	21901.6402	&	0.0023	\\
16	&	23206.6188	&	0.0020	&	23111.0739	&	0.0020	&	23021.5926	&	0.0020	&	&	22072.3706	&	0.0025	&	21976.8284	&	0.0021	&	21887.3465	&	0.0023	\\
17	&	23199.1595	&	0.0021	&	23098.6486	&	0.0020	&	23004.1868	&	0.0020	&	&	22067.2019	&	0.0077	&	21966.6867	&	0.0024	&	21872.2285	&	0.0025	\\
18	&	23190.3816	&	0.0069	&	23085.0532	&	0.0020	&	22985.7426	&	0.0020	&	&		&		&	&		&	21856.2001	&	0.0026	\\
19	&		&		&	23070.1815	&	0.0034	&	22966.1715	&	0.0021	&	&		&		&	21943.1912	&	0.0118	&	21839.1697	&	0.0102	\\
20	&		&		&		&		&	22945.3652	&	0.0035	&	&		&		&		&		&		&		\\

\hline
&\multicolumn{6}{c}{$1-3$ band}&&\multicolumn{6}{c}{$1-4$ band}\\\cline{2-7}\cline{9-14}
$J$&$R_{11ee}$&U&$Q_{11fe}$&U&$P_{11ee}$&U&&$R_{11ee}$&U&$Q_{11fe}$&U&$P_{11ee}$&U\\\hline

0	&	20927.5626	&	0.0047	&		&		&		&		&	&		&		&		&		&		&		\\
1	&	20933.1199	&	0.0028	&	20921.3941	&	0.0056	&		&		&	&		&		&		&		&		&		\\
2	&	20938.3691	&	0.0026	&	20920.7904	&	0.0027	&	20909.0946	&	0.0112	&	&		&		&		&		&		&		\\
3	&	20943.3023	&	0.0025	&	20919.8775	&	0.0023	&	20902.3411	&	0.0058	&	&		&		&		&		&		&		\\
4	&	20947.8985	&	0.0024	&	20918.6482	&	0.0022	&	20895.2923	&	0.0063	&	&		&		&		&		&		&		\\
5	&	20952.1415	&	0.0024	&	20917.0866	&	0.0022	&	20887.9357	&	0.0050	&	&		&		&		&		&		&		\\
6	&	20956.0146	&	0.0025	&	20915.1835	&	0.0021	&	20880.2755	&	0.0044	&	&		&		&	19822.3685	&	0.0048	&	19787.4529	&	0.0118	\\
7	&	20959.4878	&	0.0034	&	20912.9169	&	0.0024	&	20872.2813	&	0.0024	&	&	19867.5893	&	0.0107	&	19821.0272	&	0.0148	&	19780.3816	&	0.0169	\\
8	&	20962.5383	&	0.0032	&	20910.2628	&	0.0021	&	20863.9464	&	0.0025	&	&		&		&	19819.4272	&	0.0148	&	19773.1160	&	0.0131	\\
9	&	20965.1324	&	0.0024	&	20907.2000	&	0.0021	&	20855.2482	&	0.0025	&	&	19875.4830	&	0.0279	&	19817.5444	&	0.0114	&		&		\\
10	&	20967.2216	&	0.0024	&	20903.6962	&	0.0021	&	20846.1630	&	0.0024	&	&	19878.8839	&	0.0179	&	19815.3656	&	0.0089	&	19757.8424	&	0.0129	\\
11	&	20968.7881	&	0.0037	&	20899.7164	&	0.0021	&	20836.6618	&	0.0025	&	&	19881.8952	&	0.0121	&	19812.8312	&	0.0090	&	19749.7818	&	0.0277	\\
12	&	20969.7597	&	0.0046	&	20895.2263	&	0.0026	&	20826.7151	&	0.0025	&	&	19884.4520	&	0.0113	&	19809.9118	&	0.0079	&		&		\\
13	&	20970.1066	&	0.0026	&	20890.1706	&	0.0026	&	20816.2897	&	0.0025	&	&	19886.4971	&	0.0129	&	19806.5742	&	0.0096	&		&		\\
14	&	20969.7597	&	0.0046	&	20884.5071	&	0.0022	&	20805.3283	&	0.0028	&	&	19887.9844	&	0.0118	&	19802.7368	&	0.0100	&		&		\\
15	&	20968.6193	&	0.0057	&	20878.1740	&	0.0022	&	20793.7954	&	0.0043	&	&	19888.8163	&	0.0192	&	19798.3660	&	0.0059	&		&		\\
16	&	20966.6475	&	0.0088	&	20871.1039	&	0.0024	&	20781.6220	&	0.0053	&	&	19888.9323	&	0.0279	&		&		&		&		\\
17	&	20963.7347	&	0.0097	&	20863.2127	&	0.0041	&	20768.7542	&	0.0057	&	&		&		&		&		&		&		\\
18	&		&		&	20854.4204	&	0.0083	&	20755.1036	&	0.0043	&	&		&		&		&		&		&		\\
19	&		&		&		&		&	20740.6037	&	0.0104	&	&		&		&		&		&		&		\\

\hline
&\multicolumn{6}{c}{$2-1$ band\textsuperscript{b}}&&\multicolumn{6}{c}{$2-2$ band\textsuperscript{b}}\\\cline{2-7}\cline{9-14}
$J$&$R_{11ee}$&U&$Q_{11fe}$&U&$P_{11ee}$&U&&$R_{11ee}$&U&$Q_{11fe}$&U&$P_{11ee}$&U\\\hline

0	&	23874.0316	&	0.0077	&		&		&		&		&	&	22721.3902	&	0.0092	&		&		&		&		\\
1	&	23877.8617	&	0.0092	&	23867.6020	&	0.0056	&		&		&	&	22725.3632	&	0.0112	&	23715.0966	&	0.0048	&		&		\\
2	&		&		&	23864.9920	&	0.0066	&	23854.7460	&	0.0153	&	&	22728.1337	&	0.0110	&	23712.7677	&	0.0024	&	22702.5202	&	0.0066	\\
3	&	23881.5567	&	0.0066	&	23861.0880	&	0.0106	&	23845.7191	&	0.0188	&	&	22729.7348	&	0.0072	&	23709.2548	&	0.0121	&	22693.9095	&	0.0125	\\
4	&		&		&	23855.8209	&	0.0050	&		&		&	&		&		&	23704.5520	&	0.0060	&	22684.1152	&	0.0101	\\
5	&		&		&		&		&	23823.7418	&	0.0092	&	&		&		&		&		&		&		\\

\hline \hline
\end{tabular}
\begin{tablenotes}
\item[a] The total uncertainty of the measured spectral line position represents $1\sigma$ standard deviation being combinations of calibration ($\textrm{U}_{\textrm{cal.}}$) and fitting ($\textrm{U}_{\textrm{fitt.}}$) uncertainty (see Section~\ref{experimental}).
\item[b] Bands of the $2-v''$ progression are sharply cut off in the intensity of the rotational lines due to the predissociation at \A~$(v=2, J=4)$ level.
\end{tablenotes}
\end{threeparttable}
\end{table*}

\section{Extracting fine line positions  from hyperfine transitions  for pure rotational lines in AlH and AlD \XS\ state}\label{line positions  from hyperfine transitions}

In order to estimate the rotational fine structure from a collection of experimental hyperfine transitions of pure rotational lines of AlH,
the following relationship for the total hyperfine energy of AlH was assumed, previously derived by \citet{84GoCo} and shown by \citet{01GeWaxx.AlH}:
\begin{equation}
    E_{HF}=E_{Q}(Al)+E_{SR}(Al)+E_{SR}(H)
\end{equation}
where $E_{HF}$ is total hyperfine energy, $E_{Q}(Al)$ is nuclear quadrupole interaction energy for Aluminium atom and $E_{SR}$ is the spin-rotation interaction energy for both the $^{27}\text{Al}$ and $^{1}\text{H}$ (or D).

Nuclear quadrupole interaction is defined as
\begin{equation}
E_{Q}=-eQqY(J,I,F)    ,
\end{equation}
where is $e$ is electron charge, $Q$ is a nuclear quadrupole moment, $q$ is the electric field gradient and Casimer function is
\begin{equation}
\label{e:casimer_factor}
Y(J,I,F)=\frac{\frac{3}{4}C(C+1)-I(I+1)J(J+1)}{2(2J-1)(2J+3)I(2I-1)}
\end{equation}
with
\begin{equation}
\label{e:c}
C=F(F+1)-J(J+1)-I(I+1) .
\end{equation}

The energy of the spin-rotation interaction is defined as $E_{SR}=C_{\perp}C$, where $C$ is from Eq.~\eqref{e:c} and $C_{\perp}$ is nuclear magnetic coupling constant. It has been shown that $C_{\perp}(^{1}H)$ is too small (on the scale of 10~kHz \cite{01GeWaxx.AlH}) to perturb the spectra and hence is ignored in our calculations.
From the above definitions we can derive the following relationship:
\begin{equation}
\label{e:purerot}
    \nu=\nu_0-eQq[Y'(J',I,F')-Y''(J'',I,F'')]+C_{\perp}[C'-C'']
\end{equation}
where $Y'$ and $C'$ represent the final state energy, $Y''$ and $C''$ represent the initial state energy in a hyperfine transition and $\nu_0$ is the ``unperturbed'' rotational transition frequency if there were no quadrupole and spin-rotation interactions.

After solving for $\nu_0$ in all hyperfine transitions, the ``true'' $\nu_0$ is found by the method of weighted averages:
\begin{equation}
    \nu_{0\text{,weighted}}=\frac{\sum_{i}{\nu_{0,i} w_i}}{\sum_{i}{w_i}}
\end{equation}
where $\nu_{0,i}$ is individually derived rotational transition frequency using Eq.~\eqref{e:purerot} and $w_i=\frac{1}{\sigma_{0}^{2}}$ with $\sigma_{0}$ as the propagated standard deviation for each transition. A standard Gaussian error propagation was used to obtain the value of $\sigma_{0}$, with the overall expression being the following:
\begin{equation}
    \sigma_0=\sqrt{\sigma_{\text{orig}}^2+((Y'-Y'')*\sigma_{eQq})^2+((C'-C'') \sigma_{C_\perp})^2},
\end{equation}
where $\sigma_{\rm orig}$ is the originally reported uncertainty in hyperfine measurement, $\sigma_{eQq}$ is the uncertainty in $eQq$ and $\sigma_{C_\perp}$ is the uncertainty in the $C_\perp(^{27}\text{Al})$. The values for $eQq$ and $C_\perp$ used in our calculations and their respective uncertainties can be in Table \ref{t:hyperfine_spectroscopic_constants}.

\begin{table}
\label{t:hyperfine_spectroscopic_constants}
\caption{Spectroscopic constant values and uncertainties that were used to derive pure rotational transition from hyperfine data.}
\begin{tabular}{l|lll|lll}\hline \hline
&\multicolumn{3}{l|}{27AlH}& \multicolumn{3}{l}{27AlD} \\
\hline
 Parameter & value (MHz) & $\sigma$ (MHz) & source  & value (MHz) & $\sigma$ (MHz) & source \\ \hline
$eQq$  &-48.61& 0.70 & \cite{16HaZixx.AlH}&-48.69 & 0.36 & \cite{14HaZixx.AlH} \\
$C_\perp$ & 0.298  & 0.035  & \cite{16HaZixx.AlH}& 0.108& 0.022& \cite{14HaZixx.AlH}\\
\hline
\end{tabular}
\end{table}

\section{Astrophysical measurements of AlH }

\begin{figure*}
    \centering
\includegraphics[width=\textwidth]{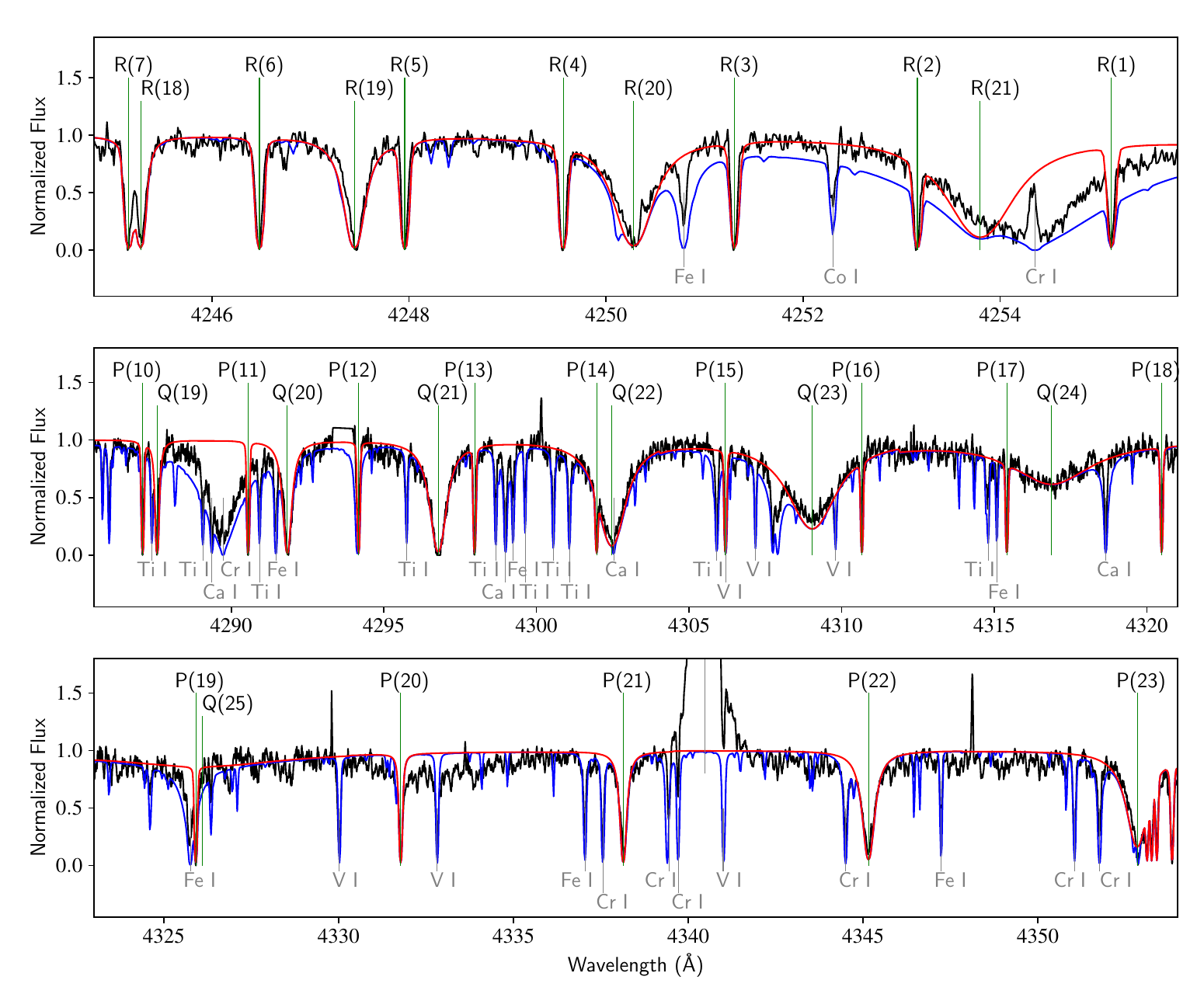}
\caption{The same as in Fig. \ref{f:prox:1}, but with atomic lines indicates:
Comparison with the observed spectrum of AlH of Proxima Cen by \citet{jt874} shown by black line for the (0-0) band spectral range.
Here and onward, we have adhered to the procedure of computation and identification of spectral features outlined by \citet{jt874}. The blue line marks the synthetic spectrum including atomic and molecular species, the red line spectrum is calculated including AlH lines only.
}
    \label{f:prox:3}
\end{figure*}

As part of this work, by using an output from the updated \Duo\ model for AlH, a number of new ro-vibronic transitions in the Proxima Cen spectrum
from the (1-1) \AX\ band of AlH have been identified, following the same procedure as described previously by \citet{jt874}. The remeasured line positions can be seen in Table \ref{tab:Proxima_cen_AlH}. Figure \ref{f:prox:3} shows the Proxima Cen spectrum, with comparison to the calculated AlH spectrum and nearby atomic lines.

\begin{table*}
\caption{Re-measured astrophysical line positions in nm and \cm from the Proxima Cen. spectrum observed by \citet{jt874} for the \AX\ system.}
\label{tab:Proxima_cen_AlH}
\begin{tabular}{c|c|c|cc|cc|cc}\hline \hline
Band & Branch & J & $\lambda$ nm & Unc. nm & $\nu$ \cm & Unc. \cm\ & Width \AA & Unc. \AA \\
\hline
1-0 & Q & 2 & 4072.572 & 0.003 & 24547.576 & 0.018 & 0.024 & 0.002 \\
1-0 & Q & 4 & 4074.981 & 0.003 & 24533.064 & 0.018 & 0.022 & 0.003 \\
1-0 & Q & 7 & 4081.455 & 0.003 & 24494.151 & 0.018 & 0.032 & 0.002 \\
1-0 & Q & 8 & 4084.468 & 0.003 & 24476.083 & 0.018 & 0.041 & 0.002 \\
1-0 & Q & 9 & 4087.974 & 0.050 & 24455.091 & 0.299 & 0.081 & 0.028 \\
1-0 & Q & 10 & 4092.080 & 0.060 & 24430.554 & 0.358 & 0.132 & 0.008 \\
1-0 & Q & 11 & 4096.721 & 0.030 & 24402.878 & 0.179 & 0.281 & 0.007 \\
1-0 & Q & 13 & 4108.593 & 0.150 & 24332.366 & 0.888 & 0.766 & 0.037 \\
1-0 & P & 4 & 4081.951 & 0.003 & 24491.175 & 0.018 & 0.021 & 0.002 \\
1-0 & P & 5 & 4085.420 & 0.003 & 24470.379 & 0.018 & 0.021 & 0.002 \\
1-0 & P & 6 & 4089.261 & 0.003 & 24447.395 & 0.018 & 0.023 & 0.002 \\
1-0 & P & 7 & 4093.490 & 0.003 & 24422.139 & 0.018 & 0.021 & 0.002 \\
1-0 & P & 8 & 4098.126 & 0.003 & 24394.512 & 0.018 & 0.023 & 0.001 \\
1-0 & P & 9 & 4103.228 & 0.024 & 24364.180 & 0.143 & 0.035 & 0.002 \\
1-0 & P & 10 & 4108.813 & 0.003 & 24331.063 & 0.018 & 0.045 & 0.003 \\
1-0 & P & 11 & 4114.950 & 0.006 & 24294.777 & 0.035 & 0.113 & 0.007 \\
1-0 & P & 12 & 4121.753 & 0.006 & 24254.679 & 0.035 & 0.232 & 0.010 \\
1-0 & P & 13 & 4129.276 & 0.080 & 24210.491 & 0.469 & 0.355 & 0.019 \\
1-0 & R & 0 & 4069.800 & 0.003 & 24564.295 & 0.018 & 0.019 & 0.002 \\
1-0 & R & 1 & 4068.391 & 0.003 & 24572.802 & 0.018 & 0.021 & 0.002 \\
1-0 & R & 3 & 4066.623 & 0.003 & 24583.485 & 0.018 & 0.024 & 0.002 \\
1-0 & R & 6 & 4066.830 & 0.003 & 24582.234 & 0.018 & 0.022 & 0.002 \\
1-0 & R & 7 & 4067.757 & 0.003 & 24576.632 & 0.018 & 0.043 & 0.002 \\
1-0 & R & 8 & 4069.190 & 0.004 & 24567.978 & 0.024 & 0.072 & 0.004 \\
1-0 & R & 9 & 4071.185 & 0.003 & 24555.939 & 0.018 & 0.104 & 0.003 \\
1-0 & R & 10 & 4073.747 & 0.080 & 24540.496 & 0.482 & 0.246 & 0.008 \\
1-0 & R & 11 & 4077.140 & 0.010 & 24520.074 & 0.060 & 0.375 & 0.019 \\
1-0 & R & 12 & 4081.352 & 0.034 & 24494.769 & 0.204 & 0.504 & 0.045 \\
0-0 & R & 3 & 4251.300 & 0.003 & 23515.596 & 0.017 & 0.022 & 0.002 \\
0-0 & R & 4 & 4249.562 & 0.003 & 23525.214 & 0.017 & 0.028 & 0.002 \\
0-0 & R & 5 & 4247.955 & 0.003 & 23534.113 & 0.017 & 0.023 & 0.002 \\
0-0 & R & 6 & 4246.477 & 0.003 & 23542.304 & 0.017 & 0.028 & 0.002 \\
0-0 & R & 9 & 4242.978 & 0.003 & 23561.718 & 0.017 & 0.028 & 0.001 \\
0-0 & R & 16 & 4242.435 & 0.003 & 23564.734 & 0.017 & 0.025 & 0.002 \\
0-0 & R & 17 & 4243.632 & 0.003 & 23558.087 & 0.017 & 0.027 & 0.002 \\
0-0 & R & 18 & 4245.278 & 0.003 & 23548.953 & 0.017 & 0.045 & 0.002 \\
0-0 & R & 19 & 4247.462 & 0.010 & 23536.845 & 0.055 & 0.076 & 0.002 \\
0-0 & R & 20 & 4250.273 & 0.004 & 23521.278 & 0.022 & 0.172 & 0.004 \\
0-0 & Q & 9 & 4264.448 & 0.003 & 23443.095 & 0.016 & 0.040 & 0.003 \\
0-0 & Q & 10 & 4265.689 & 0.003 & 23436.275 & 0.016 & 0.034 & 0.002 \\
0-0 & Q & 11 & 4267.102 & 0.003 & 23428.515 & 0.016 & 0.032 & 0.002 \\
0-0 & Q & 13 & 4270.533 & 0.003 & 23409.692 & 0.016 & 0.031 & 0.001 \\
0-0 & Q & 14 & 4272.590 & 0.003 & 23398.422 & 0.016 & 0.027 & 0.002 \\
0-0 & Q & 17 & 4280.475 & 0.003 & 23355.321 & 0.016 & 0.032 & 0.002 \\
0-0 & Q & 18 & 4283.797 & 0.003 & 23337.210 & 0.016 & 0.042 & 0.002 \\
0-0 & Q & 19 & 4287.564 & 0.003 & 23316.706 & 0.016 & 0.061 & 0.003 \\
0-0 & Q & 20 & 4291.854 & 0.008 & 23293.400 & 0.043 & 0.153 & 0.009 \\
0-0 & Q & 21 & 4296.784 & 0.004 & 23266.675 & 0.022 & 0.261 & 0.005 \\
0-0 & Q & 22 & 4302.466 & 0.006 & 23235.948 & 0.032 & 0.385 & 0.006 \\
0-0 & Q & 23 & 4309.019 & 0.009 & 23200.612 & 0.048 & 0.752 & 0.018 \\
0-0 & Q & 24 & 4316.792 & 0.150 & 23158.837 & 0.805 & 1.016 & 0.040 \\
0-0 & P & 2 & 4264.069 & 0.003 & 23445.179 & 0.016 & 0.018 & 0.002 \\
0-0 & P & 4 & 4269.154 & 0.003 & 23417.254 & 0.016 & 0.022 & 0.002 \\
0-0 & P & 16 & 4310.653 & 0.003 & 23191.818 & 0.016 & 0.026 & 0.002 \\
0-0 & P & 17 & 4315.409 & 0.003 & 23166.259 & 0.016 & 0.026 & 0.002 \\
0-0 & P & 18 & 4320.477 & 0.003 & 23139.085 & 0.016 & 0.029 & 0.002 \\
0-0 & P & 19 & 4325.911 & 0.003 & 23110.019 & 0.016 & 0.043 & 0.002 \\
0-0 & P & 20 & 4331.781 & 0.003 & 23078.704 & 0.016 & 0.063 & 0.003 \\
0-0 & P & 21 & 4338.153 & 0.003 & 23044.806 & 0.016 & 0.102 & 0.003 \\
0-0 & P & 22 & 4345.156 & 0.005 & 23007.665 & 0.026 & 0.180 & 0.005 \\
0-0 & P & 23 & 4352.994 & 0.100 & 22966.238 & 0.528 & 0.305 & 0.010 \\
\hline
\hline
\end{tabular}
\end{table*}


\bsp	
\label{lastpage}
\end{document}